\documentclass[11pt,a4paper]{article}
\usepackage[T1]{fontenc}
\usepackage[utf8]{inputenc}
\usepackage{lmodern,microtype}
\usepackage[margin=26mm]{geometry}
\usepackage{amsmath,amssymb,bm,mathtools}
\usepackage{xcolor}
\usepackage{tikz}
\usepackage{float}
\usepackage{booktabs}
\usepackage{pgfplots}
\pgfplotsset{compat=1.18}
\definecolor{aximassblue}{HTML}{244E80}
\definecolor{aximassorange}{HTML}{B54B2E}
\definecolor{aximassgreen}{HTML}{338367}
\usetikzlibrary{decorations.pathmorphing}
\usepackage[numbers,sort&compress]{natbib}
\usepackage[colorlinks=true,linkcolor=blue!55!black,citecolor=blue!55!black,urlcolor=blue!55!black]{hyperref}
\allowdisplaybreaks
\newcommand{\dd}{\mathrm d}
\newcommand{\GeV}{\mathrm{GeV}}
\newcommand{\mpl}{\overline{M}_{\mathrm{Pl}}}
\newcommand{\tr}{\operatorname{tr}}
\newcommand{\Tr}{\operatorname{Tr}}
\newcommand{\diag}{\operatorname{diag}}
\title{\large\bfseries Hidden $E_6$ Portals for Collisionless and Self-Interacting Dark Matter\\[5pt]
  {\normalsize\mdseries Vacuum Structure and the Symmetry-Protected Axi-Higgs}}
\author{Claudio Corian\`o\\[5pt]
\small Dipartimento di Matematica e Fisica ``Ennio De Giorgi'', Universit\`a del Salento,\\
\small INFN, Sezione di Lecce, and CNR NANOTEC, Lecce, Italy}
\date{}
\begin{document}
\maketitle
\begin{abstract}
We study a hidden $E_{6,D}\times U(1)_X$ gauge theory coupled to visible matter through a generalized $B-L$ portal. We investigate its large scalar sector by combining an analytic study of the Higgs alignment with numerical tests of the full scalar mass matrix. Ordinary quartic interactions preclude an isolated minimum at the proposed parallel alignment of the antisymmetric Higgs multiplets. Adding a neutral symmetric Higgs multiplet gives a bounded tree-level potential with distinct locally stable vacua. A search outside the symmetry sector of the original benchmark finds a lower-energy vacuum with a different unbroken gauge group and light fermion spectrum. The detailed spectrum, portal rates and cosmological estimates for the original benchmark are conditional on its selection and persistence. The gauge-invariant axi-Higgs combines the relative phase of the two antisymmetric Higgs fields with the St\"uckelberg field. An anomaly-free global symmetry keeps this mode massless. An independent interaction that breaks this symmetry can give it an ultralight mass. The broader programme asks whether a small self-interacting component of an effectively collisionless halo can undergo gravothermal collapse through heat transport among its own particles, potentially producing black-hole seeds for Little Red Dots. Testing this connection requires determining the populations of both components and following halo collapse and accretion. The strategy of using symmetry to locate vacuum candidates and testing their stability in the full scalar field space can be extended to other portals with large scalar representations.
\end{abstract}
\noindent\textbf{Keywords:} hidden $E_6$; scalar vacuum; Goldstone symmetry; Higgs--St\"uckelberg mixing; generalized $B-L$ portal; dark radiation.

\clearpage
\begingroup
\small
\pdfbookmark[1]{\contentsname}{front-contents}
\tableofcontents
\endgroup
\clearpage

\section{Introduction}
\label{sec:scope}
Dark matter reveals its presence through gravity. Establishing whether its particles also scatter from one another requires following how their distribution evolves. In a halo, scattering can transport heat and change the central density \cite{TulinYuReview}. The effect depends on the fraction of dark matter that scatters, on how the cross section varies with velocity, and on the halo's assembly history. Interpreting a change in the density profile also requires accounting for the gravity of gas and stars \cite{TulinYuZurek,YangYuTransport,FengYuSeeds}.

The compact red sources revealed by the James Webb Space Telescope (JWST), known as Little Red Dots, give this question a concrete astrophysical setting. Broad emission lines in some of these objects support an interpretation in terms of accreting black holes \cite{MattheeLRD}. Gas can power the observed emission once a black hole is present. Building that black hole from a light seed takes time. For example, growth from $100 M_\odot$ to $10^7 M_\odot$ takes about half a billion years of uninterrupted gas accretion at the Eddington limit for a radiative efficiency of ten per cent \cite{MadauSupercritical}. The available growth time begins at seed formation, and interruptions in the gas supply prolong the interval required. A mechanism that forms massive seeds early could shorten the subsequent growth.

The proposed dark-matter seeding mechanism requires two populations with different scattering rates \cite{PollackSeeds,RobertsEarlySeeds,RobertsLRDSeeds}. The effectively collisionless majority builds the extended halo and provides the gravitational potential in which the smaller population evolves. The smaller population must scatter often enough to transport heat among its own particles. If it forms a self-gravitating core, outward heat flow can drive gravothermal collapse and produce an early black-hole seed. A portal gives this proposal a particle-physics setting by specifying the interactions through which visible matter can produce or heat the hidden populations. It allows one to test whether the sectors shared a temperature and to constrain the radiation and relic abundances that result. A separate force within the hidden sector can set the self-scattering rate.

Here we study a hidden $E_{6,D}\times U(1)_X$ effective gauge theory. The Abelian generator acts as ordinary $B-L$ on visible fields and uniformly on each hidden multiplet. Connecting the portal to a halo calculation requires knowing which particles survive at low energies and how they interact. This starts with finding a vacuum and checking its stability throughout the large Higgs representations.

The construction generalizes Refs.~\cite{CorianoPortal,GatewayPortal}. Reference~\cite{CorianoPortal} develops an anomalous Abelian portal with a charged hidden Higgs and derives the physical phase and its interactions. Reference~\cite{GatewayPortal} studies a hidden-$SU(2)$ realization with confinement-generated curvature and conditional coherent and confined abundances. The Higgs--St\"uckelberg effective-action framework and the chiral $\mathbf{27}$ and antisymmetric $\mathbf{351}$ content used here are established in Refs.~\cite{CIK,CF,CFIT}. Our new results concern the complete gauge orbit, the obstruction to vacuum isolation for the proposed two-multiplet alignment, an explicit neutral-$\mathbf{351}^{\prime}$ scalar completion, the exact relative symmetry, the residual fermion spectrum and the production relations for the physical phase.

We use the desired residual gauge symmetry to select candidate expectation values, then check the curvature in every physical scalar direction. This exposes an instability or a continuous degeneracy for the proposed alignment of the antisymmetric Higgs multiplets. Adding a neutral symmetric multiplet gives a bounded potential and a locally stable benchmark after the prescribed phase interaction is included. The benchmark minimizes the energy within its residual symmetry sector. A search over the full field space finds a lower tree-level vacuum with a different symmetry (Sec.~\ref{sec:lower-vacuum}). The physical realization of the higher-energy branch requires cosmological selection and a sufficiently long lifetime.

An anomaly-free global rephasing is preserved by the gauge, Yukawa and ordinary scalar interactions. Its spontaneous breaking leaves an exact Nambu--Goldstone field. An ultralight mass requires an explicit symmetry-breaking interaction with an independent coefficient. The higher-energy branch leaves 69 hidden Weyl fermions massless at tree level. We determine their residual representations, gauge running and radiation density, and give the conditions on reheating for that branch. The lower vacuum gives a different light spectrum.

Section~\ref{sec:two-dirac-sector} develops a singlet extension with independently adjustable scattering rates and specifies the particle populations for an illustrative halo. Predicting seed masses, formation times or source abundances requires deriving those populations and following their subsequent halo evolution and gas accretion.

The main text presents the portal, matter content, exact symmetry, scalar vacuum and production vertices, followed by the two-component halo benchmark. The appendices give the anomaly counterterms, conditional coherent abundances and tensor calculations supporting the full vacuum analysis. Appendix~\ref{app:full-orbit} gives the exact orbit construction and the two-multiplet obstruction. The formulas for the antisymmetric multiplets describe the core construction. The completed vacuum uses the additional expectation values and mass eigenstates specified below.

\section{The generalized \texorpdfstring{$B-L$}{B-L} portal}
\label{sec:portal-definition}
The gauge group is
\begin{equation}
 G=G_{\rm SM}\times E_{6,D}\times U(1)_X,
 \qquad
 G_{\rm SM}=SU(3)_c\times SU(2)_w\times U(1)_Y.
 \label{eq:portal-group}
\end{equation}
Visible fields are singlets of $E_{6,D}$, while all hidden fields are singlets of the Standard Model. The two sectors nevertheless carry the same Abelian gauge charge $X$ according to Eq.~\eqref{eq:generalized-X}. The shared generator is
\begin{equation}
 X=(B-L)_{\rm vis}+D_{\rm hid}.
 \label{eq:generalized-X}
\end{equation}
The dark charge vanishes on visible fields, and the visible baryon and lepton numbers vanish on hidden fields. The external generator commutes with $E_{6,D}$ and has one charge on each complete hidden multiplet. It is distinct from the internal $B-L$ generator of an $SO(10)$ subgroup, which acts differently on components of a $\mathbf{27}$. We use $\hbar=c=1$ and canonical real fields of mass dimension one. The St\"uckelberg parameter $M$, the Abelian coupling $g_X$ and the hidden non-Abelian coupling $g_G$ are independent inputs. The reduced gravitational mass is $\mpl=(8\pi G)^{-1/2}\simeq2.4\times10^{18}\,\GeV$.

The ordinary Standard Model fields have their conventional $B-L$ values,
\begin{equation}
\begin{array}{c|ccccccc}
 f&Q_L&u_R&d_R&L_L&e_R&\nu_R&H_{\rm SM}\\ \hline
 X_f&\frac13&\frac13&\frac13&-1&-1&-1&0 .
\end{array}
 \label{eq:visible-BL-charges}
\end{equation}
The overall normalization of these charges can always be absorbed into $g_X$; Eq.~\eqref{eq:visible-BL-charges} fixes the convention used in the paper. There is one right-handed neutrino $\nu_R$ per family. The hidden assignments are given in Sec.~\ref{sec:GUT}.

The complete matter and scalar assignments are collected in Table~\ref{tab:portal-matter}, including the singlet extension of Sec.~\ref{sec:two-dirac-sector}. Hypercharge is normalized by $Q_{\rm em}=T_3+Y$. Right-handed fields carry their ordinary charges in the table; their left-handed conjugates used in anomaly sums carry the opposite charges. Table~\ref{tab:portal-vectors} lists the gauge connections.

\begin{table}[!htbp]
\centering
\small
\setlength{\tabcolsep}{4pt}
\renewcommand{\arraystretch}{1.18}
\begin{tabular}{@{}lccccccccc@{}}
\toprule
Field & Spin & Copies & $SU(3)_c$ & $SU(2)_w$ & $Y$
 & $E_{6,D}$ & $(B-L)_{\rm vis}$ & $D_{\rm hid}$ & $X$\\
\midrule
\multicolumn{10}{@{}l}{Visible matter and Higgs}\\[2pt]
$Q_L$ & $\tfrac12$ & $3$ & $\mathbf3$ & $\mathbf2$ & $\tfrac16$
 & $\mathbf1$ & $\tfrac13$ & $0$ & $\tfrac13$\\
$u_R$ & $\tfrac12$ & $3$ & $\mathbf3$ & $\mathbf1$ & $\tfrac23$
 & $\mathbf1$ & $\tfrac13$ & $0$ & $\tfrac13$\\
$d_R$ & $\tfrac12$ & $3$ & $\mathbf3$ & $\mathbf1$ & $-\tfrac13$
 & $\mathbf1$ & $\tfrac13$ & $0$ & $\tfrac13$\\
$L_L$ & $\tfrac12$ & $3$ & $\mathbf1$ & $\mathbf2$ & $-\tfrac12$
 & $\mathbf1$ & $-1$ & $0$ & $-1$\\
$e_R$ & $\tfrac12$ & $3$ & $\mathbf1$ & $\mathbf1$ & $-1$
 & $\mathbf1$ & $-1$ & $0$ & $-1$\\
$\nu_R$ & $\tfrac12$ & $3$ & $\mathbf1$ & $\mathbf1$ & $0$
 & $\mathbf1$ & $-1$ & $0$ & $-1$\\
$H_{\rm SM}$ & $0$ & $1$ & $\mathbf1$ & $\mathbf2$ & $\tfrac12$
 & $\mathbf1$ & $0$ & $0$ & $0$\\
\midrule
\multicolumn{10}{@{}l}{Hidden chiral matter and Higgs multiplets}\\[2pt]
$\Psi_i$ & $\tfrac12$ & $3$ & $\mathbf1$ & $\mathbf1$ & $0$
 & $\mathbf{27}$ & $0$ & $+1$ & $+1$\\
$A^{(1)}$ & $0$ & $1$ & $\mathbf1$ & $\mathbf1$ & $0$
 & $\mathbf{351}$ & $0$ & $-2$ & $-2$\\
$A^{(2)}$ & $0$ & $1$ & $\mathbf1$ & $\mathbf1$ & $0$
 & $\mathbf{351}$ & $0$ & $+2$ & $+2$\\
$\Sigma$ & $0$ & $1$ & $\mathbf1$ & $\mathbf1$ & $0$
 & $\mathbf{351}^{\prime}$ & $0$ & $0$ & $0$\\
\midrule
\multicolumn{10}{@{}l}{Optional vectorlike fermions and real force mediator}\\[2pt]
$\psi_{1L},\psi_{1R}$ & $\tfrac12$ & $1$ Dirac & $\mathbf1$ & $\mathbf1$ & $0$
 & $\mathbf1$ & $0$ & $+1$ & $+1$\\
$\psi_{2L},\psi_{2R}$ & $\tfrac12$ & $1$ Dirac & $\mathbf1$ & $\mathbf1$ & $0$
 & $\mathbf1$ & $0$ & $+1$ & $+1$\\
$S$ (real) & $0$ & $1$ & $\mathbf1$ & $\mathbf1$ & $0$
 & $\mathbf1$ & $0$ & $0$ & $0$\\
\midrule
\multicolumn{10}{@{}l}{St\"uckelberg field}\\[2pt]
$b$ & $0$ & $1$ & $\mathbf1$ & $\mathbf1$ & $0$
 & $\mathbf1$ & $0$ & --- & shift\\
\bottomrule
\end{tabular}
\caption{Matter and scalar fields of the portal theory before symmetry breaking. The copy counts give the number of families or independent multiplets. Both chiralities in each Dirac row have the displayed charges. The $\mathbf{351}$ is the antisymmetric representation with the conjugation convention fixed by the Yukawa contraction in the text. The field $b$ transforms by an additive gauge shift. The neutral symmetric $\Sigma$ supplies the scalar-vacuum completion in Sec.~\ref{sec:stable-extended-vacuum}. The vectorlike fermions and real $S$ form the optional halo extension.}
\label{tab:portal-matter}
\end{table}

\begin{table}[!htbp]
\centering
\small
\setlength{\tabcolsep}{8pt}
\renewcommand{\arraystretch}{1.18}
\begin{tabular}{@{}lcccccc@{}}
\toprule
Gauge field & Spin & $SU(3)_c$ & $SU(2)_w$ & $Y$ & $E_{6,D}$ & $X$\\
\midrule
$G_\mu^a$ & $1$ & $\mathbf8$ & $\mathbf1$ & $0$ & $\mathbf1$ & $0$\\
$W_\mu^i$ & $1$ & $\mathbf1$ & $\mathbf3$ & $0$ & $\mathbf1$ & $0$\\
$Y_\mu$ & $1$ & $\mathbf1$ & $\mathbf1$ & $0$ & $\mathbf1$ & $0$\\
$\mathcal G_\mu^A$ & $1$ & $\mathbf1$ & $\mathbf1$ & $0$ & $\mathbf{78}$ & $0$\\
$B_\mu$ & $1$ & $\mathbf1$ & $\mathbf1$ & $0$ & $\mathbf1$ & $0$\\
\bottomrule
\end{tabular}
\caption{Gauge connections of the same theory. Non-Abelian vectors are in the adjoint representation of their own group. The zero Abelian entries refer to the homogeneous adjoint charges. The connections also have inhomogeneous gauge transformations; for example, $B_\mu\to B_\mu+\partial_\mu\omega_X$. The generator indices run over $a=1,\ldots,8$, $i=1,2,3$ and $A=1,\ldots,78$.}
\label{tab:portal-vectors}
\end{table}

The St\"uckelberg field $b$ transforms additively as $b\to b+M\omega_X$. Its dual tensor description does not add another independent degree of freedom. The neutral vector $N_\mu$ used below is a component of the $E_{6,D}$ connection. The selected fields $H_r$, their radial and phase fluctuations, and the surviving $\chi$ are components of these multiplets after symmetry breaking. The optional global numbers of the Dirac extension are $(N_1,N_2)=(1,0)$ for $\psi_1$, $(0,1)$ for $\psi_2$ and zero for the other fields.

The reason for adding the three $\nu_R$ fields follows directly from the anomalies. The sums are easiest to evaluate when every fermion is written as a left-handed Weyl field. Thus $u_R^c,d_R^c,e_R^c,\nu_R^c$ carry the opposite charges to the corresponding right-handed fields in Eq.~\eqref{eq:visible-BL-charges}. For one family,
\begin{align}
 {\cal A}_{SU(3)^2X}&\propto
 2\left(\frac13\right)-\frac13-\frac13=0,\notag\\
 {\cal A}_{SU(2)^2X}&\propto
 3\left(\frac13\right)-1=0,\notag\\
 {\cal A}_{\mathrm{grav}^2X}&\propto
 6\left(\frac13\right)-3\left(\frac13\right)
 -3\left(\frac13\right)-2+1+1=0,\notag\\
 {\cal A}_{X^3}&\propto
 6\left(\frac13\right)^3-3\left(\frac13\right)^3
 -3\left(\frac13\right)^3-2+1+1=0 .
 \label{eq:visible-BL-anomalies}
\end{align}
The mixed Abelian sums $Y^2X$ and $YX^2$ also vanish. The visible $B-L$ current is therefore anomaly free. The final $+1$ in the last two lines is the contribution of $\nu_R^c$; without it the gravitational and cubic anomalies would remain.

The Standard Model Higgs has $X=0$. Consequently the usual charged-fermion Yukawa interactions and the Dirac neutrino coupling
\begin{equation}
 {\cal L}_{\nu,\rm Dirac}
 =-y^\nu_{ij}\,\overline L_i\widetilde H_{\rm SM}\nu_{Rj}
 +\mathrm{h.c.}
 \label{eq:Dirac-neutrino}
\end{equation}
are gauge invariant. A bare polynomial Majorana mass for $\nu_R$ carries two units of $B-L$ and is forbidden. The hidden Higgs multiplets are not $E_{6,D}$ singlets, so they cannot couple directly to $\nu_R\nu_R$ in a renormalizable $E_{6,D}$-invariant operator. The displayed perturbative renormalizable Lagrangian therefore contains Dirac neutrino masses.

This statement does not prove that the neutrinos are exactly Dirac in the ultraviolet theory. Since $b\to b+M\omega_X$, an ultraviolet matching interaction or a charged nonperturbative event can generate the gauge-invariant dressed operator
\begin{equation}
 {\cal L}_{\nu,\rm dressed}
 =-\frac12 {\cal M}_{ij}\,
 \nu_{Ri}^{\rm T}C\nu_{Rj}\,
 e^{2ig_Xb/M}+\mathrm{h.c.}
 \label{eq:dressed-Majorana}
\end{equation}
in the charge and gauge-transformation convention used below. Its existence and coefficient depend on the compact periodicity and on the charged events of the ultraviolet completion, just as for the dressed scalar operators discussed in Sec.~\ref{sec:quality}. If it is present, the neutrinos acquire Majorana or pseudo-Dirac masses. An ordinary alternative is an additional $E_{6,D}$-singlet scalar of charge $+2$; that completion can introduce another scalar phase and is not assumed in the phase count below.

Let $Y_\mu$ and $B_\mu$ be the hypercharge and $X$ gauge fields. Before diagonalizing their kinetic terms, the current interactions are
\begin{equation}
 {\cal L}_{\rm currents}
 =g_YY_\mu J_Y^\mu
 +g_XB_\mu\left(J_{B-L}^\mu+J_D^\mu\right),
 \label{eq:uncanonical-currents}
\end{equation}
where $J_D^\mu$ contains the hidden fields only. The most general renormalizable Abelian kinetic terms also contain \cite{Holdom}
\begin{equation}
 {\cal L}_{\rm mix}=-\frac{\epsilon}{2}F^Y_{\mu\nu}F_X^{\mu\nu},
 \qquad |\epsilon|<1 .
 \label{eq:kinetic-portal}
\end{equation}
The parameter $\epsilon$ is a renormalized coupling. Visible fermions carry both $Y$ and $X=B-L$ and generate its running. In the charge convention of Eq.~\eqref{eq:visible-BL-charges}
\begin{equation}
 \sum_{\text{one family}}YX=\frac83 .
 \label{eq:YX-trace}
\end{equation}
The mixed vacuum polarization therefore generates kinetic mixing under renormalization-group evolution even when a boundary value is chosen to vanish.

Set $r_\epsilon=\sqrt{1-\epsilon^2}$ and define canonical fields by
\begin{equation}
 Y_\mu=\widehat Y_\mu-\frac{\epsilon}{r_\epsilon}\widehat B_\mu,
 \qquad
 B_\mu=\frac{\widehat B_\mu}{r_\epsilon}.
 \label{eq:portal-canonical-fields}
\end{equation}
Substitution cancels the mixed kinetic term and gives
\begin{equation}
 {\cal L}_{\rm currents}
 =g_Y\widehat Y_\mu J_Y^\mu
 +\widehat B_\mu
 \left[
 \frac{g_X}{r_\epsilon}\left(J_{B-L}^\mu+J_D^\mu\right)
 -\frac{\epsilon g_Y}{r_\epsilon}J_Y^\mu
 \right].
 \label{eq:portal-currents}
\end{equation}
This equation is the definition of the portal used in the remainder of the paper. The same field $\widehat B_\mu$ gauges both parts of Eq.~\eqref{eq:generalized-X}; there is no second $B-L$ vector. Even for $\epsilon=0$, it couples directly to visible baryon and lepton number. Its large mass is generated by the Stückelberg term and the hidden Higgs vevs discussed below, while kinetic mixing adds a hypercharge component to its visible coupling.

The Stückelberg term becomes
\begin{equation}
 \frac12\left(\partial_\mu b-\frac{M}{r_\epsilon}\widehat B_\mu\right)^2 .
\end{equation}
The hidden Higgs coupling to $\widehat B_\mu$ and the Stückelberg coefficient are rescaled by the same factor. Their ratio is unchanged, so the gauge-invariant angular coordinate derived later is independent of the choice of canonical Abelian basis.

The visible anomaly cancellation in Eq.~\eqref{eq:visible-BL-anomalies} has an important consequence. The Wess--Zumino terms are required only for the anomalous hidden contribution to the $X$ current. No additional visible $bF_Y\widetilde F_Y$, $b\,\mathrm{tr}(F_w\widetilde F_w)$ or $b\,\mathrm{tr}(F_c\widetilde F_c)$ counterterm is generated by the anomaly-free $B-L$ assignment. A change of gauge-field basis does not alter this conclusion.

Gauge invariance also permits a coupling between the magnitude of the visible Higgs and the magnitudes of the hidden Higgs fields. Here $H_{\rm SM}$ is the Standard Model Higgs doublet, while $A^{(1)}$ and $A^{(2)}$ are the two complex hidden Higgs multiplets in the antisymmetric $\mathbf{351}$ representation of $E_{6,D}$, both singlets of the Standard Model gauge group. The label $r=1,2$ distinguishes these two multiplets. Their contribution to the scalar potential is
\begin{equation}
 V_{\rm mix}=\sum_{r=1}^2\lambda_{Hr}
 (H_{\rm SM}^\dagger H_{\rm SM})
 (A^{(r)\dagger}\!\cdot A^{(r)}).
 \label{eq:induced-scalar-portals}
\end{equation}
The dagger denotes Hermitian conjugation, and the dot contracts all hidden representation indices: $A^{(r)\dagger}\!\cdot A^{(r)}$ is the invariant sum of the squared magnitudes of the multiplet components. Each real, dimensionless coefficient $\lambda_{Hr}$ measures the strength of the coupling between the visible Higgs magnitude and one hidden Higgs magnitude. These products are independent of the scalar phases. We set both couplings to zero at the chosen renormalization scale in the minimal portal considered here. Their smallness is an additional assumption needed to preserve the electroweak hierarchy. In the selected vacuum, $v_r/\sqrt2$ denotes the magnitude of the vacuum expectation value of $A^{(r)}$, so that $\langle A^{(r)}\rangle^\dagger\!\cdot\langle A^{(r)}\rangle=v_r^2/2$. Nonzero portal couplings would therefore shift the visible Higgs mass parameter by
\begin{equation}
 \Delta m_H^2=\frac12\sum_r\lambda_{Hr}v_r^2 .
 \label{eq:visible-Higgs-threshold}
\end{equation}
The completion in Sec.~\ref{sec:stable-extended-vacuum} also permits an independent coupling $\lambda_{H\Sigma}(H_{\rm SM}^\dagger H_{\rm SM})\tr(\Sigma\Sigma^\dagger)$. We set it to zero at the same reference scale. A nonzero value would add $\lambda_{H\Sigma}(s^2+w^2)/2$ to the visible Higgs mass parameter. The hidden vacuum benchmark consequently retains the electroweak hierarchy assumption discussed here.
For $v_r$ near $10^{15}\,\GeV$, maintaining the electroweak scale requires extremely small matched values of $\lambda_{Hr}$ or a separate ultraviolet mechanism that protects the Higgs mass. The heavy vector gives a second threshold problem. Kinetic mixing couples $H_{\rm SM}$ to the massive Abelian vector with strength of order $\epsilon g_Y$, and a one-loop matching correction scales as
\begin{equation}
 \Delta m_H^2\sim
 \frac{\epsilon^2g_Y^2}{16\pi^2}\,m_{Z'}^2,
 \label{eq:vector-Higgs-threshold}
\end{equation}
up to an order-one coefficient that depends on the heavy spectrum and the renormalization choice. For $\epsilon=10^{-3}$, $g_Y=0.35$ and the physical visible Higgs mass $m_h\simeq1.3\times10^2\,\GeV$, the estimate becomes
\begin{equation}
 \frac{|\Delta m_H^2|}{m_h^2}\sim4.6\times10^{18}
 \left(\frac{\epsilon}{10^{-3}}\right)^2
 \left(\frac{g_Y}{0.35}\right)^2
 \left(\frac{m_{Z'}}{10^{16}\,\GeV}\right)^2.
 \label{eq:visible-Higgs-tuning-number}
\end{equation}
It reaches about $4.6\times10^{22}$ at $10^{18}\,\GeV$. The direct $X$ coupling to the top quark also communicates a heavy threshold through a two-loop diagram with the top Yukawa interaction,
\begin{equation}
 \Delta m_H^2\big|_{tX}\sim
 \frac{N_c y_t^2(g_Xx_t)^2}{(16\pi^2)^2}m_{Z'}^2,
 \qquad N_c=3,\qquad x_t=\frac13.
 \label{eq:top-vector-Higgs-threshold}
\end{equation}
Its coefficient depends on the complete heavy spectrum. Thus even an exactly vanishing kinetic-mixing boundary condition would not isolate the visible Higgs from all heavy thresholds. Because Eq.~\eqref{eq:YX-trace} regenerates $\epsilon$, setting only its boundary value to zero does not remove this sensitivity. With a vector mass near the hidden or St\"uckelberg scale, the four-dimensional construction is electroweak-unnatural unless an ultraviolet symmetry, cancellation or compositeness mechanism protects $m_H^2$. We treat the observed Higgs mass as a tuned low-energy input. This hierarchy problem does not change the angular phase projection.

\section{The grand-unified matter content and its scalar potential}
\label{sec:GUT}
The hidden fields transform under $E_{6,D}\times U(1)_X$ and are singlets of $G_{\rm SM}$. Their $X$ charge is the dark component $D_{\rm hid}$ in Eq.~\eqref{eq:generalized-X}. The $E_{6,D}$-charged sector consists of three left-handed fermion multiplets and two complex scalar multiplets:
\begin{equation}
 \Psi_i\sim\mathbf{27}_{+1}\quad(i=1,2,3),\qquad
 A^{(1)}\sim\mathbf{351}_{-2},\qquad A^{(2)}\sim\mathbf{351}_{+2}.
 \label{eq:fields}
\end{equation}
We adopt this spectrum \cite{CF,CFIT}. A representation dimension counts internal field components, while the subscript denotes the additional Abelian charge. Once the visible convention in Eq.~\eqref{eq:visible-BL-charges} has been fixed, the relative normalization of $D_{\rm hid}$ cannot be absorbed into the single coupling $g_X$. Our explicit choice is
\begin{equation}
 \frac{X_\Psi}{|X_\ell|}=1,
 \qquad \frac{X_\Psi}{X_Q}=3,
 \qquad \frac{X_{A^{(1)}}}{|X_\ell|}=-2,
 \qquad \frac{X_{A^{(2)}}}{|X_\ell|}=+2.
 \label{eq:hidden-visible-charge-ratios}
\end{equation}
These charge ratios define the portal. A different relative normalization would change the hidden anomaly coefficients, vector masses and all visible-to-hidden amplitudes, and would define a different model. The conjugation convention for naming the tensor representation is fixed here by its invariant Yukawa contraction. There is also a different symmetric 351-dimensional representation, conventionally distinguished by a prime. The two core Higgs fields $A^{(1)},A^{(2)}$ use the antisymmetric tensor; the vacuum completion in Sec.~\ref{sec:stable-extended-vacuum} adds the symmetric $\Sigma$.

\subsection{Dark-sector Lagrangian and covariant derivatives}
The gauge and matter terms can now be collected in one expression. We use the original Abelian field $B_\mu$, before the kinetic transformation in Eq.~\eqref{eq:portal-canonical-fields}, and denote the 78 hidden $E_6$ vectors by $\mathcal G_\mu^a$. The core hidden-sector Lagrangian is
\begin{equation}
 \begin{aligned}
 \mathcal L_D={}&-\frac14\mathcal F^a_{\mu\nu}\mathcal F^{a\mu\nu}
 -\frac14F^X_{\mu\nu}F_X^{\mu\nu}
 +\frac12(\partial_\mu b-MB_\mu)(\partial^\mu b-MB^\mu)\\
 &+\sum_{i=1}^3 i\Psi_i^\dagger\bar\sigma^\mu D_\mu\Psi_i
 +\sum_{r=1}^2(D_\mu A^{(r)})^\dagger\!\cdot D^\mu A^{(r)}\\
 &-V_{\rm inv}(A^{(1)},A^{(2)})
 -V_{\rm br}(A^{(1)},A^{(2)},b)
 +\mathcal L_Y+\mathcal L_{\rm WZ}.
 \end{aligned}
 \label{eq:dark-core-lagrangian}
\end{equation}
Here $\bar\sigma^\mu=(\mathbf1,-\boldsymbol\sigma)$ acts on the two-component spinor indices. The scalar dot product uses a canonically normalized basis of independent antisymmetric components, so each independent complex scalar has kinetic term $|D_\mu A_{IJ}|^2$ with $I<J$. The phase-independent invariant potential $V_{\rm inv}$ contains the self-interactions and mixed contractions discussed below; its restriction to the chosen magnitudes gives $V_{\rm rad}$ in Eq.~\eqref{eq:radial}. The explicitly symmetry-breaking dressed potential $V_{\rm br}$ is an optional addition, set to zero in the symmetry-preserving theory, and reduces to $V_p$ in Eq.~\eqref{eq:Vp}. The Yukawa term $\mathcal L_Y$ is specified in Eq.~\eqref{eq:Yukawa}, and the anomaly-compensating term $\mathcal L_{\rm WZ}$ is given in Eq.~\eqref{eq:WZ}. These potentials are independent inputs: the gauge kinetic terms do not determine their coefficients.

For a field $\Phi$ in an $E_6$ representation $R$ with external charge $q_\Phi$, our derivative convention is
\begin{equation}
 D_\mu\Phi=\left(\partial_\mu-ig_G\mathcal G_\mu^aT_R^a
 -ig_Xq_\Phi B_\mu\right)\Phi,
 \qquad q_\Psi=1,\quad q_1=-2,\quad q_2=+2.
 \label{eq:dark-covariant-derivative}
\end{equation}
The matrices $T_R^a$ act only on internal representation indices; they do not act on the Lorentz spinor index of $\Psi_i$. Explicitly, the fundamental and dual antisymmetric tensor derivatives are
\begin{align}
 (D_\mu\Psi_i)^I={}&\partial_\mu\Psi_i^I
 -ig_G\mathcal G_\mu^a(T_{27}^a)^I{}_J\Psi_i^J
 -ig_XB_\mu\Psi_i^I,
 \label{eq:dark-fermion-derivative}\\
 (D_\mu A^{(r)})_{IJ}={}&\partial_\mu A^{(r)}_{IJ}
 +ig_G\mathcal G_\mu^a\left[(T_{27}^a)^K{}_I A^{(r)}_{KJ}
 +(T_{27}^a)^K{}_J A^{(r)}_{IK}\right]\notag\\
 &-ig_Xq_rB_\mu A^{(r)}_{IJ}.
 \label{eq:dark-tensor-derivative}
\end{align}
The opposite non-Abelian sign in the second expression follows because both tensor indices are dual to the upper fundamental indices. The field strengths in Eq.~\eqref{eq:dark-core-lagrangian} are
\begin{equation}
 \mathcal F^a_{\mu\nu}=\partial_\mu\mathcal G_\nu^a
 -\partial_\nu\mathcal G_\mu^a+g_Gf^{abc}\mathcal G_\mu^b\mathcal G_\nu^c,
 \qquad F^X_{\mu\nu}=\partial_\mu B_\nu-\partial_\nu B_\mu,
 \label{eq:dark-field-strengths}
\end{equation}
where $[T^a,T^b]=if^{abc}T^c$ defines the structure constants.

In this convention the local Abelian transformations are
\begin{equation}
 \begin{aligned}
 B_\mu&\longrightarrow B_\mu+\partial_\mu\omega_X,
 &b&\longrightarrow b+M\omega_X,\\
 \Psi_i&\longrightarrow e^{ig_X\omega_X}\Psi_i,
 &A^{(r)}&\longrightarrow e^{ig_Xq_r\omega_X}A^{(r)}.
 \end{aligned}
 \label{eq:dark-X-transformations}
\end{equation}
Thus $b$ is an $E_6$ singlet with the additive $X$ transformation $b\to b+M\omega_X$. The shared gauge parameter rotates visible quarks and leptons according to their $B-L$ charges and shifts $b$. The shift makes $\partial_\mu b-MB_\mu$ invariant and cancels the hidden contribution to the gauge anomaly through $\mathcal L_{\rm WZ}$.

Although Eq.~\eqref{eq:dark-core-lagrangian} collects the hidden fields, $B_\mu$ is shared with the visible sector. The visible current coupling and kinetic portal are those of Eqs.~\eqref{eq:uncanonical-currents}--\eqref{eq:kinetic-portal}; the scalar magnitude portal is Eq.~\eqref{eq:induced-scalar-portals}. The covariant derivatives already contain the hidden current coupling, which must not be added a second time. If the two additional Dirac singlets and real mediator are included, their separate contribution is $\mathcal L_{\psi S}$ in Eq.~\eqref{eq:two-dirac-lagrangian}. Those fields are distinct from the chiral $\Psi_i$ in Eq.~\eqref{eq:fields}.

\subsection{Charges, Yukawa contractions and anomaly traces}
The scalar multiplets are represented by complex antisymmetric tensors $A^{(r)}_{IJ}=-A^{(r)}_{JI}$, with $I,J=1,\ldots,27$. With the appropriate conjugate index convention, an invariant fermion coupling has the form
\begin{equation}
 \mathcal L_Y=-\frac12 y_{ij}A^{(1)}_{IJ}
 \epsilon_{\alpha\beta}\Psi_i^{I\alpha}\Psi_j^{J\beta}
 +\mathrm{h.c.}
 \label{eq:Yukawa}
\end{equation}
The Greek indices here are two-component spinor indices; the Roman indices are internal. Its total $X$ charge is $-2+1+1=0$. The analogous coupling with $A^{(2)}$ has charge $+4$ and is absent at this renormalizable level. This explains the asymmetric role of the two scalar copies without assuming that the second one has no gauge interactions.

Fermion statistics also constrains the family tensor. Define $\mathcal B_{ij}^{IJ}=\epsilon_{\alpha\beta}\Psi_i^{I\alpha}\Psi_j^{J\beta}$. Anticommuting the fermions and then interchanging the spinor indices gives $\mathcal B_{ij}^{IJ}=\mathcal B_{ji}^{JI}$. Contracting with antisymmetric $A_{IJ}$ therefore makes the result antisymmetric in $i,j$. Only $y_{[ij]}$ contributes. A three-by-three antisymmetric family matrix has zero determinant and leaves at least one family combination massless.

The chiral traces can be computed independently of the scalar potential. In the standard normalization $T(\mathbf{27})=3$, the mixed non-Abelian, cubic Abelian and mixed gravitational coefficients are proportional to
\begin{align}
 \mathcal A_{E_6^2X}&=\sum_i X_iT(\mathbf{27})=3\cdot3=9,\notag\\
 \mathcal A_{X^3}&=\sum_i\dim(\mathbf{27})X_i^3=3\cdot27=81,\notag\\
 \mathcal A_{\mathrm{grav}^2X}&=\sum_i\dim(\mathbf{27})X_i=81.
 \label{eq:trace-values}
\end{align}
These charge traces enter the normalized anomaly coefficients $\mathfrak a_i$ in Eq.~\eqref{eq:anomaly-variation}. Scalars do not contribute to four-dimensional chiral triangle anomalies. The pure $E_6$ local gauge anomaly vanishes, while a trace with one $E_6$ generator and two $X$ insertions vanishes because $\tr T^a=0$. The nonzero mixed trace is $E_6^2X$. The shifting-field counterterms are therefore essential for the stated fermion assignment.

In the two-multiplet core theory, a common scalar rephasing with a compensating fermion rephasing leaves Eq.~\eqref{eq:Yukawa} invariant:
\begin{equation}
 A^{(1)}\to e^{i\beta}A^{(1)},\qquad
 A^{(2)}\to e^{i\beta}A^{(2)},\qquad
 \Psi_i\to e^{-i\beta/2}\Psi_i.
 \label{eq:PQ}
\end{equation}
This is the formal Peccei--Quinn-type rephasing of the construction. It differs from $X$, which assigns opposite charges to the two scalar copies. At the alignment below, its action on the scalar vacuum is a broken non-Abelian gauge direction. Section~\ref{sec:mixing} gives the resulting Goldstone projection. This global direction does not solve the visible strong-CP problem, because the dark multiplets have no visible QCD charge.

\subsection{Exact relative symmetry}
The symmetry-preserving interactions have a continuous global $U(1)_Q$. In a family basis with the nonzero antisymmetric Yukawa block in families 1 and 2, choose
\begin{equation}
 Q(A^{(1)})=-1,\qquad Q(A^{(2)})=+1,\qquad
 Q(\Sigma)=Q(b)=0,\qquad
 Q(\Psi_1,\Psi_2,\Psi_3)=\left(\frac12,\frac12,-1\right).
 \label{eq:relative-global-symmetry}
\end{equation}
Visible fields and the optional vectorlike singlets have zero $Q$ charge. Every gauge-invariant polynomial made from $A^{(1)},A^{(2)},\Sigma$ and their conjugates has $Q=0$, since their scalar charges satisfy $Q=X/2$. The Yukawa term also preserves $Q$. The relevant fermion traces are
\begin{equation}
 \mathcal A_{E_6^2Q}=3\sum_iQ_i=0,\qquad
 \mathcal A_{X^2Q}=27\sum_iQ_i=0,\qquad
 \mathcal A_{\mathrm{grav}^2Q}=27\sum_iQ_i=0.
 \label{eq:exact-Q-anomalies}
\end{equation}
The Wess--Zumino terms preserve this rephasing because $b$ is $Q$ neutral. Thus the complete displayed symmetry-preserving effective action has an exact $Q$ Ward identity. On the chosen vacuum the gauge-invariant angle obeys $\Theta\to\Theta+2\beta$, so that $\chi\to\chi+2F\beta$. Spontaneous breaking of $Q$ leaves one exactly massless physical field. Its potential is constant to every perturbative order and in every $E_6$ topological sector as long as $Q$ is preserved. Saturating additional fermion zero modes or adding further vacuum components while preserving $Q$ cannot generate a mass for this Goldstone direction.

The dressed bilinear $Be^{-4ig_Xb/M}$, with $B=A^{(1)\dagger}\!\cdot A^{(2)}$, has $Q=2$. Its coefficient and the harmonic coefficients below explicitly break $Q$. They are independent effective inputs. Treating such coefficients as parameters that transform oppositely under $Q$ shows that their radiative corrections vanish when all $Q$ breaking is removed. The ultraviolet completion must also suppress every other allowed $Q$-breaking interaction, including any gravitational contribution. The construction here specifies the symmetry and its possible explicit breaking; it supplies no microscopic value for that breaking.

\subsection{Invariant contractions and radial vacuum choices}
The scalar potential assigns an energy density to each configuration of the Higgs fields. It must give the same energy to configurations related by a gauge transformation. We therefore build it by contracting Higgs indices to form gauge singlets. For example, $I_2(A)=A_{IJ}\bar A^{IJ}$ is quadratic in the fields, while $I_2(A)^2$ and $A_{IJ}\bar A^{JK}A_{KL}\bar A^{LI}$ are quartic. The symmetric invariant tensor $d_{IJK}$ of $E_6$ gives further quartic contractions. Collecting them into a basis $\mathcal I_s(A,\bar A)$ gives
\begin{equation}
 V_r(A^{(r)})=m_r^2 I_2(A^{(r)})+
 \sum_s h_{rs}\mathcal I_s(A^{(r)},\bar A^{(r)}).
 \label{eq:full-radial}
\end{equation}
The energy depends on both the field magnitude and the distribution of that magnitude among the components of the multiplet. A vacuum expectation value therefore requires a choice of the coefficients and a check of which direction minimizes the energy. The potential can also contain independent mixed couplings between the two multiplets. Their restrictions to the chosen directions contribute to the energy of the two magnitudes.

To separate these questions, first choose unit tensors $e_1,e_2$ that specify the internal directions, and write $A^{(r)}=(r_r/\sqrt2)e_r$. The real variables $r_r$ are the magnitudes in this restriction. If $\mathcal I_s(e_r,\bar e_r)=c_{rs}$, the corresponding quartic contributes $h_{rs}c_{rs}r_r^4/4$, so the effective radial coefficient is $u_r=\sum_sh_{rs}c_{rs}$. This calculation gives the energy along the chosen directions. Stability against turning toward other components is tested separately in Sec.~\ref{sec:full-multiplet-stability}.

For the phase analysis it is helpful to display the branching under a rank-six subgroup:
\begin{align}
 \mathbf{351}\longrightarrow{}&
 (\mathbf1,\overline{\mathbf3},\mathbf3)
 +(\mathbf1,\overline{\mathbf3},\overline{\mathbf6})
 +(\mathbf1,\mathbf6,\mathbf3)
 +(\mathbf3,\mathbf3,\mathbf1)
 +(\mathbf3,\overline{\mathbf6},\mathbf1)\notag\\
 &+(\mathbf3,\mathbf3,\mathbf8)
 +(\overline{\mathbf3},\mathbf1,\overline{\mathbf3})
 +(\overline{\mathbf3},\mathbf1,\mathbf6)
 +(\overline{\mathbf3},\mathbf8,\overline{\mathbf3})\notag\\
 &+(\overline{\mathbf6},\mathbf3,\mathbf1)
 +(\mathbf6,\mathbf1,\overline{\mathbf3})
 +(\mathbf8,\overline{\mathbf3},\mathbf3),
 \label{eq:branching}
\end{align}
where the three entries refer to the hidden factors $SU(3)_{C,D},SU(3)_{L,D},SU(3)_{H,D}$. We suppress the dark subscript on these internal subgroup labels below; none denotes observed colour or weak isospin. Their dimensions sum to $9+18+18+9+18+72+9+18+72+18+18+72=351$. The first three pieces are colour singlets and have 45 complex components. Under the upper $SU(2)_L$ subgroup, $\overline{\mathbf3}_L=\mathbf2+\mathbf1$ and $\mathbf6_L=\mathbf3+\mathbf2+\mathbf1$. Those pieces contain twelve doublets, twelve singlets and three triplets, with $24+12+9=45$ components. The multiplicities follow directly from the branching in Eq.~\eqref{eq:branching}. We select singlets of the internal $SU(2)_L$ subgroup to make its neutral phase block explicit. All these Higgs components are visible Standard Model singlets. The choice specifies the pattern of dark symmetry breaking.

We choose an explicit illustrative alignment in $S^{(r)}\sim(\mathbf1,\overline{\mathbf3},\mathbf3)_{q_r}$:
\begin{equation}
 \langle S^{(r)}{}_{\bar a}{}^\alpha\rangle
 =\frac{v_r}{\sqrt2}\delta_{a3}\delta^{\alpha3},
 \qquad q_1=-2,\quad q_2=+2,\quad v_r>0.
 \label{eq:vacuum}
\end{equation}
The upper dark $SU(2)_L$ and an upper dark $SU(2)_H$ preserve this direction. For $T_8=\diag(1,1,-2)/(2\sqrt3)$ the selected component has eigenvalues $+1/\sqrt3$ under $T_{8L}$ and $-1/\sqrt3$ under $T_{8H}$. The sum preserves it, while the difference changes the phases of both copies by the same amount. This alignment produces a broken neutral GUT generator in addition to $X$, which is needed to reduce the three phases to one physical phase.

The connected stabilizer of Eq.~\eqref{eq:vacuum} has Lie algebra
\begin{equation}
 \mathfrak h=\mathfrak{su}(3)_{C,D}\oplus\mathfrak{su}(2)_{L,D}
 \oplus\mathfrak{su}(2)_{H,D}\oplus\mathfrak u(1).
 \label{eq:full-E6-stabilizer}
\end{equation}
This statement includes all 78 generators. To obtain it, let $e$ be the unit antisymmetric tensor selecting the vev and normalize the fundamental generators by $\tr_{27}(T_aT_b)=3\delta_{ab}$ \cite{E6Tensors}. Their action on an antisymmetric matrix is $T_aA+AT_a^{\rm T}$. The real orbit Gram matrix $Q_{ab}=\operatorname{Re}[(T_ae)^\dagger\!\cdot(T_be)]$ has the exact spectrum
\begin{equation}
 \operatorname{spec}Q=
 \left\{0^{\times 15},\left(\frac16\right)^{\times 24},
 \left(\frac14\right)^{\times 32},\left(\frac23\right)^{\times 7}\right\},
 \label{eq:full-E6-orbit-spectrum}
\end{equation}
Here $x^{\times n}$ means that the eigenvalue $x$ appears $n$ times in the spectrum. For example, $0^{\times 15}$ lists fifteen zero eigenvalues, one for each unbroken generator, and $(1/6)^{\times 24}$ lists twenty-four copies of $1/6$. Thus all 54 generators outside $SU(3)^3$ are broken, together with nine generators inside that subgroup. Appendix~\ref{app:full-orbit} gives the construction and checks.

With $v^2=v_1^2+v_2^2$, the complete hidden vector squared masses at vanishing Abelian kinetic mixing are
\begin{equation}
 \operatorname{spec}\mathcal M_{\rm gauge}^2=
 \left\{0^{\times 15},\left(\frac{g_G^2v^2}{6}\right)^{\times 24},
 \left(\frac{g_G^2v^2}{4}\right)^{\times 32},
 \left(\frac{2g_G^2v^2}{3}\right)^{\times 6},m_{V,+}^2,m_{V,-}^2\right\}.
 \label{eq:full-gauge-spectrum}
\end{equation}
The last two entries follow from Eq.~\eqref{eq:vector-masses} with the unified normalization $g_Nt_N=\sqrt{2/3}\,g_G$. The corresponding generator is $(T_{8L}-T_{8H})/\sqrt2$. The St\"uckelberg term makes the total gauge rank 64 for nonzero $M,v_1,v_2,g_X,g_G$. Finite Abelian kinetic mixing changes the canonical neutral block as described in Sec.~\ref{sec:portal-vertices}. This gauge calculation fixes the residual algebra at the assumed alignment; scalar stability is tested separately in Sec.~\ref{sec:full-multiplet-stability}.

\subsection{Fermion thresholds and the residual light sector}
\label{sec:residual-light-spectrum}
The spectrum and thermal conditions in this subsection apply to the alignment in Eq.~\eqref{eq:vacuum}, which is retained by the original completed benchmark. Its cosmology assumes selection and persistence of that branch in the presence of the lower vacuum in Sec.~\ref{sec:lower-vacuum}. In a unitary family basis the only nonzero entries of the antisymmetric Yukawa matrix are $y_{12}=-y_{21}=y$. The mass matrix is $\mathcal M_\Psi=y_{ij}\langle A^{(1)}_{IJ}\rangle$. The family factor has rank two and the antisymmetric tensor $e$ has rank six, so
\begin{equation}
 \operatorname{rank}\mathcal M_\Psi=12,\qquad
 N_{\rm Weyl}^{\rm massless}=81-12=69,\qquad
 m_Y=\frac{|y|v_1}{\sqrt6}.
 \label{eq:chiral-mass-rank}
\end{equation}
The twelve massive Weyl components form two Dirac colour triplets with the common tree-level mass $m_Y$. Adding the neutral $\Sigma$ leaves this Yukawa matrix unchanged: $\Psi\Psi\Sigma$ has $X=2$ and is absent at the renormalizable level.

We denote the unbroken Abelian generator by $U=(T_{8L}+T_{8H})/\sqrt2$ and use the integer label $q=2\sqrt6\,U$ in the representation assignments below. The canonical generator obeys $\tr_{27}U^2=3$, so the coupling multiplying $U$ equals $g_G$ at the unified matching scale. Under $SU(3)_{C,D}\times SU(2)_{L,D}\times SU(2)_{H,D}\times U(1)_U$, one $\mathbf{27}$ decomposes as
\begin{equation}
 \begin{aligned}
 \mathbf{27}={}&(\mathbf3,\mathbf2,\mathbf1)_{-1}
 +(\overline{\mathbf3},\mathbf1,\mathbf2)_{1}
 +(\mathbf1,\mathbf2,\mathbf2)_{0}
 +(\mathbf1,\mathbf2,\mathbf1)_{3}\\
 &+(\mathbf1,\mathbf1,\mathbf2)_{-3}
 +(\mathbf1,\mathbf1,\mathbf1)_{0}
 +(\mathbf3,\mathbf1,\mathbf1)_{2}
 +(\overline{\mathbf3},\mathbf1,\mathbf1)_{-2}.
 \end{aligned}
 \label{eq:residual-27-branching}
\end{equation}
The massless spectrum contains three copies of the first six terms and one copy of each of the last two terms. Its dimensions are $3(6+6+4+2+2+1)+3+3=69$. These particles are additional to the optional Dirac singlets. Their masses and relic densities must be included whenever a cosmological population is assigned to the hidden sector. The colour doublets and the charged colour-singlet doublets account for 48 Weyl components without conjugate partners under the residual group. Giving all light components bilinear masses therefore requires additional symmetry breaking or matter content; confinement offers a separate dynamical possibility.

Below the massive chiral-fermion, vector and charged-scalar thresholds, the sums of fermion indices are $(7,9,9,7)$ for $(SU(3)_C,SU(2)_L,SU(2)_H,U(1)_U)$. With $\mu\,\mathrm dg_a/\mathrm d\mu=-b_a g_a^3/(16\pi^2)$, this gives
\begin{equation}
 (b_C,b_L,b_H,b_U)=\left(\frac{19}{3},\frac43,\frac43,-\frac{14}{3}\right).
 \label{eq:residual-running-coefficients}
\end{equation}
Thresholds above this interval require the corresponding massive representations. The positive non-Abelian coefficients establish their one-loop growth toward the infrared. Their eventual confinement or an infrared fixed point depends on the coupled gauge dynamics and on any further fermion masses; the Abelian factor remains unconfined in the displayed theory.

For a common hidden temperature $T_D$, the full relativistic sector has fifteen vectors and 69 Weyl fermions. If they remain unconfined, their radiation contribution is
\begin{equation}
 \Delta N_{\rm eff}
 =\left(\frac{120}{7}+69\right)\left(\frac{T_D}{T_\nu}\right)^4
 =\frac{603}{7}\left(\frac{T_D}{T_\nu}\right)^4.
 \label{eq:full-light-dark-radiation}
\end{equation}
Past thermal contact followed by decoupling above the electroweak scale and no later hidden entropy release gives $T_D/T_\nu=(43/427)^{1/3}\simeq0.47$ and $\Delta N_{\rm eff}\simeq4.0$. This history is excluded by the radiation measurement~\cite{Planck2018}. A prescribed radiation budget $\Delta N_{\rm eff}^{\max}$ instead requires
\begin{equation}
 \frac{T_D}{T_\nu}<\left(\frac{7\Delta N_{\rm eff}^{\max}}{603}\right)^{1/4}
 \simeq0.22\left(\frac{\Delta N_{\rm eff}^{\max}}{0.20}\right)^{1/4}.
 \label{eq:full-light-temperature-bound}
\end{equation}
The value $0.20$ illustrates an adopted radiation budget. If both sectors possess separate thermal distributions above the electroweak scale, their initial temperature ratio must then be below about $0.47$. Out-of-equilibrium distributions require the energy-density version of Eq.~\eqref{eq:full-light-dark-radiation}.

A consistent conditional history has hidden symmetry breaking before inflation, negligible inflaton branching into the chiral hidden sector and a maximum post-inflation temperature below the restoration and heavy-particle thresholds. Visible reheating can then produce only a small hidden radiation density through off-shell portal exchange. For $m_X\gg T$, the rate scales as $\Gamma_{\rm vis\to D}/H\propto g_X^4\mpl T^3/m_X^4$, with the species multiplicities and phase-space factors fixed by the current. For example, $g_X=0.10$, $m_X\simeq10^{16}\,\GeV$ and $T\leq10^{10}\,\GeV$ give a factor of about $2.4\times10^{-20}$ before these finite multiplicities. Such a history readily avoids equilibration. It also requires the visible right-handed neutrinos and any light force mediator to satisfy their own production bounds. The light-sector radiation constraint can therefore be met by a cold, weakly populated sector; its initial conditions are independent cosmological inputs. For example, a final ratio $T_D/T_\nu=0.10$ gives $\Delta N_{\rm eff}\simeq0.0086$ and a present relativistic relic density $\Omega_Dh^2\simeq4.8\times10^{-8}$, using $\Omega_\gamma h^2\simeq2.5\times10^{-5}$. This is a possible radiation boundary condition. A calculation of reheating and the optional particle abundances is required to generate it.

The breaking pattern also admits topological defects. For a simply connected $E_6$ and the connected stabilizer, the global embedding inferred from Eq.~\eqref{eq:residual-27-branching} is $H_0=[SU(3)_C\times SU(2)_L\times SU(2)_H\times U(1)_U]/\mathbb Z_6$. The kernel is generated by $(e^{-2\pi i/3}\mathbf1_3,-\mathbf1_2,-\mathbf1_2,e^{i\pi/3})$. Consequently $\pi_2(E_6/H_0)=\pi_1(H_0)=\mathbb Z$, and a post-inflation transition can produce monopoles. The assumed pre-inflation transition dilutes these defects and requires subsequent temperatures to stay below the restoration scale. A classification of strings involving the external $U(1)_X$ also requires the compact period of $b$ and the full global gauge-group quotient. Those data are not fixed by the local action. The broken relative global symmetry can produce strings in a post-inflation transition; the explicit angular term and its periodicity govern their later evolution.

A pure Yang--Mills approximation would require all charged matter of the relevant factor to decouple. The distinction between pure glueballs and bound states containing light fermions is also explicit in Ref.~\cite{DevGehrleinSenguptaSoni2026DarkSU2}, Sec.~V.2. The chiral residual spectrum above prevents imposing a pure-gauge limit without further dynamics.

\subsection{Radial restriction of the core potential}
For the two-multiplet core, hold both fields along the same internal direction $e$ and vary their magnitudes, $A^{(r)}=r_re/\sqrt2$. In the example below the negative quadratic terms favour nonzero fields and positive quartic terms can stabilize their growth. The mixed coefficient $\lambda_{12}$ shifts the preferred magnitudes together. The phase-independent restriction is
\begin{equation}
 V_{\rm rad}=-\frac12\mu_1^2r_1^2-\frac12\mu_2^2r_2^2
 +\frac14u_1r_1^4+\frac14u_2r_2^4+\frac12\lambda_{12}r_1^2r_2^2.
 \label{eq:radial}
\end{equation}
The nonzero stationary magnitudes satisfy $\mu_1^2=u_1v_1^2+\lambda_{12}v_2^2$ and $\mu_2^2=\lambda_{12}v_1^2+u_2v_2^2$. With $\Delta=u_1u_2-\lambda_{12}^2\ne0$,
\begin{equation}
 v_1^2=\frac{u_2\mu_1^2-\lambda_{12}\mu_2^2}{\Delta},\qquad
 v_2^2=\frac{u_1\mu_2^2-\lambda_{12}\mu_1^2}{\Delta},\qquad
 \mathcal M_h^2=2\begin{pmatrix}
 u_1v_1^2&\lambda_{12}v_1v_2\\
 \lambda_{12}v_1v_2&u_2v_2^2
 \end{pmatrix}.
 \label{eq:vevs}
\end{equation}
Positive $v_r^2$, $u_r$ and $\Delta$ give a radial minimum. The symmetric choice $u_1=u_2=u$, $\mu_1^2=\mu_2^2=\mu^2$ has $v_1=v_2=v_0=\mu/\sqrt{u+\lambda_{12}}$ and canonical eigenvectors $h_\pm=(h_1\pm h_2)/\sqrt2$, with $m_\pm^2=2v_0^2(u\pm\lambda_{12})$. These are restrictions to two radial directions. The full-multiplet obstruction and the completed vacuum are established in Secs.~\ref{sec:full-multiplet-stability} and \ref{sec:stable-extended-vacuum}.

\section{Canonical phase mixing and the surviving physical field}
\label{sec:mixing}
The large radial vevs do not themselves tell us which angular motion is physical. The relevant Higgs variables are the particular complex components that acquire the vevs in Eq.~\eqref{eq:vacuum}. More explicitly, in each multiplet we choose
\begin{equation}
 H_r\equiv S^{(r)}{}_{\bar3}{}^3,
 \qquad S^{(r)}\sim(\mathbf1,\overline{\mathbf3},\mathbf3)_{q_r},
 \qquad A^{(r)}=H_re_r+A_\perp^{(r)},
 \label{eq:selected-phase-components}
\end{equation}
where $e_r$ is the normalized internal tensor selecting that component and $A_\perp^{(r)}$ contains the other components of the same $\mathbf{351}$. The two $e_r$ are copies of the same internal direction in the illustrative alignment. The symbols $H_1,H_2$ denote these selected components of the original hidden Higgs multiplets. The following calculation concerns the neutral phase block associated with these vevs; the other components still belong to the full theory.

To separate a change of magnitude from a change of phase, write
\begin{equation}
 H_r=\frac{v_r+h_r}{\sqrt2}e^{i\eta_r/v_r}.
 \label{eq:polar}
\end{equation}
The real variable $h_r$ changes the magnitude, whereas $\eta_r/v_r$ changes the angle in the complex $H_r$ plane. Expanding about the real vacuum makes the distinction explicit:
\begin{equation}
 H_r=\frac1{\sqrt2}(v_r+h_r+i\eta_r)
 +O\!\left(\frac{h_r\eta_r}{v_r},\frac{\eta_r^2}{v_r}\right).
 \label{eq:linear-higgs-phase}
\end{equation}
Thus the Higgs contribution to the axi-Higgs comes from angular fluctuations, which are the imaginary fluctuations at linear order about this vacuum. The radial fluctuations $h_r$ do not enter this phase combination. At $h_r=0$ each $\eta_r$ has a canonical kinetic coefficient $1/2$. We use $\langle H_r\rangle=v_r/\sqrt2$ throughout. Substituting Eq.~\eqref{eq:polar} into the scalar kinetic terms of Eq.~\eqref{eq:dark-core-lagrangian} determines both the phase normalization and its coupling to the vectors.

Let $N_\mu$ be the neutral vector associated with the broken GUT generator identified above, with common coupling-charge product $g_Nt_N=\sqrt{2/3}\,g_G$ on the selected components. We retain the product $g_Nt_N$ in the formulas to display the neutral projection independently of generator normalization. The phase kinetic sector is
\begin{equation}
 \mathcal L_{\rm ph}=\frac12\sum_{r=1}^2
 (\partial_\mu\eta_r-g_Xq_rv_rB_\mu-g_Nt_Nv_rN_\mu)^2
 +\frac12(\partial_\mu b-MB_\mu)^2.
 \label{eq:phase-kinetic}
\end{equation}
The three canonical coordinates $\boldsymbol a=(\eta_1,\eta_2,b)$ change along two independent gauge vectors,
\begin{equation}
 \boldsymbol k_X=(-2g_Xv_1,2g_Xv_2,M),\qquad
 \boldsymbol k_N=g_Nt_N(v_1,v_2,0),\qquad
 \delta\boldsymbol a=\boldsymbol k_X\omega_X+\boldsymbol k_N\omega_N.
 \label{eq:gauge-vectors}
\end{equation}
A physical combination must remain unchanged under both transformations. Write this physical field as a linear combination with initially unknown real coefficients:
\begin{equation}
 \chi=c_1\eta_1+c_2\eta_2+c_b b,
 \qquad c_1^2+c_2^2+c_b^2=1.
 \label{eq:chi-unknown-coefficients}
\end{equation}
The last condition gives a canonical kinetic term. Applying the two gauge transformations in Eq.~\eqref{eq:gauge-vectors} gives two linear conditions:
\begin{equation}
 v_1c_1+v_2c_2=0,
 \qquad -2g_Xv_1c_1+2g_Xv_2c_2+Mc_b=0.
 \label{eq:chi-invariance-conditions}
\end{equation}
The first condition eliminates the common Higgs phase shifted by the broken GUT generator. Substituting its solution for $c_2$ into the second fixes the relative amount of $b$:
\begin{equation}
 c_2=-\frac{v_1}{v_2}c_1,
 \qquad c_b=\frac{4g_Xv_1}{M}c_1.
 \label{eq:chi-coefficient-ratios}
\end{equation}
Only the overall sign and normalization remain. Choosing $c_1=-F/v_1$ with $F>0$ yields
\begin{equation}
 \chi=F\left(-\frac{\eta_1}{v_1}
 +\frac{\eta_2}{v_2}-\frac{4g_Xb}{M}\right).
 \label{eq:chi-direct-projection}
\end{equation}
Its normalization fixes $F$, whose explicit value is given in Eq.~\eqref{eq:F}. The angle appearing in the periodic potential is the dimensionless combination $\Theta=\chi/F$; $\chi$ itself is the canonically normalized field. Although $b$ shifts under $X$, its shift cancels that of the relative Higgs phase in this expression. Consequently $\chi$ is invariant under the shared gauge symmetry that acts as $B-L$ on visible fields. The Higgs vevs, gauge coupling and St\"uckelberg mass fix all its coefficients. The orthogonal rotations below also identify the two combinations absorbed by the vectors.

To count the massive dark gauge directions, collect the terms quadratic in the original fields $B,N$. Their mass matrix is the Gram matrix of the gauge vectors. At $\epsilon=0$ it is also the canonical vector mass matrix; for nonzero kinetic mixing its canonical form is given in Sec.~\ref{sec:portal-vertices}:
\begin{equation}
 \mathcal M_V^2=\begin{pmatrix}
 M^2+4g_X^2v^2&2g_Xg_Nt_N(v_2^2-v_1^2)\\
 2g_Xg_Nt_N(v_2^2-v_1^2)&g_N^2t_N^2v^2
 \end{pmatrix},\qquad v^2=v_1^2+v_2^2.
 \label{eq:vector-masses}
\end{equation}
For a symmetric matrix with entries $(A,B;B,D)$, the two eigenvalues follow from $(A-\lambda)(D-\lambda)-B^2=0$:
\begin{equation}
 m_{V,\pm}^2=\frac12\left[A+D\pm\sqrt{(A-D)^2+4B^2}\right].
\end{equation}
Here its determinant is
\begin{equation}
 \det\mathcal M_V^2=g_N^2t_N^2\left[M^2v^2+16g_X^2v_1^2v_2^2\right]>0.
 \label{eq:vector-determinant}
\end{equation}
Together with a positive trace this establishes a positive dark mass form. The kinetic matrix is also positive for $|\epsilon|<1$, so canonical normalization preserves two massive dark directions. The vectors generally mix, but their two longitudinal polarizations require only the span of the two gauge vectors. The scalar coordinates spanning that subspace need not already be aligned with $m_{V,\pm}$.

The orthogonalization can be carried out in two steps. First separate the common Higgs phase from the relative one:
\begin{equation}
 G_1=\frac{v_1\eta_1+v_2\eta_2}{v},\qquad
 R=\frac{-v_2\eta_1+v_1\eta_2}{v}.
 \label{eq:relative-phase}
\end{equation}
Both coefficient rows have length one and their scalar product is zero. The $N$ transformation shifts $G_1$ and leaves $R$ unchanged. Under $X$, direct substitution of Eq.~\eqref{eq:gauge-vectors} gives
\begin{equation}
 \delta_XR=\frac{4g_Xv_1v_2}{v}\omega_X\equiv\kappa\omega_X,
 \qquad \delta_Xb=M\omega_X.
\end{equation}
Thus in the remaining $(R,b)$ plane one combination shifts and the orthogonal combination does not. With $\Sigma_X=\sqrt{M^2+\kappa^2}$, define
\begin{equation}
 G_2=\frac{\kappa R+Mb}{\Sigma_X},\qquad
 \chi=\frac{MR-\kappa b}{\Sigma_X}.
 \label{eq:physical-chi}
\end{equation}
The cancellation $M\delta_XR-\kappa\delta_Xb=0$ proves $X$ invariance; $N$ invariance is already built into $R$ and $b$. The full matrix can be displayed explicitly:
\begin{equation}
 \begin{pmatrix}G_1\\G_2\\\chi\end{pmatrix}
 =O\begin{pmatrix}\eta_1\\\eta_2\\b\end{pmatrix},\qquad
 O=\begin{pmatrix}
 v_1/v&v_2/v&0\\
 -\kappa v_2/(v\Sigma_X)&\kappa v_1/(v\Sigma_X)&M/\Sigma_X\\
 -Mv_2/(v\Sigma_X)&Mv_1/(v\Sigma_X)&-\kappa/\Sigma_X
 \end{pmatrix}.
 \label{eq:O}
\end{equation}
The two elementary rotations ensure $OO^{\rm T}=1$. In particular, the sum of squared coefficients in the last row is $(M^2+\kappa^2)/\Sigma_X^2=1$. Removing the two gauge coordinates leaves a canonical kinetic term $\tfrac12(\partial\chi)^2$.

The degree count is now explicit: three phase coordinates minus two independent gauge directions leave one physical pseudoscalar in this block. A two-coordinate sector containing one Higgs phase and $b$, with one additional Abelian vector, instead has two minus one. If the chosen GUT vevs were singlets under every GUT generator, $\boldsymbol k_N$ would vanish; three coordinates would then leave two physical phases. The rank of the gauge action on the vacuum determines the phase count. Extra vacuum components in the full theory require a larger calculation and can contain the additional dark PQ direction.

The invariant periodic coordinate follows independently from the charges. The singlet $H_1^*H_2$ has $X$ charge four, so define
\begin{equation}
 \mathcal J=H_1^*H_2e^{-4ig_Xb/M}
 =\frac{(v_1+h_1)(v_2+h_2)}2 e^{i\Theta},\qquad
 \Theta=\frac{\eta_2}{v_2}-\frac{\eta_1}{v_1}-\frac{4g_Xb}{M}.
 \label{eq:J}
\end{equation}
The Higgs phase variation $4g_X\omega_X$ cancels the Stückelberg variation; the common $N$ phase cancels between the conjugate components. Expressing $\Theta$ in terms of $R,b$ gives
\begin{equation}
 \Theta=\frac{v}{v_1v_2}R-\frac{4g_X}{M}b
 =\frac{v\Sigma_X}{v_1v_2M}\chi\equiv\frac{\chi}{F},\qquad
 F=\frac{v_1v_2M}{v\Sigma_X},\qquad
 \frac1{F^2}=\frac1{v_1^2}+\frac1{v_2^2}+\frac{16g_X^2}{M^2}.
 \label{eq:F}
\end{equation}
The two gauge coordinates have disappeared, as they must. At fixed vevs, increasing $M$ raises $F$ because it suppresses the last positive contribution to $1/F^2$, whereas increasing $g_X$ lowers it. The increase stops at the exact vev-controlled upper bound
\begin{equation}
 F\leq F_{\max}=\frac{v_1v_2}{\sqrt{v_1^2+v_2^2}}.
 \label{eq:F-upper-bound}
\end{equation}
In the regime $M\gg g_Xv_r$ this bound is saturated. Equal vevs $v_0$ give $F_{\max}=v_0/\sqrt2$; for very unequal vevs the smaller one sets the width. The numerical example used below already lies in this regime, so increasing $M$ cannot replace an increase of the hidden Higgs vevs.

There is a second useful scale. Inverting Eq.~\eqref{eq:physical-chi} gives
\begin{equation}
 b=\frac{M}{\Sigma_X}G_2-\frac{\kappa}{\Sigma_X}\chi.
\end{equation}
The explicit Wess--Zumino term therefore contains, in unitary gauge,
\begin{equation}
 \mathcal L_{\rm WZ}\supset
 \frac{\mathfrak a_G\kappa}{M\Sigma_X}\chi\tr(F_G\widetilde F_G),
 \qquad \mathcal F_{\rm WZ}\equiv\frac{M\Sigma_X}{\kappa}
 \sim\frac{M^2}{g_Xv_0}.
 \label{eq:WZ-projection}
\end{equation}
The numerical factor in the last estimate depends on the vev ratio; for equal vevs $\kappa=2\sqrt2g_Xv_0$. The phase width is of order $v_0$, whereas this explicit gauge coefficient is suppressed by a scale of order $M^2/v_0$, together with the anomaly coefficient. Equation~\eqref{eq:WZ-projection} gives the projected counterterm. Section~\ref{sec:relative-anomaly-matching} performs the additional massive-fermion matching and verifies the full light-fermion Ward identity.

\subsection{Anomaly matching of the surviving phase}
\label{sec:relative-anomaly-matching}
The vanishing anomaly of $Q$ constrains the complete Ward identity. The local coupling of the Goldstone field must also include the determinant of the massive fermions. The matching here applies to the branch with 69 massless Weyl components. Define $\mathfrak q_a=g_a^2F^a_{\mu\nu}\widetilde F^{a\mu\nu}/(32\pi^2)$ and choose the anomaly sign by $\delta_\beta\Gamma=2\beta\sum_\psi Q_\psi T_a(\psi)\,\mathfrak q_a$. The massive twelve Weyl components have $Q$ traces $(1,0,0,1)$ under $(SU(3)_C,SU(2)_L,SU(2)_H,U(1)_U)$, while the massless traces are $(-1,0,0,-1)$. The total traces vanish.

At equal vevs and vanishing Abelian kinetic mixing, the physical field has $\theta_1=-F\chi/v^2$ and $b=-4g_XF\chi/M$. Integrating out the two massive Dirac triplets gives $-2\theta_1(\mathfrak q_C+\mathfrak q_U)$ at leading derivative order. The original shifting-field counterterm gives $-18g_Xb\sum_a\mathfrak q_a/M$. The resulting local coupling is
\begin{equation}
 \mathcal L_{\chi,\rm local}=\chi\sum_a c_a\mathfrak q_a,
 \qquad
 c_C=c_U=\frac{2F}{v^2}+\frac{72g_X^2F}{M^2},
 \qquad
 c_L=c_H=\frac{72g_X^2F}{M^2}.
 \label{eq:matched-local-chi-couplings}
\end{equation}
The omitted terms are suppressed by external momenta squared divided by $m_Y^2$. To preserve the unitary gauge under a $Q$ transformation, the light fermions transform with $\widehat Q=Q-aX$, where $a=8g_X^2F^2/M^2$. Their mixed anomaly traces are $\widehat{\mathcal A}_{C,U}=-1-7a$ and $\widehat{\mathcal A}_{L,H}=-9a$. Since $F^{-2}=2/v^2+16g_X^2/M^2$, they obey
\begin{equation}
 Fc_a+\widehat{\mathcal A}_a=0,
 \qquad
 \delta_Q\Gamma_{\rm local}+\delta_Q\Gamma_{\rm light}=0.
 \label{eq:full-relative-anomaly-ward}
\end{equation}
Thus the complete effective action preserves $Q$. The local coefficients in Eq.~\eqref{eq:matched-local-chi-couplings} can be nonzero while their variation is cancelled by the massless fermion measure. Physical finite-momentum amplitudes include the fermion contributions appropriate to the chosen field variables. A zero total anomaly coefficient alone does not set those amplitudes to zero. This separation of the Ward identity from a local axion coupling is also relevant in other chiral axion effective theories~\cite{QuevillonSmithAnomalies}. The exact symmetry nevertheless forbids a $\chi$ potential whenever its explicit breaking parameters vanish.

\section{Periodic potential and vacuum structure}
\label{sec:mass}
The Higgs fields have magnitudes, directions within their multiplets and complex phases. Section~\ref{sec:mixing} identifies a relative angle $\Theta$ that is invariant under the gauge transformations. The St\"uckelberg field enters this angle because its gauge shift compensates the shift of the relative Higgs phase. The exact global symmetry $Q$ changes this physical angle while leaving the energy unchanged. With $Q$ intact, all values of $\Theta$ have the same energy and its fluctuation is massless, even if the Higgs magnitudes are large.

Consider a limit of the radial potential in Eq.~\eqref{eq:radial} with identical parameters $\mu^2>0$, $u>0$ for the two Higgs magnitudes and with $\lambda_{12}=0$. The fields are held along the same internal direction $e$ and the angular interaction is initially absent. Completing the square gives
\begin{equation}
 V_{\rm rad}(r_1,r_2)-V_{\rm rad}(v_0,v_0)
 =\frac{u}{4}\left[(r_1^2-v_0^2)^2+(r_2^2-v_0^2)^2\right],
 \qquad v_0^2=\frac{\mu^2}{u}.
 \label{eq:worked-radial-square}
\end{equation}
The minimum is at $r_1=r_2=v_0$. A small change of either magnitude raises the energy quadratically and gives that radial mode a squared mass $2\mu^2$. Changing $\Theta$ leaves this energy fixed. This simple restriction shows how the potential can fix a large Higgs scale while leaving an exactly flat phase.

A gauge-invariant interaction that explicitly breaks $Q$ can select a preferred angle. At fixed magnitudes the minimum selects the angular vacuum, and the curvature there gives the restoring force for a small displacement. The scale $F$ converts a change in angle into a change in the canonically normalized field, $\delta\chi=F\delta\Theta$. The angular curvature $\mathcal X$ gives
\begin{equation}
 U(\Theta)=U(\Theta_0)+\frac12\mathcal X(\Theta-\Theta_0)^2+\cdots,
 \qquad m_\chi^2=\frac{\mathcal X}{F^2}.
 \label{eq:basic-mass}
\end{equation}
For a concrete example, add the positive squared difference of the two Higgs fields used in Eq.~\eqref{eq:stable-extension-phase-square}. On the aligned configurations it becomes
\begin{equation}
 \left.\Delta V_\alpha\right|_{\mathrm{aligned}}
 =\frac{\alpha}{2}\left|r_1-r_2e^{i\Theta}\right|^2
 =\frac{\alpha}{2}(r_1-r_2)^2+\alpha r_1r_2(1-\cos\Theta),
 \qquad \alpha>0.
 \label{eq:worked-aligned-phase-square}
\end{equation}
The added energy vanishes when the magnitudes agree and $\Theta=0$ modulo $2\pi$. In the radial example above this keeps $r_1=r_2=v_0$ and selects $\Theta=0$. The first term raises the cost of separating the magnitudes. The second gives the angular curvature $\alpha v_0^2$, hence $m_\chi^2=\alpha v_0^2/F^2$. At fixed $F$ a small $\alpha$ can make this mass far smaller than the radial masses. The completed potential below uses the same interaction after the other components of the multiplets have been stabilized.

More general interactions change the shape of the angular potential. The kinetic terms still determine $F$, and the explicit breaking determines $\mathcal X$. We first examine the shape with the magnitudes held fixed, then let the magnitudes and the other multiplet components vary. A compact basis for the angular potential uses the invariant $\mathcal J$ of Eq.~\eqref{eq:J}, which carries the phase $e^{i\Theta}$, and its square, which carries $e^{2i\Theta}$:
\begin{equation}
 V_p=-\left[(a_0v_G^2+a_2|H_1|^2+a_3|H_2|^2)\mathcal J
              +a_1\mathcal J^2+\mathrm{h.c.}\right].
 \label{eq:Vp}
\end{equation}
The $a_i$ are dimensionless effective coefficients, independent of the radial couplings $u_r,\lambda_{12}$. The reference scale $v_G$ has dimensions of energy. Each nonzero harmonic coefficient breaks $Q$, so these terms vanish in the symmetry-preserving theory. Their magnitudes are inputs to the vacuum and mass calculation. Section~\ref{sec:quality} quantifies the suppression required of a possible interaction at higher energies.

At the radial vacuum the projected quantities entering the potential are
\begin{align}
 H_r&=\frac{v_r}{\sqrt2}e^{i\eta_r/v_r},
 &|H_r|^2&=\frac{v_r^2}{2},
 \label{eq:potential-vacuum-fields}\\
 \mathcal J&=\frac{v_1v_2}{2}e^{i\Theta},
 &\mathcal J^2&=\frac{v_1^2v_2^2}{4}e^{2i\Theta}.
 \label{eq:potential-vacuum-invariant}
\end{align} Adding a term and its complex conjugate gives twice its real part. With the radial values fixed, subtracting the value at $\Theta=0$ gives
\begin{align}
 U(\Theta)&=C_1(1-\cos\Theta)+C_2(1-\cos2\Theta),\notag\\
 C_1&=v_1v_2\left[a_0v_G^2+\frac{a_2v_1^2+a_3v_2^2}{2}\right],
 &C_2&=\frac{a_1v_1^2v_2^2}{2}.
 \label{eq:heights}
\end{align}
The $C_i$ set the size of the angular variation in energy density. The first cosine repeats once per turn of $\Theta$, and the second repeats twice. The small-angle expansions are
\begin{equation}
 1-\cos\Theta=\frac{\Theta^2}{2}+O(\Theta^4),\qquad
 1-\cos2\Theta=2\Theta^2+O(\Theta^4).
 \label{eq:harmonic-cosine-expansions}
\end{equation}
They give
\begin{equation}
 U=\frac12(C_1+4C_2)\Theta^2+\cdots
 =\frac12\frac{C_1+4C_2}{F^2}\chi^2+\cdots,
 \qquad m_\chi^2=\frac{C_1+4C_2}{F^2}.
 \label{eq:mass}
\end{equation}
The second cosine therefore contributes four times its amplitude to the curvature. Positive $C_1,C_2$ favour the same angle, $\Theta=0$, at fixed magnitudes. An opposite sign can make the two terms favour different angles. Complex coefficients likewise shift the preferred angles of the two terms. Writing $\delta_n=\arg C_n$, the fixed-magnitude potential, up to an angular-independent constant, is
\begin{equation}
 U=-|C_1|\cos(\Theta+\delta_1)-|C_2|\cos(2\Theta+\delta_2).
 \label{eq:complex-periodic-potential}
\end{equation}
Its stationary angles solve
\begin{equation}
 |C_1|\sin(\Theta_0+\delta_1)+2|C_2|\sin(2\Theta_0+\delta_2)=0.
 \label{eq:complex-periodic-minimum}
\end{equation}
A candidate must have the lowest energy among the stationary angles. Its frozen-radial squared mass is
\begin{equation}
 m_{\chi,\mathrm{fr}}^2=\frac{|C_1|\cos(\Theta_0+\delta_1)
 +4|C_2|\cos(2\Theta_0+\delta_2)}{F^2}.
 \label{eq:complex-periodic-mass}
\end{equation} Positive curvature establishes a nondegenerate angular minimum. Vanishing curvature requires the higher-order terms. Radial mixing can change the light mass, as derived in Sec.~\ref{sec:coupled-vacuum}.

The same mass can be checked directly in the original three coordinates. Define $\boldsymbol{\ell}=(-1/v_1,1/v_2,-4g_X/M)$, so that $\Theta=\boldsymbol{\ell}\cdot\boldsymbol a$. The quadratic potential matrix is
\begin{equation}
 \mathcal M_{\rm odd}^2=(C_1+4C_2)
 \begin{pmatrix}
 v_1^{-2}&-(v_1v_2)^{-1}&4g_X/(v_1M)\\
 -(v_1v_2)^{-1}&v_2^{-2}&-4g_X/(v_2M)\\
 4g_X/(v_1M)&-4g_X/(v_2M)&16g_X^2/M^2
 \end{pmatrix}.
 \label{eq:odd-matrix}
\end{equation}
It is an outer product $(C_1+4C_2)\boldsymbol{\ell}\boldsymbol{\ell}^{\rm T}$ and hence has rank one. Both gauge vectors are in its kernel because $\boldsymbol{\ell}\cdot\boldsymbol k_X=\boldsymbol{\ell}\cdot\boldsymbol k_N=0$. Its single nonzero eigenvalue is $(C_1+4C_2)|\boldsymbol{\ell}|^2=(C_1+4C_2)/F^2$, in agreement with Eq.~\eqref{eq:mass}. The remaining two zero eigenvalues belong to gauge directions; they are not additional physical massless axions in this reduced block.

For comparable vevs and a large Stückelberg mass, the scaling takes a simple form. Take $v_1=v_2=v_G=v_0$, $M\gg g_Xv_0$. Then
\begin{equation}
 C_1=v_0^4\left[a_0+\frac{a_2+a_3}{2}\right],\qquad
 C_2=\frac12a_1v_0^4,\qquad
 m_\chi^2\simeq v_0^2(2a_0+a_2+a_3+4a_1).
 \label{eq:equal-mass}
\end{equation}
For harmonics set by a common small coefficient $a_{\rm eff}$, the scaling is $m_\chi^2\sim a_{\rm eff}v_G^2$. The hierarchy of masses comes from the weak angular restoring force. The radial squared masses can remain of order $u_rv_G^2$ while the phase is ultralight. A cancellation between harmonics is a separate way to reduce the angular curvature and can leave substantial forces on the magnitudes.

At a nondegenerate real minimum with fixed radial values, the homogeneous small fluctuation obeys
\begin{equation}
 K_\Theta=\frac12F^2\dot{\delta\Theta}^{\,2},\qquad
 U_\Theta=\frac12\mathcal X\delta\Theta^2.
 \label{eq:pendulum-angular-energies}
\end{equation}
Its linear equation is
\begin{equation}
 F^2\ddot{\delta\Theta}+\mathcal X\delta\Theta=0.
 \label{eq:pendulum-static-motion}
\end{equation}
The phase normalization and the angular curvature therefore enter the oscillation frequency separately. The following analysis determines the minimum and the regime in which fixing the radial values is justified.

\subsection{Angular minima and harmonic phases}
\label{sec:angular-vacuum}
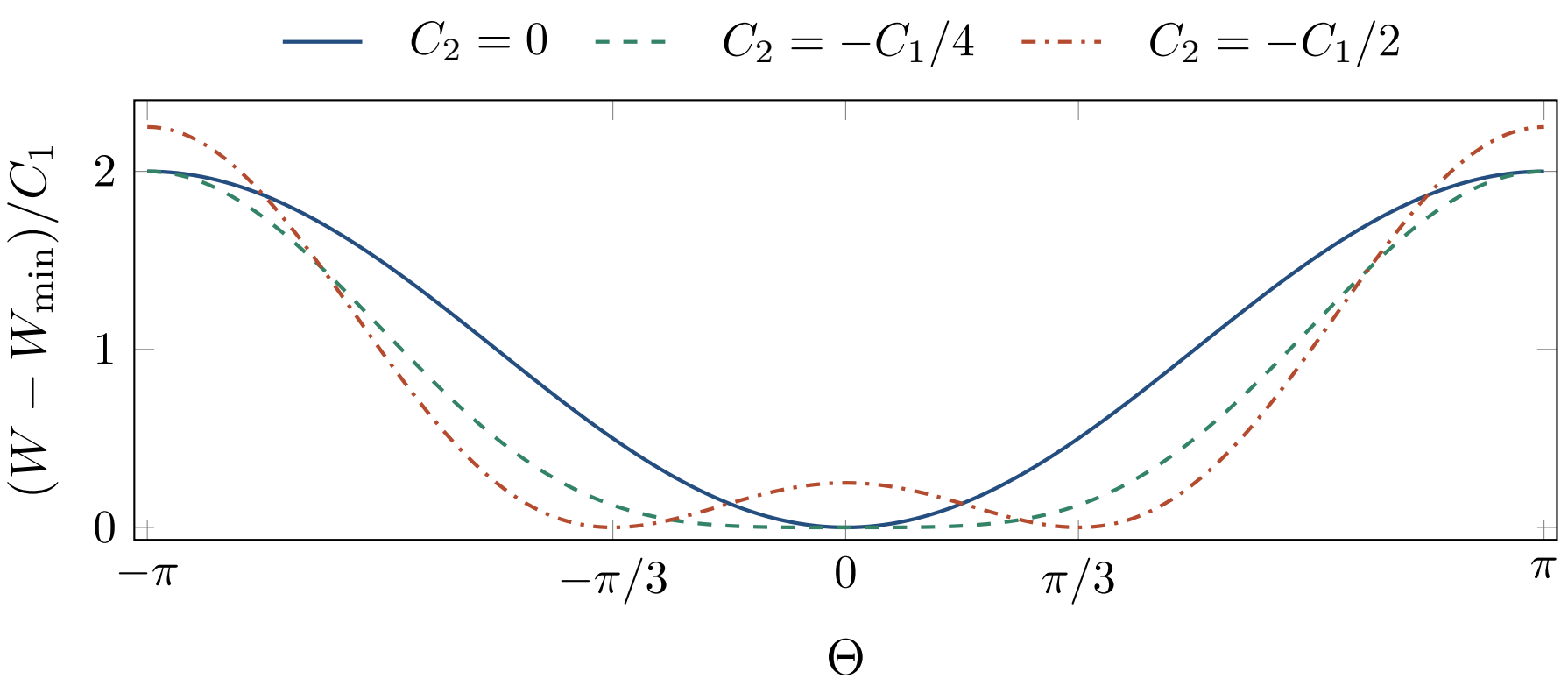
\begin{figure}[tbp]
\centering
\begin{tikzpicture}
 \begin{axis}[
  width=0.78\linewidth,height=4.9cm,
  xmin=-3.2,xmax=3.2,ymin=-0.07,ymax=2.4,
  xlabel={$\Theta$},ylabel={$(W-W_{\min})/C_1$},
  xtick={-3.14159,-1.0472,0,1.0472,3.14159},
  xticklabels={$-\pi$,$-\pi/3$,$0$,$\pi/3$,$\pi$},
  ytick={0,1,2},tick label style={font=\small},
  legend columns=3,
  legend style={at={(0.5,1.04)},anchor=south,draw=none,font=\small,column sep=0.7em},
  domain=-pi:pi,samples=181,no markers]
  \addplot[aximassblue,thick] {1-cos(deg(x))};
  \addlegendentry{$C_2=0$}
  \addplot[aximassgreen,thick,dashed] {0.5*(1-cos(deg(x)))^2};
  \addlegendentry{$C_2=-C_1/4$}
  \addplot[aximassorange,thick,dashdotted] {(cos(deg(x))-0.5)^2};
  \addlegendentry{$C_2=-C_1/2$}
 \end{axis}
\end{tikzpicture}
\caption{Angular energy at fixed Higgs magnitudes for real coefficients and $C_1>0$. Each curve is measured from its own minimum. With a single cosine, the minimum at zero has positive curvature. At $C_2=-C_1/4$, it has zero curvature and a positive quartic restoring term. At $C_2=-C_1/2$, there are two equal minima at $\Theta=\pm\pi/3$.}
\label{fig:angular-vacuum-shapes}
\end{figure}

The two cosines can reinforce each other or compete. Figure~\ref{fig:angular-vacuum-shapes} shows a simple example. Adding a negative second harmonic first flattens the original minimum and then splits it into a pair at nonzero angles. The following calculation locates the minima and determines their curvatures. For real coefficients and fixed nonzero Higgs magnitudes, all angular extrema satisfy
\begin{equation}
 \frac{\partial V_p}{\partial\Theta}
 =\sin\Theta\left(C_1+4C_2\cos\Theta\right)=0.
 \label{eq:angular-stationarity}
\end{equation}
The candidates are $0$, $\pi$ and, when $C_2\ne0$ and $|C_1|\leq4|C_2|$, a pair with $\cos\Theta_b=-C_1/(4C_2)$. To compare their energies it is convenient to use $W(\Theta)=-C_1\cos\Theta-C_2\cos2\Theta$, with the same constant for every angle. At the stationary points,
\begin{align}
 W(0)&=-C_1-C_2,&W(\pi)&=C_1-C_2,\notag\\
 W(\Theta_b)&=C_2+\frac{C_1^2}{8C_2}.
 \label{eq:angular-stationary-energies}
\end{align}
Their curvatures are
\begin{equation}
 \begin{aligned}
 K_0&=C_1+4C_2,&K_\pi&=-C_1+4C_2,\\
 K_b&=-4C_2\left(1-\frac{C_1^2}{16C_2^2}\right).
 \end{aligned}
 \label{eq:angular-branch-curvatures}
\end{equation}
These relations determine both local stability and the global angular minimum within the two-harmonic truncation.

For $C_2>0$, the off-origin stationary points are maxima whenever they exist. The global minimum is $0$ for $C_1>0$ and $\pi$ for $C_1<0$. Both endpoints are local minima when $|C_1|<4C_2$; the energy splitting is $W(\pi)-W(0)=2C_1$. At $C_1=0$ the two minima are degenerate. These statements can also be obtained by setting $z=\cos\Theta$: $W=C_2-C_1z-2C_2z^2$ is then concave on $-1\leq z\leq1$, so its lowest value lies at an endpoint.

For $C_2<0$, the same polynomial is convex. If $|C_1|<4|C_2|$, its minimum lies inside that interval and gives two degenerate angular vacua,
\begin{equation}
 \Theta_b=\pm\arccos\!\left(-\frac{C_1}{4C_2}\right),
 \qquad
 m_{\chi,\mathrm{fr}}^2=\frac{4|C_2|}{F^2}
 \left(1-\frac{C_1^2}{16C_2^2}\right).
 \label{eq:off-origin-angular-vacua}
\end{equation}
They are related by $\Theta\to-\Theta$ in the assumed $2\pi$ interval. Whether this pair represents physical spontaneous CP breaking depends on the CP transformation of the full interactions and on additional discrete identifications. The reduced potential establishes the degenerate pair, without fixing those further data. For $C_1>4|C_2|$ the minimum returns to $0$; for $C_1<-4|C_2|$ it is $\pi$.

At the boundaries $|C_1|=4|C_2|$ with $C_2<0$, the off-origin pair merges with the appropriate endpoint and the quadratic curvature vanishes. At fixed radial values the leading angular term there is
\begin{equation}
 W(\Theta_0+\delta\Theta)-W(\Theta_0)
 =\frac{|C_2|}{2}\delta\Theta^4+O(\delta\Theta^6).
 \label{eq:angular-critical-quartic}
\end{equation}
The harmonic mass approximation fails at this boundary. Radial relaxation can also change the quartic term, so its displayed positive coefficient establishes only the fixed-radial angular behaviour. For $C_2=0$, a nonzero $C_1$ gives the usual single-harmonic minimum, while $C_1=C_2=0$ leaves a flat angular direction. This classification concerns the displayed two harmonics. Higher harmonics require a new comparison of stationary energies.

Complex coefficients leave one relative harmonic phase at each fixed radial point. A constant shift $\varphi=\Theta+\delta_1$ gives
\begin{equation}
 W=-A\cos\varphi-B\cos(2\varphi+\delta),
 \qquad A=|C_1|,\quad B=|C_2|,
 \quad\delta=\delta_2-2\delta_1.
 \label{eq:relative-harmonic-phase}
\end{equation}
Only one coefficient phase can be removed by translating the angular coordinate. For a general $\delta$, solve $A\sin\varphi+2B\sin(2\varphi+\delta)=0$ over a full period, compare all stationary energies, and evaluate the curvature at the lowest one. When radial derivatives are taken, the full complex functions $C_n(r_1,r_2)$ must be retained: a radial-dependent rephasing would also change the field coordinates and their kinetic terms.

\subsection{Coupled radial vacuum and physical stability}
\label{sec:coupled-vacuum}
The preferred angle was found with the Higgs magnitudes held fixed. The interaction that chooses that angle also contains powers of the magnitudes, so it can shift them. To find the joint vacuum, we vary the angle and the magnitudes in the same energy function. For real effective coefficients it is
\begin{equation}
 \begin{aligned}
 V(r_1,r_2,\Theta)&=V_{\rm rad}(r_1,r_2)
   -C_1(r_1,r_2)\cos\Theta-C_2(r_1,r_2)\cos2\Theta,\\
 C_1(r_1,r_2)&=r_1r_2\left[a_0v_G^2
                  +\frac{a_2r_1^2+a_3r_2^2}{2}\right],\\
 C_2(r_1,r_2)&=\frac{a_1r_1^2r_2^2}{2}.
 \end{aligned}
 \label{eq:full-reduced-vacuum-potential}
\end{equation}
Here $r_1,r_2\geq0$ and $V_{\rm rad}$ is Eq.~\eqref{eq:radial}. The reference scale $v_G$ is held fixed when differentiating. Replacing the last two terms by $C_1(1-\cos\Theta)+C_2(1-\cos2\Theta)$ would add $C_1(r_1,r_2)+C_2(r_1,r_2)$ to the radial potential. Such a radial-dependent subtraction changes its stationary equations unless it is explicitly absorbed into $V_{\rm rad}$. Subtracting a constant evaluated at the final vacuum is harmless.

Denote a candidate joint minimum by $(\boldsymbol r_*,\Theta_*)$, with $r_{*1},r_{*2}>0$. Its equations are the angular stationarity condition together with
\begin{equation}
 \left.\partial_{r_i}V_{\rm rad}\right|_*
 =\left.\partial_{r_i}C_1\right|_*\cos\Theta_*
  +\left.\partial_{r_i}C_2\right|_*\cos2\Theta_* ,
 \qquad i=1,2.
 \label{eq:coupled-radial-stationarity}
\end{equation}
The needed derivatives are explicit:
\begin{equation}
 \begin{aligned}
 \partial_{r_1}C_1&=r_2\left[a_0v_G^2
                   +\frac{3a_2r_1^2+a_3r_2^2}{2}\right],\\
 \partial_{r_2}C_1&=r_1\left[a_0v_G^2
                   +\frac{a_2r_1^2+3a_3r_2^2}{2}\right],\\
 \partial_{r_1}C_2&=a_1r_1r_2^2,
 \qquad \partial_{r_2}C_2=a_1r_1^2r_2.
 \end{aligned}
 \label{eq:harmonic-radial-derivatives}
\end{equation}
If a candidate approaches $r_i=0$, the polar coordinate for that field is singular and its stationarity must instead be checked in Cartesian scalar components.

The vevs $\boldsymbol v=(v_1,v_2)$ obtained in Eq.~\eqref{eq:vevs} minimize $V_{\rm rad}$ alone. Let $H_0=\mathcal M_h^2(\boldsymbol v)$ be their positive radial Hessian. Multiplying $V_p$ by a formal small parameter $\varepsilon_p$, the leading change of the magnitudes is
\begin{equation}
 \delta\boldsymbol r\equiv\boldsymbol r_*-\boldsymbol v
 =-\varepsilon_p H_0^{-1}
      \nabla_r V_p(\boldsymbol v,\Theta_0)
      +O(\varepsilon_p^2).
 \label{eq:radial-vacuum-shift}
\end{equation}
The reference angle $\Theta_0$ is a stationary angular branch at $\boldsymbol v$. For an isolated branch its change affects this radial shift only at higher order; near a degeneracy the coupled equations must be solved directly. The fixed-vev treatment requires $|\delta r_i|\ll v_i$ and small periodic corrections to the radial Hessian compared with its smallest eigenvalue. These tests use the individual harmonic amplitudes and derivatives. A small angular mass obtained by cancellation of large curvatures does not by itself justify neglecting radial motion.

At a joint stationary point define
\begin{equation}
 H_{ij}=\left.\partial_{r_i}\partial_{r_j}V\right|_* ,
 \qquad q_i=\left.\partial_{r_i}\partial_\Theta V\right|_* ,
 \qquad K=\left.\partial_\Theta^2V\right|_* .
 \label{eq:coupled-vacuum-hessian-blocks}
\end{equation}
For the real-coefficient potential,
\begin{equation}
 \begin{aligned}
 q_i&=(\partial_{r_i}C_1)_*\sin\Theta_*
       +2(\partial_{r_i}C_2)_*\sin2\Theta_*,\\
 K&=C_{1*}\cos\Theta_*+4C_{2*}\cos2\Theta_*.
 \end{aligned}
 \label{eq:radial-angular-mixing}
\end{equation}
The phase normalization at this point is $F_*^{-2}=r_{*1}^{-2}+r_{*2}^{-2}+16g_X^2/M^2$. After the two gauge coordinates have been removed, the quadratic mass matrix for canonical fluctuations $(\delta r_1,\delta r_2,F_*\delta\Theta)$ is
\begin{equation}
 \mathcal M_{\rm red}^2=
 \begin{pmatrix}
 H&\boldsymbol q/F_*\\
 \boldsymbol q^{\rm T}/F_*&K/F_*^2
 \end{pmatrix}.
 \label{eq:coupled-canonical-mass-matrix}
\end{equation}
Its positivity is equivalent to
\begin{equation}
 H>0,\qquad
 \mathcal S\equiv K-\boldsymbol q^{\rm T}H^{-1}\boldsymbol q>0.
 \label{eq:reduced-vacuum-stability}
\end{equation}
Here $H>0$ means that both radial eigenvalues are positive. When the angle changes by $\delta\Theta$, the mixed term exerts a force on the magnitudes. They relax by $-H^{-1}\boldsymbol q\,\delta\Theta$ to first order. This relaxation lowers the energy of the displaced configuration and reduces the angular curvature from $K$ to $\mathcal S$. Positive curvature in both tests establishes a nondegenerate local minimum in the reduced sector. A zero eigenvalue requires higher-order terms, including at the angular critical boundary. For real coefficients at $\Theta_*=0$ or $\pi$, $\boldsymbol q=0$ and the angular mass is exactly $K/F_*^2$ at quadratic order, evaluated at the shifted radial vevs.

The same completion of the square determines the adiabatic light action. Along the radial stationary curve, $\dd\boldsymbol r_*/\dd\Theta=-H^{-1}\boldsymbol q$. Substituting into the canonical radial kinetic terms gives
\begin{equation}
 F_{\rm eff}^2=F_*^2+\boldsymbol q^{\rm T}H^{-2}\boldsymbol q,
 \qquad
 m_{\rm light}^2\simeq\frac{\mathcal S}{F_{\rm eff}^2}.
 \label{eq:adiabatic-angular-mass}
\end{equation}
This is the leading derivative expansion for frequencies and momenta below the radial and integrated vector masses. It retains the kinetic contribution from motion of the radial minimum. When $\|H^{-1}\boldsymbol q\|\ll F_*$, it reduces to $\mathcal S/F_*^2$. For arbitrary scale separation the physical masses must instead be obtained from Eq.~\eqref{eq:coupled-canonical-mass-matrix}. The fixed-radial curvature alone is insufficient when the mixing is appreciable.

An exactly solvable first-harmonic example shows that these conditions can be satisfied. Set $u_1=u_2=u$, $\mu_1^2=\mu_2^2=\mu^2>0$, $a_1=a_2=a_3=0$, and define the positive mass-squared coefficient $\alpha=a_0v_G^2>0$. For $u>|\lambda_{12}|$, the joint vacuum is
\begin{equation}
 r_{*1}=r_{*2}=v_* ,\qquad \Theta_*=0,
 \qquad v_*^2=\frac{\mu^2+\alpha}{u+\lambda_{12}}.
 \label{eq:symmetric-joint-vacuum}
\end{equation}
It is a global minimum within the selected two-magnitude and one-angle sector. Writing $V_*=-(\mu^2+\alpha)^2/[2(u+\lambda_{12})]$, the exact identity
\begin{equation}
 \begin{aligned}
 V-V_*={}&\frac{u+\lambda_{12}}{8}
            (r_1^2+r_2^2-2v_*^2)^2
          +\frac{u-\lambda_{12}}{8}(r_1^2-r_2^2)^2\\
 &+\frac\alpha2(r_1-r_2)^2
          +\alpha r_1r_2(1-\cos\Theta)
 \end{aligned}
 \label{eq:symmetric-global-vacuum-proof}
\end{equation}
is nonnegative for $r_1,r_2\geq0$. All four terms vanish at Eq.~\eqref{eq:symmetric-joint-vacuum}, with the angle understood modulo $2\pi$. The canonical radial combinations $h_\pm=(\delta r_1\pm\delta r_2)/\sqrt2$ and the angular mode have masses
\begin{equation}
 \begin{aligned}
 m_{+,*}^2&=2(u+\lambda_{12})v_*^2,
 &m_{-,*}^2&=2(u-\lambda_{12})v_*^2+2\alpha,\\
 F_*^{-2}&=\frac2{v_*^2}+\frac{16g_X^2}{M^2},
 &m_{\chi,*}^2&=\frac{\alpha v_*^2}{F_*^2}.
 \end{aligned}
 \label{eq:symmetric-joint-vacuum-masses}
\end{equation}
For $\alpha\ll\mu^2$ the fractional change of the common vev is $\delta v/v=\alpha/(2\mu^2)+O(\alpha^2/\mu^4)$, in agreement with Eq.~\eqref{eq:radial-vacuum-shift}. This example proves existence and stability for a specified effective coefficient; its origin and smallness remain the matching problem discussed below.

For general real $a_i$, a useful sufficient boundedness test can also be stated before minimizing. The inequalities $r_1^3r_2\leq(3r_1^4+r_2^4)/4$ and $r_1r_2^3\leq(r_1^4+3r_2^4)/4$ bound the potentially negative angular quartics. Defining
\begin{equation}
 \widetilde u_1=u_1-\frac{3|a_2|+|a_3|}{2},\qquad
 \widetilde u_2=u_2-\frac{|a_2|+3|a_3|}{2},\qquad
 \widetilde\lambda=\lambda_{12}-|a_1|,
 \label{eq:angular-quartic-lower-bound-parameters}
\end{equation}
the quartic part is bounded below by $\widetilde u_1r_1^4/4+\widetilde u_2r_2^4/4+\widetilde\lambda r_1^2r_2^2/2$. Strict positivity is ensured by
\begin{equation}
 \widetilde u_1>0,\qquad \widetilde u_2>0,\qquad
 \widetilde\lambda>-\sqrt{\widetilde u_1\widetilde u_2}.
 \label{eq:reduced-potential-boundedness}
\end{equation}
These sufficient conditions apply uniformly in the angle. They can be stronger than necessary, and boundedness does not select a particular stationary point. For general coefficients the remaining task is to solve the coupled equations, compare candidate energies and include any Cartesian boundary solutions.

The reduced minimum must also be tested against changes of internal orientation in the complete multiplets. That test gives a stronger restriction than radial boundedness, as follows.

\subsection{Full multiplet directions and vacuum isolation}
\label{sec:full-multiplet-stability}
The preceding calculation describes changes in the magnitudes and relative phase along the proposed alignment. A vacuum must also be tested against fluctuations that turn the fields into other components. Moving along a gauge orbit leaves the physical configuration unchanged, so those directions are removed from the stability test. Of the 1405 real coordinates in the two complex $\mathbf{351}$ fields and $b$, 1341 are physical at the proposed alignment after removing 64 broken gauge directions. The radial and angular calculation covers three of these physical directions.

An invariant that directly tests the relative orientation is
\begin{equation}
 I_r=A^{(r)\dagger}\!\cdot A^{(r)},\qquad
 B=A^{(1)\dagger}\!\cdot A^{(2)},\qquad
 \mathcal I_{\rm align}=I_1I_2-|B|^2\geq0.
 \label{eq:full-alignment-invariant}
\end{equation}
For nonzero fields, $|B|^2/(I_1I_2)$ measures their overlap. It equals one when the tensors are parallel and decreases when one turns away from the other. A positive coefficient $\kappa$ of $\mathcal I_{\rm align}$ therefore penalizes a change in their relative orientation. The invariant vanishes for parallel tensors of any magnitude and phase, including when they turn together. For orthogonal complex perturbations $x_r$ with $e^\dagger x_r=0$, the quadratic cost is $\kappa\lVert v_2x_1-v_1x_2\rVert^2/2$. The reduced potential probes parallel configurations, which explains why it cannot determine this coefficient or the associated masses.

A complete example with the original field content is
\begin{equation}
 \begin{aligned}
 V_{\rm ex}={}&-\mu^2(I_1+I_2)+u(I_1^2+I_2^2)
       +2\lambda_{12}I_1I_2+\kappa\mathcal I_{\rm align}\\
 &-2\alpha\operatorname{Re}\!\left[B\exp\!\left(-\frac{4ig_Xb}{M}\right)\right],
 \qquad \alpha>0.
 \end{aligned}
 \label{eq:full-vector-example-potential}
\end{equation}
For $u>|\lambda_{12}|$, $\mu^2>0$ and $\kappa\geq0$, Cauchy--Schwarz bounds this potential below by the symmetric potential in Eq.~\eqref{eq:symmetric-global-vacuum-proof}. Parallel fields with equal magnitudes $v_*/\sqrt2$ attain the bound. Consequently the selected configuration is a global minimum of this full example. Its two radial and one phase masses are those in Eq.~\eqref{eq:symmetric-joint-vacuum-masses}. The 350 complex relative-orientation modes have
\begin{equation}
 m_{\rm rel}^2=2\alpha+\kappa v_*^2
 \qquad\text{with 700 real modes}.
 \label{eq:full-relative-orientation-mass}
\end{equation}
The Hessian has 702 zero eigenvalues before gauge removal and 638 physical zero eigenvalues afterwards. The common internal orientation is undetermined because the scalar potential has an accidental common $U(351)$ invariance. Gauge and Yukawa interactions break that larger invariance and can lift these directions through quantum corrections; such corrections have not been calculated here.

The ordinary $E_6$ quartic interactions also admit an exact test. Keep arbitrary fixed nonzero $v_1,v_2$, and vary both fields together as $A^{(r)}=v_r n/\sqrt2$, where $n^\dagger n=1$. Every ordinary charge-conserving quartic, including mixed contractions, restricts to a combination of the six balanced single-$\mathbf{351}$ invariants $J_i(n)$ defined in Appendix~\ref{app:full-orbit}. The displayed phase potential depends only on norms and the dressed bilinear and is constant under this common variation. Write the orientation-dependent restriction as $\sum_i c_iJ_i(n)$; the real $c_i$ include the fixed vev factors and have dimension four.

There are two unit tangents $q_a,q_b$, orthogonal to the full gauge orbit and to $e$, for which the canonical directional curvatures satisfy
\begin{equation}
 \mathcal K_a=\frac{16\mathcal C}{3v^2},\qquad
 \mathcal K_b=-\frac{8\mathcal C}{3v^2},\qquad
 \mathcal C=c_2+c_3+c_5,
 \qquad \mathcal K_a+2\mathcal K_b=0.
 \label{eq:full-orientation-obstruction}
\end{equation}
The curvatures are the squared-mass tests along two physical changes of orientation. For either nonzero sign of $\mathcal C$, one curvature is negative and the energy decreases along that direction. Setting $\mathcal C=0$ makes the potential exactly constant along both normalized curves $n=(e+s q_{a,b})/\sqrt{1+s^2}$. The configurations on each curve are physically distinct because individual gauge-invariant contractions change. Thus the parallel alignment either has an instability or is continuously degenerate for the ordinary degree-four potential and the displayed phase potential. The exact polynomial identities are given in Eq.~\eqref{eq:exact-orientation-polynomials}.

This restriction concerns the proposed alignment and the stated class of potentials. A different alignment of the same multiplets, higher-dimensional operators, additional Higgs fields or quantum corrections can change the conclusion and require their own vacuum calculation. The reduced phase and production formulas remain conditional on realizing the assumed vevs in such a completion.

\subsection{A stable vacuum with a neutral symmetric Higgs multiplet}
\label{sec:stable-extended-vacuum}
The obstacle is the relation between the curvatures for changes of orientation. Adding a neutral symmetric Higgs multiplet $\Sigma\sim\mathbf{351}^{\prime}$ allows new couplings that depend on how the antisymmetric fields are oriented relative to $\Sigma$. These couplings can stabilize the chosen orientations while preserving $Q$. In the explicit example below, the phase remains massless until a separate interaction breaks that symmetry.

The symmetric tensor $\Sigma_{IJ}=\Sigma_{JI}$ uses the same two lower indices as $A^{(r)}_{IJ}$ in Eq.~\eqref{eq:dark-tensor-derivative}. Its $E_6$ action is $\Sigma\mapsto U^*\Sigma U^\dagger$, where $U$ acts on the fundamental. The irreducible tensor obeys $d^{IJK}\Sigma_{JK}=0$ and has canonical norm $z=\tr(\Sigma\Sigma^\dagger)$ \cite{E6Tensors,E6NovelIntermediate}. Its kinetic term is $\tr[(D_\mu\Sigma)^\dagger D^\mu\Sigma]$. The covariant derivative is $D_\mu\Sigma=\partial_\mu\Sigma+ig_G(\mathcal G_\mu^{\rm T}\Sigma+\Sigma\mathcal G_\mu)$, with $\mathcal G_\mu=\mathcal G_\mu^a T_{27}^a$. We retain the canonical antisymmetric norms $I_r=\tr(A^{(r)}A^{(r)\dagger})/2$. The additional multiplet contains an $SO(10)$ singlet and a $\mathbf{54}$ component whose Pati--Salam singlet can acquire a vev. Let their orthonormal tensors be $P$ and $T$, defined explicitly in Appendix~\ref{app:stable-benchmark}. The proposed configuration is
\begin{equation}
 \langle A^{(1)}\rangle=\langle A^{(2)}\rangle=\frac v{\sqrt2}e,
 \qquad \langle\Sigma\rangle=\frac{sP+wT}{\sqrt2},
 \qquad (v,s,w)=(1,4,1)E_0,
 \label{eq:stable-extension-vevs}
\end{equation}
where $E_0$ is an arbitrary common energy scale. The connected residual algebra remains Eq.~\eqref{eq:full-E6-stabilizer}.

An explicit potential can be given with real coefficients. Define $R_r=A^{(r)}A^{(r)\dagger}$, $R_\Sigma=\Sigma\Sigma^\dagger$, and
\begin{equation}
 \begin{aligned}
 T_2&=\tr(R_\Sigma^2),\\
 T_3&=d^{\mu\nu\lambda}d_{\xi\eta\lambda}
       \Sigma_{\mu\sigma}\Sigma_{\nu\tau}
       \bar\Sigma^{\xi\sigma}\bar\Sigma^{\eta\tau},\\
 T_4&=\sum_{\alpha\beta}|\mathcal N_{\alpha\beta}(\Sigma)|^2,\\
 C_A&=d^{\mu\nu\lambda}d^{\sigma\tau\rho}
       A^{(1)}_{\mu\sigma}A^{(2)}_{\nu\tau}\Sigma_{\lambda\rho},\\
 C_\Sigma&=d^{\mu\nu\lambda}d^{\sigma\tau\rho}
       \Sigma_{\mu\sigma}\Sigma_{\nu\tau}\Sigma_{\lambda\rho}.
 \end{aligned}
 \label{eq:extended-potential-invariants}
\end{equation}
Here $\mathcal N$ uses Eq.~\eqref{eq:full-quartic-covariants}. The potential is
\begin{equation}
 \begin{aligned}
 V_0={}&m_A^2(I_1+I_2)+m_\Sigma^2z
 +u_A(I_1^2+I_2^2)+u_\Sigma z^2+\lambda_A I_1I_2
 +\eta(I_1+I_2)z\\
 &+\kappa_A\mathcal I_{\rm align}
 +a_A\sum_r\tr(R_r^2)+b_\Sigma T_2+c_\Sigma T_3+d_\Sigma T_4\\
 &+\beta\sum_r\tr(R_rR_\Sigma)
 +\rho\operatorname{Re}\tr\!\left(A^{(1)}\Sigma^\dagger
                               A^{(2)}\Sigma^\dagger\right)
 +\tau\operatorname{Re}C_A+\sigma\operatorname{Re}C_\Sigma .
 \end{aligned}
 \label{eq:stable-extension-potential}
\end{equation}
Under $E_6$, the matrices $R_r$, $R_\Sigma$ and $A^{(r)}\Sigma^\dagger$ all transform by the same similarity transformation with $U^*$ on the left and $U^{\rm T}$ on the right. Their trace contractions are invariant, and the two invariant $d$ tensors make $C_A$ invariant. The norm terms in the first line depend on the magnitudes alone. The tensor contractions in the later lines also respond to changes in the internal directions. For instance, $\tr(R_rR_\Sigma)$ is the sum of the squared absolute values of the entries of $\Sigma^\dagger A^{(r)}$. It can change when the two antisymmetric tensors turn together relative to $\Sigma$, even though their mutual overlap stays fixed. The full stability calculation uses the combined curvature from all these terms.

The parameters $m_A^2,m_\Sigma^2$ are signed quadratic coefficients, $\tau,\sigma$ have mass dimension one, and the other coefficients are dimensionless. We specify a tree-level benchmark and set the remaining coefficients to zero at the chosen renormalization scale. Quantum stability requires a calculation of the radiative corrections. Appendix~\ref{app:stable-benchmark} gives the numerical inputs, the exact stationarity prescription and the checks of every scalar component.

The quartic form has a uniform positive lower bound. With $W=I_1+I_2$, the benchmark satisfies
\begin{equation}
 V_4\geq0.081\,(W^2+z^2).
 \label{eq:stable-extension-boundedness}
\end{equation}
The cubic terms grow at most as $(W+z)^{3/2}$, so the complete potential grows without bound at large field norm. Direct differentiation at Eq.~\eqref{eq:stable-extension-vevs} gives a stationary point. Its canonical $2106\times2106$ real scalar Hessian has 2042 positive eigenvalues and 64 zero eigenvalues. The zeros consist of 63 broken $E_6$ gauge directions and one relative rephasing. The smallest positive squared mass is
\begin{equation}
 m_{\rm min,heavy}^2\simeq1.6\times10^{-5}E_0^2.
 \label{eq:stable-extension-gap}
\end{equation}
This check includes every component of the three multiplets. The benchmark has the lowest energy throughout the complete complex subspace invariant under its residual group $H$. We refer to it as the $H$-preserving benchmark. Section~\ref{sec:lower-vacuum} gives a lower local minimum of the same tree-level potential outside that subspace.

An additional interaction can give the remaining phase a small mass without shifting the chosen expectation values. We use a positive multiple of the square of the difference between the two antisymmetric fields, with the St\"uckelberg factor making the comparison gauge invariant:
\begin{equation}
 \Delta V_\alpha=\alpha\left\lVert
 A^{(1)}-e^{-4ig_Xb/M}A^{(2)}\right\rVert^2
 =\alpha\left[I_1+I_2
       -2\operatorname{Re}\!\left(Be^{-4ig_Xb/M}\right)\right],
 \qquad \alpha>0.
 \label{eq:stable-extension-phase-square}
\end{equation}
It vanishes at the chosen vacuum and adds a positive semidefinite Hessian. The ordinary cubic interactions fix the two additional physical phase combinations associated with the $\Sigma$ vevs. After including $b$ and removing all 64 broken gauge directions, the completed three-multiplet Higgs--St\"uckelberg sector has 2043 positive physical scalar squared masses. There are no physical flat directions at this point. The surviving light field is
\begin{equation}
 \chi=F\left(-\frac{\eta_1}{v}+\frac{\eta_2}{v}
                       -\frac{4g_Xb}{M}\right),\qquad
 F^{-2}=\frac2{v^2}+\frac{16g_X^2}{M^2},\qquad
 m_\chi^2=\frac{\alpha v^2}{F^2}.
 \label{eq:stable-extension-chi}
\end{equation}
For $g_X=0.1$ and $M=7E_0$, $F\simeq0.71E_0$. At $E_0=10^{16}\,\GeV$, the reference mass $3\times10^{-19}\,\mathrm{eV}$ used in Appendix~\ref{sec:coherent-abundance} requires $\alpha/E_0^2\simeq4.5\times10^{-88}$. The relative symmetry and the ultraviolet input represented by this coefficient are discussed in Sec.~\ref{sec:quality}.

The additional vevs also give a definite full vector spectrum. At zero Abelian kinetic mixing,
\begin{equation}
 \begin{aligned}
 \operatorname{spec}\mathcal M_{\rm gauge,ext}^2=
 \biggl\{&0^{\times 15},
 \left[g_G^2\left(\frac{v^2}{3}+\frac{5w^2}{12}\right)\right]^{\times 24},
 \left[g_G^2\left(\frac{v^2+s^2}{2}+\frac{w^2}{4}\right)\right]^{\times 32},\\
 &\left(\frac{4g_G^2v^2}{3}\right)^{\times 6},
 \frac{2g_G^2}{3}(2v^2+4s^2+w^2),\ M^2+8g_X^2v^2\biggr\}.
 \end{aligned}
 \label{eq:stable-extension-vector-spectrum}
\end{equation}
The last two entries are the neutral $N$ and $X$ masses. Their mixing vanishes for equal $A$ vevs because $\Sigma$ is $X$ neutral. The original two-multiplet formulas remain useful for the core theory; applications to the completed scalar sector use the new thresholds and its scalar mass eigenstates.

\subsection{A lower vacuum with a different residual symmetry}
\label{sec:lower-vacuum}
The potential of Eq.~\eqref{eq:stable-extension-potential}, with the same benchmark couplings, has a lower-energy local minimum. We located it numerically on real configurations with equal antisymmetric tensors and refined it by solving the stationarity equations in all real scalar directions. The imaginary gradients vanish at the real configuration because the coefficients are real and the potential is invariant under complex conjugation. Evaluating the full potential gives the energy difference
\begin{equation}
 V_{\rm low}-V_{H}\simeq-4.1E_0^4.
 \label{eq:lower-vacuum-energy-gap}
\end{equation}
In a separate search, all 2106 real coordinates were varied from 16 random starts. All runs reached the same energy as the lower point to within $10^{-6}E_0^4$. The gauge generators that annihilate the refined vacuum close on $\mathfrak{su}(4)\oplus\mathfrak u(1)$. Both the symmetric and the antisymmetric Higgs fields take values outside the original $H$-invariant subspace. Each antisymmetric multiplet has a single $H$-singlet direction of rank six. At the lower minimum the two antisymmetric tensors remain parallel, but each has rank twelve. A gauge transformation multiplies $A^{(r)}$ by an Abelian phase and acts on its two indices as $A^{(r)}\mapsto U^* A^{(r)}U^\dagger$, where $U$ acts on the fundamental. These operations preserve matrix rank, so the change in alignment cannot be removed by a gauge transformation. The symmetric field provides a second rank test. The two $H$-singlet tensors $P$ and $T$ in Eq.~\eqref{eq:benchmark-symmetric-tensors} have ranks one and ten, so any $aP+bT$ has rank at most eleven. At the lower minimum $\Sigma$ has rank 27. Since its gauge transformation also preserves rank, it lies outside all gauge images of the $H$-singlet subspace. The new symmetric alignment increases $T_4$, lowering the contribution $d_\Sigma T_4$ to the energy.

At $\alpha=0$, the Hessian of the 2106 real Higgs coordinates before restoring $b$ has 2043 positive eigenvalues and 63 zeros, with smallest positive squared mass about $1.7\times10^{-2}E_0^2$. The zeros correspond to the 62 broken $E_6$ directions and the relative rephasing of the two antisymmetric tensors. Restoring $b$ and adding the positive interaction in Eq.~\eqref{eq:stable-extension-phase-square} leaves the energy unchanged at both minima because the antisymmetric tensors align and their invariant relative angle vanishes. At the lower minimum it gives a positive squared mass to the physical relative phase and leaves 63 gauge zero directions. For the same $g_X$ and $M$ as at the $H$-preserving benchmark, the phase width is $F\simeq0.72E_0$.

The rank increase gives 24 massive and 57 massless Weyl components at tree level with the same family matrix. If the 16 massless vectors and the light Weyl fermions are relativistic and unconfined, they contribute
\begin{equation}
 \Delta N_{\rm eff}^{\rm low}=\frac{527}{7}\left(\frac{T_D}{T_\nu}\right)^4.
 \label{eq:lower-vacuum-radiation}
\end{equation}
This gives about $3.5$ for the same high-temperature decoupling assumptions as in Sec.~\ref{sec:residual-light-spectrum}. The full residual representations, thresholds and matching for this branch require a separate calculation. The lower stationary point establishes that the $H$-preserving benchmark is a false vacuum of the stated classical potential. The absolute minimum, the tunnelling action and cosmological selection remain open.

\subsection{Explicit angular breaking and the instanton selection rule}
The first-harmonic choice $a_1=a_2=a_3=0$ gives $m_\chi^2=a_0v_G^2v_1v_2/F^2$. At equal vevs $v_1=v_2=v_G=v_0$,
\begin{equation}
 m_\chi^2=2a_0v_0^2\left(1+\frac{8g_X^2v_0^2}{M^2}\right),
 \qquad
 a_0=\frac{m_\chi^2}{2v_0^2\left(1+8g_X^2v_0^2/M^2\right)}.
 \label{eq:coefficient-for-fixed-mass}
\end{equation}
This expresses a chosen symmetry-breaking coefficient through a target mass. Changing $M$ or the vevs rescales that input through the kinetic normalization. We use no instanton estimate to assign $a_0$ or $\alpha$.

A unit $E_6$ instanton has action $S_I=8\pi^2/g_G^2$ and supplies six left-handed zero modes per chiral $\mathbf{27}$ \cite{tHooft}. The null family of the antisymmetric Yukawa matrix leaves its six modes unsaturated in a closed bosonic amplitude with the displayed interactions. At any topological charge the exact $Q$ Ward identity in Eq.~\eqref{eq:exact-Q-anomalies} forbids a contribution to a $Q$-charged dressed operator. A microscopic calculation of the symmetry-breaking coefficient requires a specified interaction that breaks $Q$.

For completeness, a Higgs monomial evaluated in the two-component core vacuum, with signed phase powers $n_r$, satisfies
\begin{equation}
 t_N(n_1+n_2)=0,\qquad -2n_1+2n_2=0
 \quad\Longrightarrow\quad n_1=n_2=0.
 \label{eq:general-phase-selection}
\end{equation}
In the completed vacuum the nonzero symmetric-Higgs components have signed powers $n_0,n_{54}$ and instead require
\begin{equation}
 n_1+n_2-2n_0+n_{54}=0,\qquad n_1=n_2.
 \label{eq:extended-phase-selection}
\end{equation}
The ordinary cubic interactions can fix common phases, while the relative $Q$ direction remains flat. Stückelberg dressing allows harmonics $e^{in\Theta}$ by gauge invariance; their nonzero coefficients explicitly violate $Q$. The constant-mass coherent solution in Appendix~\ref{sec:coherent-abundance} assumes that this extra breaking and an initial displacement have been supplied.

\section{Gauge-invariant operators and ultraviolet protection}
\label{sec:quality}
The exact symmetry in Eq.~\eqref{eq:relative-global-symmetry} forbids a Goldstone potential throughout the symmetry-preserving theory. Gauge invariance separately permits the dressed invariant $\mathcal J$. We now use an illustrative parent interaction to quantify the explicit $Q$ breaking needed for a target mass. The local tensor--scalar duality of Appendix~\ref{sec:stueckelberg} fixes neither a coefficient nor its ultraviolet suppression.

A scalar-parent realization gives a direct test. Suppose that the shifting field is the phase of an $E_6$-singlet scalar $S_0$ with $X=+1$,
\begin{equation}
 S_0=\frac{v_S+\rho_S}{\sqrt2}e^{ib/v_S},
 \qquad M=g_Xv_S,
 \qquad v_S<\Lambda,
 \label{eq:quality-scalar-parent}
\end{equation}
where $\Lambda$ is the cutoff of this illustrative parent theory. Integrating out $\rho_S$ gives the St\"uckelberg kinetic term, as in the scalar construction of Ref.~\cite{CorianoPortal}. Gauge invariance also permits the dimension-six potential term
\begin{equation}
 \Delta V=-\left[\frac{c_6}{\Lambda^2}(S_0^\dagger)^4
          A^{(1)\dagger}\!\cdot A^{(2)}+\mathrm{h.c.}\right].
 \label{eq:quality-parent-operator}
\end{equation}
Its total $X$ charge is $-4+2+2=0$, and the hidden indices form an $E_6$ singlet. On the aligned background the coefficient of its first harmonic is
\begin{equation}
 \Delta U=-\Delta C_1\cos(\Theta+\arg c_6),
 \qquad \Delta C_1=\frac{|c_6|v_S^4v_1v_2}{4\Lambda^2}.
 \label{eq:quality-parent-curvature}
\end{equation}
This local interaction can be a matching coefficient of the parent theory; its occurrence does not require the particular $E_6$ instanton studied above. It gives an explicit realization of the dressing already allowed in the low-energy scalar description. The example adds a heavy radial field but no additional surviving phase, since the phase of $S_0$ is the existing field $b$. The neutral scalar $S$ in the optional fermion extension is a different field.

To keep the magnitude of this extra curvature below a fraction $\delta$ of a target $m_\chi^2$, a sufficient condition without cancellations is
\begin{equation}
 |c_6|\leq
 \frac{4\delta\,m_\chi^2F^2\Lambda^2}{v_S^4v_1v_2}.
 \label{eq:quality-parent-bound}
\end{equation}
This condition bounds the absolute curvature amplitude and is independent of the relative phase. It does not control a cancellation between large harmonics. The magnitude of the suppression follows from the chosen vacuum scales, cutoff and target mass; gauge invariance alone supplies no such bound on $c_6$. A fundamental tensor construction may have different allowed charged events and no scalar parent of this form. It must still specify the global charge and periodicity data that decide which dressed operators occur and how their coefficients are suppressed.

For $v_S=10^{17}\,\GeV$, $\Lambda=\mpl$, equal $A$ vevs, large $M$ and the reference mass $m_\chi=3\times10^{-19}\,\mathrm{eV}$, Eq.~\eqref{eq:quality-parent-bound} gives $|c_6|\lesssim1.1\times10^{-86}$ at $\delta=1$. This is an explicit hierarchy required of the parent interaction. A gravitational source of periodic curvature also breaks $Q$ and must satisfy the same curvature budget. Matching the harmonic spectra considered in Ref.~\cite{RezaeiAkbarieh2026Wormhole} to this charge assignment and vacuum requires a specified gravitational completion. We retain the mass as a free effective parameter.

The common rephasing in Eq.~\eqref{eq:PQ} acts on the two core vevs along a broken non-Abelian gauge direction, so it supplies no second physical axion at this alignment. In the completed vacuum the ordinary cubic interactions fix the two additional common-phase combinations, while the relative field in Eq.~\eqref{eq:stable-extension-chi} remains protected by $Q$. Residual strong dynamics that preserves $Q$ cannot gap the associated Goldstone direction; further condensates can change its mixture of fields. A different vacuum or additional explicit breaking requires its own phase and anomaly calculation.

\section{Quantum scope and the effective cutoff}
\label{sec:quantum-cutoff}
The boundedness and the two local stability results apply to the displayed classical potential. The coefficients of additional symmetry-allowed scalar operators are set to zero at the chosen scale and can be generated by running. The exact relative symmetry protects its phase potential in the symmetric limit. Quantum stability of the heavy directions and compatibility of any weak-coupling benchmark with the portal and cosmology require separate calculations.

Quantum stability requires a separate calculation of the effective potential and its shifted tadpoles. The large representation dimensions make the loop expansion important even for quartics of order unity \cite{MachacekVaughnQuartic}. For example, write $z=\varphi_a\varphi_a/2$ with $a=1,\ldots,702$. The contribution of $u_\Sigma z^2$ alone to its one-loop beta function is
\begin{equation}
\left.\mu\frac{\dd u_\Sigma}{\dd\mu}\right|_{u_\Sigma^2}
=\frac{702+8}{8\pi^2}u_\Sigma^2,
\qquad
\frac{1}{u_\Sigma}
\left.\mu\frac{\dd u_\Sigma}{\dd\mu}\right|_{u_\Sigma^2}
\simeq19.
\label{eq:benchmark-scalar-loop-diagnostic}
\end{equation}
This is one contribution to the coupled beta functions. Other scalar, gauge and Yukawa terms and the split thresholds must be included before assigning a perturbative ultraviolet interval to the displayed point. The small physical gap also makes positivity sensitive to loop corrections projected onto its light modes. No quantum-stable vacuum is established by the tree-level Hessian.

An exact continuous family of classical examples is obtained by replacing $V_0$ by $\zeta V_0$ with $\zeta>0$. The vevs and stationary directions remain fixed; the boundedness coefficient, scalar squared masses and vacuum-energy differences acquire the same factor $\zeta$. The ratios defining the local parameter neighbourhood remain unchanged. At a renormalization scale proportional to $\sqrt\zeta E_0$, pure-scalar one-loop terms scale as $\zeta^2$, whereas the tree potential scales as $\zeta$. Taking $g_G^2$ and the squared Yukawa couplings of order $\zeta$, and $g_X^2$ of order $\zeta^2$ at fixed $M/E_0$, also suppresses gauge and fermion loop terms relative to the classical potential. The stronger Abelian scaling controls the threshold proportional to $g_X^2M^2$. This supplies a parametric weak-coupling family. These scalings apply to the displayed effective action. Higher-dimensional operators must be suppressed consistently at the matching cutoff. A numerical member requires its own threshold matching and quantum stability calculation.

Above every symmetry-breaking and particle threshold, the completed field content has
\begin{equation}
b_0=\frac{11}{3}C_2(E_6)-\frac23\sum_{\rm Weyl}T(R_f)
-\frac13\sum_{\rm complex}T(R_s)=-40,
\qquad
\alpha_G^{-1}(\mu)=\alpha_G^{-1}(\mu_h)
-\frac{40}{2\pi}\ln\frac\mu{\mu_h}.
\label{eq:completed-ultraviolet-running}
\end{equation}
Here $\mu_h$ lies above all masses included in the fully active coefficient. The formal one-loop pole is
\begin{equation}
\mu_L=\mu_h\exp\!\left[\frac{2\pi\alpha_G^{-1}(\mu_h)}{40}\right].
\label{eq:completed-gauge-pole}
\end{equation}
For $\alpha_G^{-1}(\mu_h)=33$, it lies at approximately $1.8\times10^2\mu_h$. Applying this coefficient from $10^{15}\,\mathrm{GeV}$ would give approximately $1.8\times10^{17}\,\mathrm{GeV}$. A different, explicitly chosen weak gauge boundary condition, $\alpha_G^{-1}(10E_0)=100$, gives $\alpha_G^{-1}(100E_0)\simeq85$. These are gauge-running diagnostics; the scalar loop condition must be checked independently.

Below the complete heavy spectrum, the appropriate couplings are those of the residual factors. In a step-function decoupling scheme their leading logarithms have the form
\begin{equation}
\alpha_a^{-1}(\mu)=\alpha_a^{-1}(\mu_0)
+\frac{b_a^{\rm light}}{2\pi}\ln\frac\mu{\mu_0}
+\sum_j\frac{\Delta b_{a,j}}{2\pi}
\ln\frac\mu{m_j}\,\Theta(\mu-m_j),
\label{eq:residual-threshold-running}
\end{equation}
where $\mu_0$ is below the thresholds displayed in the sum and finite threshold constants are additional matching terms. The spectrum in Eq.~\eqref{eq:stable-extension-vector-spectrum} and the full scalar Hessian supply the masses; a completed running analysis must also assign each mass eigenstate to its residual representation. Using the two-multiplet coefficient $-12$ for this three-multiplet spectrum would omit the neutral symmetric multiplet.

The physical domain is restricted by a cutoff $\Lambda_{\rm EFT}$ lying above every retained mass and below the scale where any coupled interaction loses perturbative control. In the unscaled benchmark the scalar masses extend to about $8.4E_0$, the largest Higgs vev is $4E_0$, and for $M=7E_0$ and $g_X=0.1$ the Abelian vector mass is about $7.0E_0$. A choice of cutoff must therefore accommodate these thresholds. The coefficient $M$ enters a physical vector mass and cannot be placed beyond that cutoff while retaining the vector as a propagating field. Formal large-$M$ limits describe the normalization of the low-energy phase after the heavy vector is removed.

The Wess--Zumino terms impose additional conditions on the anomalous effective theory \cite{PreskillAnomalies,CIK,AnastasopoulosBDK}. With the coefficients defined by Eq.~\eqref{eq:WZ}, dimension-five power counting requires
\begin{equation}
\frac{|\mathfrak a_G|\Lambda_{\rm EFT}}{4\pi M}\ll1,
\qquad
\frac{|\mathfrak a_X|\Lambda_{\rm EFT}}{4\pi M}\ll1.
\label{eq:wz-eft-power-counting}
\end{equation}
These are necessary expansion criteria; a channel-dependent scattering analysis determines the actual unitarity limit. Canonically normalized gravitational fluctuations similarly introduce the parameter $|\mathfrak a_R|\Lambda_{\rm EFT}^3/(4\pi M\overline M_{\rm Pl}^2)$. The gauge, scalar and anomaly restrictions jointly determine the cutoff. None of the periodic coefficients is inferred by extending an instanton integral beyond this domain.

\section{Physical mediator couplings and production vertices}
\label{sec:portal-vertices}
The generalized $B-L$ assignment sets the visible current of the Abelian vector. This vector also receives the hypercharge contribution induced by kinetic mixing. At the $H$-preserving alignment, a neutral $E_6$ generator mixes with it when $v_1\ne v_2$. The rates below describe that alignment and its scalar completion in Sec.~\ref{sec:stable-extended-vacuum}.

\subsection{Canonical neutral-vector spectrum}
For compact notation define
\begin{equation}
 A_V=M^2+4g_X^2v^2,\qquad
 D_V=g_N^2t_N^2v^2,\qquad
 E_V=2g_Xg_Nt_N(v_2^2-v_1^2),\qquad
 v^2=v_1^2+v_2^2.
\end{equation}
After Eq.~\eqref{eq:portal-canonical-fields}, the hidden neutral mass matrix in the canonical basis $\boldsymbol V=(\widehat B,N)$ is
\begin{equation}
 \widehat{\mathcal M}_V^2=
 \begin{pmatrix}
 A_V/r_\epsilon^2&E_V/r_\epsilon\\
 E_V/r_\epsilon&D_V
 \end{pmatrix}.
 \label{eq:portal-vector-matrix}
\end{equation}
Its determinant is $D_V\Sigma_X^2/r_\epsilon^2$, where $\Sigma_X^2=M^2+\kappa^2$ and $\kappa=4g_Xv_1v_2/v$. It is positive for the selected vacuum and $|\epsilon|<1$. An orthogonal transformation gives two massive neutral eigenstates $Z'_I$.

For a visible fermion chirality $f_\alpha$, define
\begin{equation}
 g_{f_\alpha}^{X,\rm eff}
 =g_Xx_f-\epsilon g_YY_{f_\alpha},
 \qquad
 \boldsymbol g_{f_\alpha}
 =\left(\frac{g_{f_\alpha}^{X,\rm eff}}{r_\epsilon},0\right)^{\rm T},
 \label{eq:visible-coupling-vector}
\end{equation}
where $x_f=B-L$ is given in Eq.~\eqref{eq:visible-BL-charges}. The first term is the direct generalized $B-L$ interaction; the second is the kinetic-mixing correction. Because $B-L$ is vectorlike for each charged Dirac fermion, $x_{f_L}=x_{f_R}$, while the two hypercharges generally differ.

Electroweak breaking adds a small mass mixing generated by the hypercharge part of the coupling. Let $\langle H_{\rm SM}\rangle=(0,v_w/\sqrt2)^{\rm T}$, $g_Z=\sqrt{g_2^2+g_Y^2}$, $s_w=g_Y/g_Z$, and $Z_0=(g_2W^3-g_Y\widehat Y)/g_Z$. Since $X(H_{\rm SM})=0$, its kinetic term gives
\begin{equation}
 \frac{v_w^2}{8}
 \left(g_2W^3-g_Y\widehat Y+
       \frac{\epsilon g_Y}{r_\epsilon}\widehat B\right)^2
 =\frac12m_{Z0}^2(Z_0+\zeta\widehat B)^2,
 \quad
 m_{Z0}^2=\frac{g_Z^2v_w^2}{4},
 \quad
 \zeta=\frac{\epsilon s_w}{r_\epsilon}.
\end{equation}
The photon remains massless. The other neutral fields have the mass matrix
\begin{equation}
 \mathcal M_{\rm neutral}^2=
 \begin{pmatrix}
 m_{Z0}^2&\zeta m_{Z0}^2&0\\
 \zeta m_{Z0}^2&A_V/r_\epsilon^2+\zeta^2m_{Z0}^2&E_V/r_\epsilon\\
 0&E_V/r_\epsilon&D_V
 \end{pmatrix}_{(Z_0,\widehat B,N)}.
 \label{eq:electroweak-portal-matrix}
\end{equation}
The direct $B-L$ interaction changes the current coupled to the heavy eigenstates but introduces no additional Higgs-induced mass entry because the visible Higgs has zero $B-L$. Equations below are written in the unbroken electroweak theory; amplitudes below $v_w$ must use the final neutral-vector rotation.

\subsection{The vector--radial--axi-Higgs vertex}
Expand the hidden scalar kinetic terms before fixing the radial modes. Their factors $(1+h_r/v_r)^2$ multiply the phase squares in Eq.~\eqref{eq:phase-kinetic}. The terms containing one vector, one radial field and one phase are
\begin{equation}
 \mathcal L_{Vh\eta}=-2\sum_{r=1}^2h_r
 \left(g_Xq_rB_\mu+g_Nt_NN_\mu\right)\partial^\mu\eta_r .
 \label{eq:unprojected-associated-vertex}
\end{equation}
In unitary gauge Eq.~\eqref{eq:O} gives $\eta_1=-Mv_2\chi/(v\Sigma_X)$ and $\eta_2=Mv_1\chi/(v\Sigma_X)$. Let $h_A=\sum_rR_{Ar}h_r$ be the radial mass eigenstates of Eq.~\eqref{eq:vevs}. Substitution gives
\begin{align}
 \mathcal L_{Vh\chi}&=\sum_A
  (d_{XA}B_\mu+d_{NA}N_\mu)h_A\partial^\mu\chi,\notag\\
 d_{XA}&=-\frac{4g_XM}{v\Sigma_X}
           (v_2R_{A1}+v_1R_{A2}),\notag\\
 d_{NA}&=-\frac{2g_Nt_NM}{v\Sigma_X}
           (-v_2R_{A1}+v_1R_{A2}).
 \label{eq:projected-associated-vertex}
\end{align}
The visible charge assignment does not enter these coefficients: they follow entirely from the hidden scalar projection.

For exchange against a conserved visible fermion current, use the identity
\begin{equation}
 h_A\partial_\mu\chi
 =\frac12(h_A\partial_\mu\chi-\chi\partial_\mu h_A)
 +\frac12\partial_\mu(h_A\chi).
 \label{eq:associated-current-identity}
\end{equation} The total derivative does not contribute after contraction with that current. The canonical vertex vector is therefore
\begin{equation}
 \boldsymbol G_A=
 \left(\frac{d_{XA}}{2r_\epsilon},\frac{d_{NA}}2\right)^{\rm T},
 \qquad
 \mathcal J_A^\mu=h_A\partial^\mu\chi-\chi\partial^\mu h_A .
 \label{eq:scalar-production-current}
\end{equation}

At energies below both heavy neutral-vector masses, eliminating $\boldsymbol V$ produces a contact interaction for each visible chirality,
\begin{align}
 \mathcal L_{\rm eff}
 &\supset\sum_{A,f,\alpha}
 \mathcal C_{Af_\alpha}
 \bigl(\overline f\gamma_\mu P_\alpha f\bigr)
 \mathcal J_A^\mu,\notag\\
 \mathcal C_{Af_\alpha}
 &=-\boldsymbol g_{f_\alpha}^{\rm T}
 (\widehat{\mathcal M}_V^2)^{-1}\boldsymbol G_A\notag\\
 &=-\frac{g_Xx_f-\epsilon g_YY_{f_\alpha}}{2\Sigma_X^2}
 \left(d_{XA}-\frac{E_V}{D_V}d_{NA}\right).
 \label{eq:portal-production-operator}
\end{align}
Eliminating the neutral-vector mixing gives a closed expression at unequal vevs. With $v^2=v_1^2+v_2^2$, substitution of Eq.~\eqref{eq:projected-associated-vertex} into Eq.~\eqref{eq:portal-production-operator} yields
\begin{equation}
 \mathcal C_{Af_\alpha}=
 \frac{4g_XM\bigl(g_Xx_f-\epsilon g_YY_{f_\alpha}\bigr)}
      {v^3\Sigma_X^3}
 \left(v_2^3R_{A1}+v_1^3R_{A2}\right).
 \label{eq:unequal-vev-production}
\end{equation}
The cancellation of $g_Nt_N$ follows from integrating out both massive neutral directions. This limit requires external momenta below both masses; it cannot be continued to a vanishing second-vector mass at fixed momentum. Orthogonality of the two-dimensional radial rotation then gives
\begin{equation}
 \sum_{A=1}^2|\mathcal C_{Af_\alpha}|^2=
 \frac{16g_X^2M^2|g_Xx_f-\epsilon g_YY_{f_\alpha}|^2}
      {v^6\Sigma_X^6}(v_1^6+v_2^6).
 \label{eq:radial-production-sum}
\end{equation}
Thus radial mixing redistributes the contact strength between the two scalar states without changing this sum. Their different masses and production thresholds still enter the physical rates. The result assumes the closed two-radial sector specified in Eq.~\eqref{eq:vevs}; mixing with additional components of the full $E_6$ multiplets requires extending the scalar rotation.

The two contributions are now explicit. The term proportional to $g_Xx_f$ remains when $\epsilon=0$; the term proportional to $\epsilon g_YY_{f_\alpha}$ is the usual kinetic-mixing correction. Near a mediator pole the contact expression must be replaced by
\begin{equation}
 \mathcal C_{Af_\alpha}(s)=
 \boldsymbol g_{f_\alpha}^{\rm T}
 [s\boldsymbol1-\widehat{\mathcal M}_V^2]^{-1}
 \boldsymbol G_A,
 \qquad
 \mathcal C_{Af_\alpha}(0)=\mathcal C_{Af_\alpha}.
 \label{eq:full-portal-propagator}
\end{equation}

The equal-vev limit gives a simpler result. For $v_1=v_2=v_0$ and symmetric radial parameters, $h_\pm=(h_1\pm h_2)/\sqrt2$, $E_V=0$, $\kappa=2\sqrt2g_Xv_0$, and
\begin{equation}
 d_{X+}=-\frac{4g_XM}{\Sigma_X},\qquad d_{X-}=0,
 \qquad
 \mathcal C_{+f_\alpha}
 =\frac{2g_XM}{\Sigma_X^3}
 \left(g_Xx_f-\epsilon g_YY_{f_\alpha}\right),
 \qquad
 \mathcal C_{-f_\alpha}=0 .
 \label{eq:symmetric-portal-coefficient}
\end{equation}
The symmetric radial mode is therefore produced with $\chi$ through the generalized $B-L$ vector. The antisymmetric radial mode does not couple through this channel in the symmetric vacuum. The direct interaction is nevertheless bounded by the symmetry-breaking scale. At $\epsilon=0$,
\begin{equation}
 \left|\frac{\mathcal C_{+f}}{x_f}\right|
 =\frac{2g_X^2M}{(M^2+8g_X^2v_0^2)^{3/2}}
 \leq\frac{1}{6\sqrt3\,v_0^2},
 \label{eq:portal-strength-bound}
\end{equation}
where the maximum occurs at $M=2g_Xv_0$. In the regime $M\gg2\sqrt2g_Xv_0$ that saturates $F\simeq v_0/\sqrt2$, the coefficient is smaller, $\mathcal C_{+f}/x_f\simeq2g_X^2/M^2$. The generalized charge assignment therefore fixes a genuine portal but does not produce an observable low-energy interaction at a GUT-scale $v_0$.

The relation to the physical phase width can be made exact in this symmetric limit. Define the dimensionless fraction $z=F/F_{\max}=M/\Sigma_X$, with $0<z<1$. At $\epsilon=0$,
\begin{equation}
 \frac{\mathcal C_{+f}}{x_f}
 =\frac{z(1-z^2)}{4v_0^2},
 \qquad F_{\max}=\frac{v_0}{\sqrt2}.
 \label{eq:phase-portal-correlation}
\end{equation}
Differentiating $z(1-z^2)$ places the maximum at $z=1/\sqrt3$, equivalent to $M=2g_Xv_0$ and the bound in Eq.~\eqref{eq:portal-strength-bound}. Approaching the largest allowed phase width, $z\to1$, suppresses the visible contact amplitude. Increasing $M$ therefore cannot independently maximize both quantities. This correlation follows from the same vacuum and gauge projection used to define the physical field.

For a massless Dirac fermion, the partonic cross section is
\begin{equation}
 \sigma_{f\bar f\to h_A\chi}(s)=
 \frac{|\mathcal C_{Af_L}(s)|^2+
       |\mathcal C_{Af_R}(s)|^2}{96\pi N_c}
 \frac{\lambda^{3/2}(s,m_A^2,m_\chi^2)}{s^2},
 \qquad
 \lambda(s,x,y)=(s-x-y)^2-4xy .
 \label{eq:portal-associated-cross-section}
\end{equation}
The reaction is possible only for $\sqrt{s}\geq m_A+m_\chi\simeq m_A$. Since the hidden radial mass is naturally of order the hidden breaking scale, the process requires visible energies near $10^{15}\,\GeV$ for the chosen parameters. The direct $B-L$ coupling removes the additional suppression by $\epsilon$, but it does not remove this threshold.

The required abundance provides a stronger distinction between particle production and a coherent field. The present number-to-entropy ratio of a component with density $\Omega_\chi h^2$ is
\begin{equation}
 Y_\chi^{\rm req}\equiv\frac{n_\chi}{s_0}
 =\frac{\Omega_\chi h^2}{m_\chi}\frac{\rho_c/h^2}{s_0}
 \simeq4.4\times10^{21}
 \left(\frac{\Omega_\chi h^2}{0.12}\right)
 \left(\frac{10^{-22}\,\mathrm{eV}}{m_\chi}\right).
 \label{eq:required-yield}
\end{equation}
Here $\rho_c/h^2\simeq1.1\times10^{-5}\,\GeV\,\mathrm{cm}^{-3}$ and $s_0\simeq2.9\times10^3\,\mathrm{cm}^{-3}$, while $\Omega_{\rm DM}h^2\simeq0.12$ \cite{Planck2018}. This number counts nonrelativistic quanta, or the adiabatic occupation of an oscillating coherent field. It does not convert an arbitrary relativistic gas into matter. For a thermal relativistic real scalar with temperature $T_\chi$ and negligible chemical potential,
\begin{equation}
 Y_\chi^{\rm th}=\frac{45\zeta(3)}{2\pi^4g_{*s}}
                 \left(\frac{T_\chi}{T}\right)^3,
 \qquad \langle p_\chi\rangle\simeq2.7T_\chi,
 \qquad \Delta N_{\rm eff}^{\chi,\rm hot}
              =\frac47\left(\frac{T_\chi}{T_\nu}\right)^4
 \label{eq:hot-axi-Higgs-population}
\end{equation}
as long as it is relativistic. Here $g_{*s}$ counts visible entropy degrees of freedom. An ordinary decoupled bath or dilute freeze-in population has nowhere near the required cold occupation. Moreover, typical redshifted momenta from high-energy production greatly exceed an ultralight mass even today, so that component is hot radiation and is constrained through its energy density and free streaming. The radial decay calculated below can therefore populate dark radiation without providing coherent dark matter. Its contribution depends on the parent population and branching fractions, which are not fixed here. A specially prepared, extremely occupied low-momentum nonthermal state is not excluded by a universal bound on $Y$; the present calculation explicitly adopts coherent misalignment as its source of cold ultralight matter.

For the stable scalar completion of Sec.~\ref{sec:stable-extended-vacuum}, the equal-$A$-vev limit retains Eq.~\eqref{eq:stable-extension-chi} and has no $X$--$N$ mass mixing. The $X$ production vertex couples to $h_+=(h_1+h_2)/\sqrt2$. This radial field mixes with the two real $\Sigma$ singlet amplitudes. If $R_{a+}$ is its projection onto the three corresponding canonical mass eigenstates $H_a$, their direct production coefficients obey
\begin{equation}
 C_{f_\alpha,a}=R_{a+}C_{f_\alpha,+},\qquad
 \sum_{a=1}^3|C_{f_\alpha,a}|^2=|C_{f_\alpha,+}|^2,
 \label{eq:extended-production-sum}
\end{equation}
where $C_{f_\alpha,+}$ is the equal-vev core coefficient above. Thus the stable completion distributes that strength among three scalar mass eigenstates. The unequal-vev two-field expressions require a fresh projection when the additional vevs are retained.

\subsection{Radial decay and the low-energy parameter set}
The same projected kinetic sector gives a source entirely within the dark theory. For the equal-vev vacuum and a common radial displacement $r=v_0+h_+/\sqrt2$, eliminating the heavy gauge directions gives
\begin{equation}
 F(r)=\frac{rM}{\sqrt{2\bigl(M^2+8g_X^2r^2\bigr)}},\qquad
 \mathcal L_{\chi,\rm kin}=\frac12\frac{F^2(r)}{F^2(v_0)}(\partial\chi)^2.
 \label{eq:radial-dependent-phase-width}
\end{equation}
Expanding the ratio to first order yields
\begin{equation}
 \mathcal L_{h_+\chi\chi}=\frac{c_+}{2}h_+(\partial\chi)^2,
 \qquad c_+=\frac{\sqrt2M^2}{v_0\Sigma_X^2},
 \qquad \Sigma_X^2=M^2+8g_X^2v_0^2.
 \label{eq:hidden-radial-decay-vertex}
\end{equation}
The field $\chi$ in these equations is canonical at the vacuum. For a parent of mass $m_+$ the derivative vertex gives
\begin{equation}
 |\mathcal M|=\left|\frac{c_+}{2}(m_+^2-2m_\chi^2)\right|.
 \label{eq:radial-decay-amplitude}
\end{equation} Including the identical-daughter factor in two-body phase space gives
\begin{equation}
 \Gamma_{h_+\to\chi\chi}=
 \frac{c_+^2(m_+^2-2m_\chi^2)^2}{128\pi m_+}
 \sqrt{1-\frac{4m_\chi^2}{m_+^2}}
 \simeq\frac{M^4m_+^3}{64\pi v_0^2\Sigma_X^4}.
 \label{eq:hidden-radial-decay-width}
\end{equation}
In terms of the same phase-width fraction, the small-daughter-mass result is
\begin{equation}
 \Gamma_{h_+\to\chi\chi}\simeq
 \frac{z^4m_+^3}{64\pi v_0^2}.
 \label{eq:phase-radial-decay-correlation}
\end{equation}
At fixed $v_0$ and $m_+$ this partial width remains finite as $z\to1$, even though the visible contact amplitude vanishes. A populated dark radial mode can consequently produce axi-Higgs particles when visible associated production is strongly suppressed. Its population must be specified independently. The last form of Eq.~\eqref{eq:hidden-radial-decay-width} neglects $m_\chi/m_+$. Phase-potential contributions to the vertex are suppressed by the ultralight curvature in this limit. This derivative description requires $m_+$ and the external momenta to lie below the integrated vector scales; otherwise the full gauge and scalar amplitudes must be retained. The parent can also have competing dark decay channels. The parent population and branching fractions are additional inputs to the abundance calculation.

For the completed equal-vev vacuum, let $H_a=R_{a+}h_++R_{a0}h_0+R_{a54}h_{54}$ be the three mixed real singlets. The physical phase width is independent of the $\Sigma$ magnitudes along this equal-vev path, so
\begin{equation}
 c_a=R_{a+}c_+,\qquad
 \Gamma_{H_a\to\chi\chi}
 =\frac{|R_{a+}|^2c_+^2(m_a^2-2m_\chi^2)^2}{128\pi m_a}
 \sqrt{1-\frac{4m_\chi^2}{m_a^2}}.
 \label{eq:extended-radial-decay}
\end{equation}
This formula assumes that $m_a$ lies below every vector integrated out in obtaining the derivative interaction. Associated production has its own threshold $s\geq(m_a+m_\chi)^2$ for each $H_a$ and uses the coefficients in Eq.~\eqref{eq:extended-production-sum}. Thus the inclusive coupling sum does not remove the individual scalar thresholds or predict parent populations.

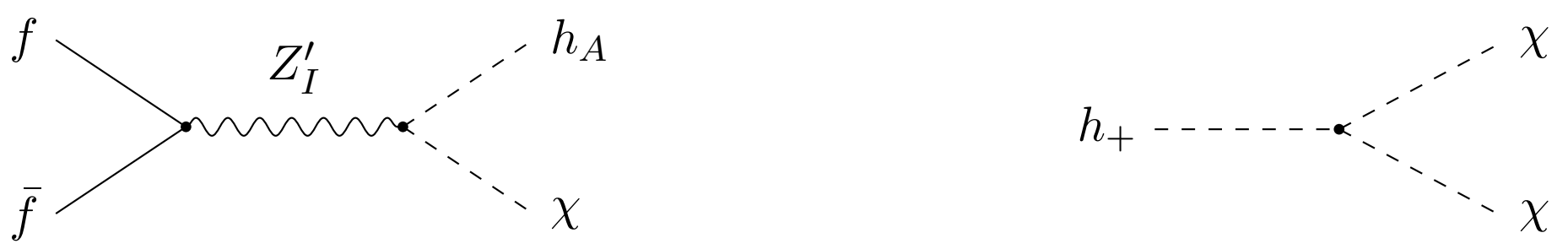
\begin{figure}[t]
\centering
\begin{minipage}{0.49\textwidth}\centering
\begin{tikzpicture}[scale=0.85]
\draw (-1.2,0.8)--(0,0)--(-1.2,-0.8);
\draw[decorate,decoration={snake,amplitude=0.7mm,segment length=2.5mm}] (0,0)--(2,0);
\draw[dashed] (2,0)--(3.2,0.8);
\draw[dashed] (2,0)--(3.2,-0.8);
\node[left] at (-1.2,0.8) {$f$};\node[left] at (-1.2,-0.8) {$\bar f$};
\node[above] at (1,0.15) {$Z'_I$};
\node[right] at (3.2,0.8) {$h_A$};\node[right] at (3.2,-0.8) {$\chi$};
\fill (0,0) circle (1.4pt);\fill (2,0) circle (1.4pt);
\end{tikzpicture}
\end{minipage}
\begin{minipage}{0.49\textwidth}\centering
\begin{tikzpicture}[scale=0.85]
\draw[dashed] (-1,0)--(0.7,0);
\draw[dashed] (0.7,0)--(2.2,0.8);
\draw[dashed] (0.7,0)--(2.2,-0.8);
\node[left] at (-1,0) {$h_+$};
\node[right] at (2.2,0.8) {$\chi$};\node[right] at (2.2,-0.8) {$\chi$};
\fill (0.7,0) circle (1.4pt);
\end{tikzpicture}
\end{minipage}
\caption{Microscopic production channels. Left: a visible fermion pair produces a hidden radial mode and the axi-Higgs through the heavy generalized $B-L$ vectors. The visible vertex contains the direct $B-L$ charge and the hypercharge correction from kinetic mixing; the hidden vertex follows from the phase projection. Right: a populated hidden radial mode decays through its derivative coupling. These amplitudes describe particle production, whereas the cosmologically relevant ultralight abundance is carried by a coherent field displacement.}
\label{fig:hidden-portal-vertices}
\end{figure}

Together, the vector masses and current projections, the radial spectrum, Eq.~\eqref{eq:portal-associated-cross-section}, and the dark partial width specify calculable amplitudes for rare particle processes. Hidden fermion interactions follow from Eq.~\eqref{eq:Yukawa} and remain entirely within the dark sector. The portal therefore fixes the relative charges and the matrix elements that would govern energy exchange after reheating. A common temperature requires an energy-transfer rate larger than the Hubble rate, including off-shell exchange and any accessible on-shell channels. The reheating temperature and initial populations determine whether this condition is met. These quantities do not provide the coherent displacement required by Eq.~\eqref{eq:required-yield}.

With a dominant first harmonic, the local potential is
\begin{equation}
 U=\frac12m_\chi^2\chi^2-\frac{m_\chi^2}{24F^2}\chi^4+\cdots.
 \label{eq:first-harmonic-local-potential}
\end{equation}
The two-harmonic relation is given in Eq.~\eqref{eq:heights}. These coefficients enter the coherent evolution calculated in Appendix~\ref{sec:coherent-abundance}. That section gives both the resulting dark-matter fraction and the Lyman-$\alpha$ limit.

\section{Collisionless and self-interacting dark matter in halos}
\label{sec:two-dirac-sector}
The proposed route to black-hole seeds for Little Red Dots assigns two roles to the dark matter. An effectively collisionless population builds the extended halo and supplies most of its gravitational potential. A smaller self-interacting population can carry heat out of its own core through collisions among its particles. If that core becomes self-gravitating, the outward heat flow can drive gravothermal contraction. We illustrate these roles with two Dirac singlets. The weakly scattering $\psi_2$ supplies the extended halo, and $\psi_1$ self-scatters through a light real scalar. The axi-Higgs can contribute a separate coherent fraction. The singlet masses, couplings and initial populations are inputs to the extension. The calculation determines their scattering rates and specifies the initial conditions to be tested in a dynamical halo calculation.

\subsection{Fields and independent hierarchies}
The two Dirac fermions are singlets of the visible and hidden non-Abelian groups, with equal left- and right-handed $X$ charges. Their Lagrangian is
\begin{equation}
 \begin{gathered}
 \mathcal L_{\psi S}=\sum_{i=1}^2\bar\psi_i(i\gamma^\mu D_\mu-m_i)\psi_i
 -\sum_{i=1}^2y_iS\bar\psi_i\psi_i+\frac12(\partial S)^2-V_S,\\
 D_\mu=\partial_\mu-ig_XB_\mu,\qquad
 X_{\psi_{iL}}=X_{\psi_{iR}}=1,\qquad X_S=0.
 \end{gathered}
 \label{eq:two-dirac-lagrangian}
\end{equation}
Each species separately contributes zero to the cubic and gravitational $X$ anomaly traces. The chiral $E_6$ anomalies and instanton zero-mode count are unchanged. We impose independent vector numbers $\psi_i\to e^{i\beta_i}\psi_i$, with all other fields neutral. They forbid flavour mixing and number-violating masses, making each species stable as the lightest carrier of its number. Annihilation and pair conversion still conserve these numbers and enter the abundance calculation.

Gauge invariance alone allows dressed Majorana operators,
\begin{equation}
 \mathcal L_{\Delta N}=-\frac12\sum_{i=1}^2\sum_{A=L,R}
 \delta m_{iA}\,\psi_i^{\rm T}CP_A\psi_i\,e^{-2ig_Xb/M}+\mathrm{h.c.},
 \qquad P_{L,R}=\frac{1\mp\gamma_5}{2}.
 \label{eq:psi-dressed-Majorana}
\end{equation}
The asymmetric example assumes that high-energy interactions preserve both vector numbers. If they are approximate, sufficient washout and oscillation conditions are
\begin{equation}
 \Gamma_{\Delta N}\ll H\quad\hbox{during asymmetry production},\qquad
 \int_{t_{\rm asy}}^{t_0}\Gamma_{\Delta N}\,\dd t\ll1,\qquad
 \delta m_i(t_0-t_{\rm asy})\ll1.
 \label{eq:psi-number-survival}
\end{equation}
For a constant physical splitting over the age of the Universe, the last condition gives $\delta m_i\ll1.5\times10^{-33}\,\mathrm{eV}$. Matter effects and decoherence require a density-matrix treatment if this sufficient condition is relaxed. Number violation can restore antiparticles and invalidate the conserved asymmetry used below.

The real scalar has no angular coordinate. Its general magnitude portals include the completing symmetric Higgs:
\begin{equation}
 \begin{aligned}
 V_S={}&t_SS+\frac12m_{S0}^2S^2+\frac{\kappa_S}{3!}S^3+\frac{\lambda_S}{4!}S^4\\
 &+\sum_{\Phi=H_{\rm SM},A^{(1)},A^{(2)},\Sigma}
 \left(\mu_{S\Phi}S+\frac12\lambda_{S\Phi}S^2\right)\Phi^\dagger\!\cdot\Phi .
 \end{aligned}
 \label{eq:real-mediator-potential}
\end{equation}
We choose the renormalized tadpole so that $\langle S\rangle=0$ and impose small scalar mixing. At fixed renormalized vevs these magnitude portals preserve the relative angular symmetry and the phase normalization; no direct axi-Higgs Yukawa coupling to the singlets is added. The physical force-mediator mass $m_S$ introduces a further hierarchy. A single hidden portal gives
\begin{equation}
 \delta m_S^2=\frac12\lambda_{SA^{(r)}}v_r^2,\qquad
 |\lambda_{SA^{(r)}}|\lesssim2\times10^{-36}
 \left(\frac{m_S}{1\,\mathrm{MeV}}\right)^2
 \left(\frac{10^{15}\,\GeV}{v_r}\right)^2
 \label{eq:light-scalar-hierarchy}
\end{equation}
when larger contributions do not cancel. The symmetric-Higgs portal similarly requires $|\lambda_{S\Sigma}|(s^2+w^2)/2\lesssim m_S^2$. The singlet-fermion correction is of order $y_1^2m_1^2/(8\pi^2)$ and is comparable with $m_S^2$ in the example below. We also choose
\begin{equation}
 \zeta_{Sh}\simeq\frac{\mu_{SH}^{\rm eff}v_w}{m_h^2-m_S^2},\qquad
 |\sin\zeta_{Sh}|\leq10^{-12},\qquad
 |\mu_{SH}^{\rm eff}|\lesssim6.9\times10^{-11}\,\GeV
 \label{eq:scalar-mixing-reference}
\end{equation}
for $m_h\simeq1.3\times10^2\,\GeV$. Visible scalar searches, supernova emission and cosmological limits depend on the visible branching fraction, dark decays and transport \cite{DevScalarSN,CoxScalarInvisible,FradetteScalarCosmology}; the chosen mixing suppresses visible production, while its persistence requires control of heavy thresholds.

These independent requirements accompany the angular bound $|\Delta\mathcal X|\lesssim m_\chi^2F^2$, including Eq.~\eqref{eq:quality-parent-bound}, and the visible-Higgs thresholds from $\lambda_{Hr}$ and $\lambda_{H\Sigma}$. Each hierarchy requires small matching coefficients or compensating counterterms. The relative $Q$ symmetry protects the angular direction. Number conservation, the mediator hierarchy and quantum stability of the heavy vacuum have their separate conditions, the last discussed in Sec.~\ref{sec:quantum-cutoff}.

\subsection{Scattering and conditional primordial populations}
Scalar exchange produces the attractive Yukawa potential and Born amplitude \cite{TulinYuZurek,RobertsSIDMModel}
\begin{equation}
 \begin{gathered}
 V_{ij}(r)=-\alpha_{ij}\frac{e^{-m_Sr}}{r},\qquad
 \alpha_{ij}=\frac{y_iy_j}{4\pi},\qquad \mu_{ij}=\frac{m_im_j}{m_i+m_j},\\
 \mathfrak f_{ij}(\vartheta)=\frac{2\mu_{ij}\alpha_{ij}}
 {m_S^2+4\mu_{ij}^2u_{\rm rel}^2\sin^2(\vartheta/2)}.
 \end{gathered}
 \label{eq:dirac-born-amplitude}
\end{equation}
For distinguishable species $\dd\sigma/\dd\Omega=|\mathfrak f|^2$. Identical unpolarized fermions require the spin-singlet and spin-triplet weights, including the identical-final-state factor:
\begin{equation}
 \begin{gathered}
 \frac{\dd\sigma_{ii}}{\dd\Omega}=\frac18|\mathfrak f_{ii}(\vartheta)+\mathfrak f_{ii}(\pi-\vartheta)|^2
 +\frac38|\mathfrak f_{ii}(\vartheta)-\mathfrak f_{ii}(\pi-\vartheta)|^2,\\
 \sigma_V^{ii}=\frac32\int\dd\Omega\,\sin^2\vartheta\,\frac{\dd\sigma_{ii}}{\dd\Omega}.
 \end{gathered}
 \label{eq:dirac-identical-scattering}
\end{equation}
This normalized viscosity cross section is suitable for heat transport \cite{YangYuTransport}. In the contact Born regime, $\alpha_{ij}\mu_{ij}/m_S\ll1$ and $\mu_{ij}u_{\rm rel}/m_S\ll1$, it gives
\begin{equation}
 \sigma_V^{ii}\simeq\frac{2\pi\alpha_{ii}^2m_i^2}{m_S^4},\qquad
 \sigma_T^{12}=\int\dd\Omega(1-\cos\vartheta)|\mathfrak f_{12}|^2
 \simeq\frac{16\pi\alpha_{12}^2\mu_{12}^2}{m_S^4}.
 \label{eq:dirac-mixed-contact}
\end{equation}
We select the independent inputs
\begin{equation}
 \begin{gathered}
 m_1=100\,\mathrm{MeV},\qquad m_2=100\,\mathrm{keV},\qquad m_S=1\,\mathrm{MeV},\\
 \alpha_{11}=10^{-3},\qquad y_1=\sqrt{4\pi\alpha_{11}}\simeq0.11,\qquad y_2=10^{-6}.
 \end{gathered}
 \label{eq:two-dirac-transport-point}
\end{equation}
Here $\alpha_{11}m_1/m_S=0.1$. Numerical integration of the exchange amplitudes at $u_{\rm rel}=30\,\mathrm{km\,s^{-1}}$ gives
\begin{equation}
 \begin{gathered}
 \frac{\sigma_V^{11}}{m_1}\simeq1.4\times10^2\,\mathrm{cm^2\,g^{-1}},\qquad
 \frac{\sigma_V^{22}}{m_2}\simeq8.7\times10^{-22}\,\mathrm{cm^2\,g^{-1}},\\
 \sigma_T^{12}\simeq1.6\times10^{-38}\,\mathrm{cm^2}.
 \end{gathered}
 \label{eq:two-dirac-scattering-strengths}
\end{equation}
For an initially unsegregated one-percent first component and comparable pair speeds, the mixed encounter rate relative to the first-species self-encounter rate is at most $6.3\times10^{-11}$. Energy relaxation additionally depends on the transfer per collision. The centre-of-mass kinetic energy is $m_1u_{\rm rel}^2/4\simeq0.25\,\mathrm{eV}$, below real-scalar emission threshold. At $1000$ and $4500\,\mathrm{km\,s^{-1}}$, the full angular integrals give $\sigma_V^{11}/m_1\simeq1.2\times10^2$ and $4.1\times10^1\,\mathrm{cm^2\,g^{-1}}$, respectively. Cluster constraints require the interacting component's own distribution and optical depth. For a shared initial spatial profile, a one-pass removed-mass estimate is
\begin{equation}
 \tau_1=f_1\Sigma_{\rm DM}\frac{\sigma_{11}}{m_1},\qquad
 \frac{\Delta M_{\rm DM}}{M_{\rm DM}}\lesssim f_1(1-e^{-\tau_1})\leq f_1,
 \label{eq:dirac-cluster-stripping}
\end{equation}
where $\sigma_{11}$ is the total cross section and every collision is counted as potentially ejecting a particle \cite{PollackSeeds}. Halo separation, subhalo depletion and the gravitational response require an evolution calculation.

The fractions obey $f_\chi+f_1+f_2=1$. Strong annihilation $\psi_1\bar\psi_1\to SS$ motivates a conditional asymmetric population, with target net yield
\begin{equation}
 \eta_1=\frac{n_{\psi_1}-n_{\bar\psi_1}}{s}
 \simeq4.4\times10^{-11}\left(\frac{f_1}{0.01}\right)
 \left(\frac{100\,\mathrm{MeV}}{m_1}\right)
 \left(\frac{\Omega_{\rm DM}h^2}{0.12}\right).
 \label{eq:dirac-asymmetric-target}
\end{equation}
The asymmetry is an initial condition. Pair conversion and mediator production must be included when deriving it from reheating. At the chosen masses $m_S<2m_e$ and $m_S>2m_2$, so prompt dark decay is open:
\begin{equation}
 \Gamma_{S\to\psi_2\bar\psi_2}=\frac{y_2^2m_S}{8\pi}
 \left(1-\frac{4m_2^2}{m_S^2}\right)^{3/2},\qquad
 \tau_S\simeq1.8\times10^{-8}\,\mathrm{s}.
 \label{eq:light-scalar-dark-decay}
\end{equation}
For comparison with the coherent sector we retain the prescribed partial-component example
\begin{equation}
 \begin{gathered}
 v_0=10^{15}\,\GeV,\quad M\gg2\sqrt2g_Xv_0,\quad
 m_\chi=2\times10^{-20}\,\mathrm{eV},\quad\Theta_i=1,\\
 f_\chi\simeq4.2\times10^{-4},\qquad f_1=0.01,\qquad
 f_2=1-f_\chi-f_1\simeq0.99.
 \end{gathered}
 \label{eq:three-component-reference-fractions}
\end{equation}
The first fraction follows from Eq.~\eqref{eq:misalignment-abundance}; $f_1$ is chosen and $f_2$ fixes the remaining matter density. The mixed harmonic analysis of Ref.~\cite{LiuMixedFDM} gives 95\% fraction limits $0.07$, $0.12$ and $0.65$ at $10^{-23}$, $10^{-22}$ and $10^{-21}\,\mathrm{eV}$, and no useful fraction limit above about $10^{-20}\,\mathrm{eV}$. This comparison assumes a sufficiently cold remaining population and requires the potential and black-hole constraints discussed in Appendix~\ref{sec:coherent-abundance}.

An explicit cold comparison distribution after production is a decoupled Fermi--Dirac shape. For $q=ap$, $a(t_0)=1$ and four spin and particle/antiparticle states,
\begin{equation}
 \begin{gathered}
 \mathfrak f_2(q)=\frac1{\exp[q/(\xi_2T_{\nu0})]+1},\qquad
 T_{\nu0}\simeq1.7\times10^{-4}\,\mathrm{eV},\qquad d_2=4,\\
 \Omega_2h^2=\frac{m_2}{93\,\mathrm{eV}}\frac{d_2}{2}\xi_2^3.
 \end{gathered}
 \label{eq:psi2-thermal-density}
\end{equation}
The required ratio is $\xi_2\simeq0.038$ for the chosen abundance. A neutrino-temperature distribution would give $\Omega_2h^2\simeq2.2\times10^3$. Reheating or dilution must therefore supply this colder distribution; a small $y_2$ does not establish a thermal shape. If a thermal $S$ bath disappears into relativistic $\psi_2$, its entropy raises their temperature by $(9/7)^{1/3}$, so $\xi_2$ is specified after that transfer.

For any final occupation the density and comoving free-streaming distance are
\begin{equation}
 \begin{gathered}
 \rho_2(a)=\frac{d_2}{2\pi^2a^4}\int_0^\infty\dd q\,q^2\sqrt{q^2+a^2m_2^2}\,\mathfrak f_2(q),\\
 \lambda_{\rm FS}=\int_{a_*}^{a}\frac{\langle v_2(a')\rangle}{a'^2H(a')}\,\dd a'.
 \end{gathered}
 \label{eq:psi2-moments-free-streaming}
\end{equation}
For the chosen shape, $\langle p_0\rangle\simeq2.0\times10^{-5}\,\mathrm{eV}$ and $\langle v_2\rangle_{\rm eq}\simeq6.9\times10^{-7}$. Using $H_0\simeq67\,\mathrm{km\,s^{-1}\,Mpc^{-1}}$, $\Omega_r=9.2\times10^{-5}$ and $\Omega_m\simeq0.32$, integration from $T_\nu=1\,\mathrm{MeV}$ to equality gives $\lambda_{\rm FS}\simeq1.3\,\mathrm{kpc}$. Earlier relativistic motion can add at most the starting comoving horizon, about $0.078\,\mathrm{kpc}$. The assumed shape is cold on Lyman-$\alpha$ scales; production determines its actual transfer function.

Scalar decay generally produces a different distribution. In the parent rest frame the daughters have $p_*\simeq0.49\,\mathrm{MeV}$, with comoving momentum $q_*=a_dp_*$; parent motion broadens this spectrum. Once inverse reactions are negligible, visible-entropy conservation gives
\begin{equation}
 Y_2^{\rm final}=Y_2^{\rm initial}+2Y_S^{\rm decay},\qquad
 Y_2^{\rm target}\simeq4.3\times10^{-6}.
 \label{eq:psi2-decay-yield-budget}
\end{equation}
The daughters must fit within this yield and replace the assumed distribution in the free-streaming integral. A short rest-frame lifetime alone does not fix the cosmic decay time or establish coldness. Visible entropy injection requires an additional dilution factor.

Relativistic $S$ and a symmetric Dirac population contribute
\begin{equation}
 \Delta N_{\rm eff}^{\rm light}=\frac47\left(\frac{T_S}{T_\nu}\right)^4
 +2\left(\frac{T_2}{T_\nu}\right)^4,
 \label{eq:light-sector-radiation}
\end{equation}
which becomes $18/7$ at a common neutrino temperature. For a common ratio $0.038$, the relativistic value is approximately $5.4\times10^{-6}$. If $S$ has disappeared, the frozen fermion energy integral instead gives $\rho_2/\rho_{\nu,1}\simeq5.7\times10^{-6}$ at $T_\nu=1\,\mathrm{MeV}$, when the fermions are partly nonrelativistic. These terms add to the full chiral-fermion and residual-vector contribution in Eq.~\eqref{eq:full-light-dark-radiation}; their temperature is an independent condition. At recombination the $100\,\mathrm{keV}$ fermions contribute matter. A viable production history must supply the chosen fractions, asymmetry and cold distribution while respecting all these radiation bounds.

\subsection{Halo relaxation and conditions for seed formation}
A dynamical calculation must establish whether the initially unsegregated one-percent $\psi_1$ population can concentrate through self-scattering until its own gravity dominates its core, starting in the potential dominated by $\psi_2$ \cite{PollackSeeds,RobertsEarlySeeds,RobertsLRDSeeds}. At the benchmark the mixed collision rate is negligible and the heat transport relevant to this test occurs within $\psi_1$. The LRD fits in Ref.~\cite{RobertsLRDSeeds} use $\sigma/m\propto m_{200}^{-4/3}$ normalized at $10^{12}M_\odot$, with approximate normalization range $8.1\times10^2$--$5.4\times10^4\,\mathrm{cm^2\,g^{-1}}$ and $f\sigma/m\sim0.5$--$10\,\mathrm{cm^2\,g^{-1}}$. Our point has $f_1\sigma_V^{11}/m_1\simeq1.4\,\mathrm{cm^2\,g^{-1}}$, a smaller individual cross section and weak velocity dependence up to $1000\,\mathrm{km\,s^{-1}}$. Its collapse and source abundance must therefore be calculated with the scalar-mediated interaction itself. The local encounter time,
\begin{equation}
 t_{\rm rel}^{-1}=\frac{\rho_1}{m_1}\langle\sigma_V^{11}u_{\rm rel}\rangle,
 \label{eq:dirac-relaxation-time}
\end{equation}
is approximately $1.1\times10^6$ years at the assumed local density $\rho_1=1\,M_\odot\,\mathrm{pc^{-3}}$ and the chosen speed. The halo evolution determines how that density changes and how many encounters are required to transport enough heat for collapse. Thermal exchange, particle depletion, baryonic gravity, angular momentum and relativistic instability affect the outcome \cite{FengYuSeeds}. A candidate history must satisfy
\begin{equation}
 \Gamma_E^{1,\mathrm{rest}}t_{\rm coll}\ll1,\qquad
 \Gamma_{\rm dep}^{1}t_{\rm coll}\ll1,\qquad
 t_{\rm coll}+t_{\rm grow}<t(z_{\rm obs})-t(z_{\rm halo}).
 \label{eq:dirac-seed-time-budget}
\end{equation}
Here $\Gamma_E^{1,\mathrm{rest}}$ is the rate of kinetic-energy exchange between $\psi_1$ and the other populations, and $\Gamma_{\rm dep}^1$ is the rate of depletion of $\psi_1$. The times $t_{\rm coll}$ and $t_{\rm grow}$ describe collapse and subsequent seed growth. Determining the seed mass requires following the late core evolution, including fermion degeneracy and relativistic gravity. Connecting that seed to an observed Little Red Dot then requires modelling gas accretion and visible emission \cite{FengYuLRD,MadauSupercritical}.

\section{Conclusions}
\label{sec:result}
The hidden $E_6$ model illustrates a strategy for finding vacua with large scalar representations. For the parallel two-$\mathbf{351}$ alignment, the exact curvature relation in Eq.~\eqref{eq:full-orientation-obstruction} rules out an isolated minimum for the stated quartic and phase interactions. Adding a neutral symmetric $\mathbf{351}^{\prime}$ gives a bounded potential with a locally stable benchmark that minimizes the energy within its residual-symmetry subspace. Searching outside that subspace finds a second local minimum with lower energy and a different residual algebra. This establishes that the spectrum and phenomenology of the $H$-preserving benchmark are conditional on the selection and lifetime of that vacuum. The ground state and quantum stability remain to be established.

The surviving relative phase is an exact Nambu--Goldstone field of the anomaly-free $Q$ symmetry when its explicit breaking vanishes. Perturbative corrections and $E_6$ instantons at any topological charge preserve that conclusion. The dressed interactions used for an ultralight mass explicitly break $Q$ and supply an independent effective coefficient. The parent-operator bound in Eq.~\eqref{eq:quality-parent-bound} quantifies one of the ultraviolet suppressions required. The physical gauge amplitudes also require the matched fermion contribution together with the Wess--Zumino counterterm; the vanishing total global anomaly is a Ward identity for their sum.

The tree-level Yukawa matrix leaves 69 hidden Weyl fermions massless in the $H$-preserving benchmark and 57 in the lower vacuum. In either case the light states would give excessive radiation under the stated assumptions of earlier visible--hidden equilibrium and an unconfined hidden sector. A colder hidden sector can satisfy a radiation-density bound, conditional on suppressed portal equilibration and no subsequent hidden entropy injection. Additional fermion masses, strong dynamics or a different reheating history must be treated explicitly if they change that spectrum.

The projection fixes the relation between the phase normalization and portal production. Its equal-vev form survives the scalar completion, while the production strength is distributed among three mixed scalar singlets with distinct thresholds. Coherent abundances require a specified breaking coefficient and initial angle and must be checked against dwarf-galaxy, structure and black-hole constraints. The seeding mechanism for Little Red Dots requires a dominant effectively collisionless population to assemble the extended halo and a smaller self-interacting population to carry heat out of its own core. The Dirac-singlet benchmark gives the transport cross sections and local relaxation rate for this configuration. The halo calculation must establish whether the initially dilute self-interacting component can form a self-gravitating core through its own scattering. Predicting black-hole seed masses and Little Red Dot abundances also requires deriving the primordial populations and following collapse and accretion. The established results are the vacuum obstruction and completion, the exact symmetry and light spectrum, and the gauge and production relations of the stated effective theory.

In planned work we will calculate the thermal contributions to the relic abundances and momentum distributions of the hidden populations, including the axion-like quanta of the axi-Higgs. The rates for these quanta will be derived from amplitudes that combine the Wess--Zumino interactions with the fermion contributions required by anomaly matching in Sec.~\ref{sec:relative-anomaly-matching}. We will use these distributions to initialize coupled kinetic equations for the two dark matter populations in their common gravitational potential. The collision terms will describe self-scattering and the portal processes that can change the particle populations or transfer energy. This calculation will test whether the interacting component develops a self-gravitating core and proceeds to gravothermal collapse. It will also quantify the contribution of the axi-Higgs population.

A separate computational study will document the numerical methods used here to construct the gauge-invariant scalar potential, find stationary configurations and test stability against all physical scalar fluctuations. It will also apply these methods to other gauge portal models.

\section*{Acknowledgments}

This work is partially supported by INFN through Iniziativa Specifica QGSKY. We constructed the $E_6$ generators and invariant tensors from the $SU(3)^3$ decomposition using the conventions documented for \texttt{E6Tensors} in Ref.~\cite{E6Tensors}.

\appendix
\section{Tensor duality, the massive vector and anomaly cancellation}
\label{sec:stueckelberg}
A four-dimensional antisymmetric two-form has one propagating degree of freedom when massless. Locally it can be described instead by a pseudoscalar. This duality relates a higher-form gauge description to the shifting scalar in the four-dimensional mass sector. It does not by itself imply that the low-energy spectrum contains a physical uneaten axion.

Let $C_{\mu\nu}$ denote the antisymmetric tensor, $B_\mu$ the canonically normalized Abelian vector, and $F^X_{\mu\nu}=\partial_\mu B_\nu-\partial_\nu B_\mu$. The three-form field strength is
\begin{equation}
 H_{\mu\nu\rho}=\partial_\mu C_{\nu\rho}
 +\partial_\nu C_{\rho\mu}+\partial_\rho C_{\mu\nu},
 \qquad \varepsilon^{0123}=+1.
\end{equation}
The tensor mass sector is written as
\begin{equation}
 \mathcal L_{\rm tensor}=-\frac14F^X_{\mu\nu}F_X^{\mu\nu}
 +\frac1{12}H_{\mu\nu\rho}H^{\mu\nu\rho}
 +\frac{M}{4}\varepsilon^{\mu\nu\rho\sigma}C_{\mu\nu}F^X_{\rho\sigma}.
 \label{eq:tensor-stueckelberg}
\end{equation}
Following the three-form formulation of Ref.~\cite{CF}, we treat $H_{\mu\nu\rho}$ as independent and introduce $b$ as the multiplier imposing its Bianchi identity. After integrating the tensor--vector coupling by parts, the first-order Lagrangian is
\begin{equation}
 \mathcal L_{\rm first}=-\frac14F^X_{\mu\nu}F_X^{\mu\nu}
 +\frac1{12}H_{\mu\nu\rho}H^{\mu\nu\rho}
 +\frac{M}{6}\varepsilon^{\mu\nu\rho\sigma}B_\mu H_{\nu\rho\sigma}
 +\frac{b}{6}\varepsilon^{\mu\nu\rho\sigma}\partial_\mu H_{\nu\rho\sigma}.
 \label{eq:first-order}
\end{equation}
Varying $b$ imposes the Bianchi identity,
\begin{equation}
 \varepsilon^{\mu\nu\rho\sigma}\partial_\mu H_{\nu\rho\sigma}=0,
 \qquad \dd H=0.
 \label{eq:tensor-bianchi-identity}
\end{equation}
Its local solution recovers the original tensor description:
\begin{equation}
 H=\dd C.
 \label{eq:tensor-bianchi-solution}
\end{equation} To obtain the scalar description instead, integrate the last term by parts. The terms involving $H$ become
\begin{equation}
 \frac1{12}H_{\mu\nu\rho}H^{\mu\nu\rho}
 -\frac16\varepsilon^{\mu\nu\rho\sigma}
 (\partial_\mu b-MB_\mu)H_{\nu\rho\sigma}.
\end{equation}
The $H$ equation is algebraic and gives
\begin{equation}
 H^{\nu\rho\sigma}=\varepsilon^{\mu\nu\rho\sigma}
 (\partial_\mu b-MB_\mu).
\end{equation}
Using the contraction identity
\begin{equation}
 \varepsilon^{\mu\nu\rho\sigma}\varepsilon_{\alpha\nu\rho\sigma}=-6\delta^\mu_\alpha,
 \label{eq:epsilon-contraction}
\end{equation}
substitution yields
\begin{equation}
 \mathcal L_{\rm St}=-\frac14(F^X)^2
 +\frac12(\partial_\mu b-MB_\mu)^2.
 \label{eq:scalar-stueckelberg}
\end{equation}
This is invariant under $B_\mu\to B_\mu+\partial_\mu\omega$ and $b\to b+M\omega$. In the gauge $b=0$ the second term is $M^2B_\mu B^\mu/2$: the pseudoscalar becomes the longitudinal polarization of the vector.

Cosmological scalar--vector--two-form systems with a topological $BF$ interaction provide another application of this local massive-vector description \cite{RezaeiAkbariehKucukakca2026BF}. That study treats scalar-dependent kinetic functions, isotropic triplets of forms, Noether symmetries and perturbative stability. The shared tensor--vector duality concerns the local mass sector. The surviving physical pseudoscalar considered here additionally requires the hidden Higgs phases and their projection against the broken $E_6$ and Abelian gauge directions.

Reversing the duality orientation reverses the mixed term without changing the spectrum. The scalar and tensor forms describe the same local degrees of freedom. Their global periods require more information about the tensor gauge field and its allowed charges. This local calculation does not fix a compactification or a string scale. D-brane models give explicit examples of the underlying tensor coupling \cite{GIIQ}.

The remaining anomaly problem is confined to the hidden current. The visible contribution is zero by Eq.~\eqref{eq:visible-BL-anomalies}, whereas the hidden chiral fermions have a nonzero anomalous $X$ current. Their quantum effective action changes under the same gauge transformation. Define the hidden anomaly coefficients by the variation itself,
\begin{equation}
 \delta_\omega\Gamma_{\rm ferm}=
 \int\dd^4x\,\omega(x)
 \left[\mathfrak a_G\tr(F_G\widetilde F_G)
       +\mathfrak a_X F_X\widetilde F_X
       +\mathfrak a_R R\widetilde R\right].
 \label{eq:anomaly-variation}
\end{equation}
This definition absorbs the gauge couplings, trace normalizations and the chosen distribution of triangle anomalies into $\mathfrak a_i$. Here $\widetilde F^{\mu\nu}=\varepsilon^{\mu\nu\rho\sigma}F_{\rho\sigma}/2$. Choosing the non-Abelian Ward identities to be preserved, the compensating terms have the schematic normalization
\begin{equation}
 \mathcal L_{\rm WZ}=-\frac{b}{M}
 \left[\mathfrak a_G\tr(F_G\widetilde F_G)
       +\mathfrak a_X F_X\widetilde F_X
       +\mathfrak a_R R\widetilde R\right].
 \label{eq:WZ}
\end{equation}
Since $\delta b/M=\omega$, the variation of this expression cancels Eq.~\eqref{eq:anomaly-variation}. Other local counterterms may be needed in a different anomaly convention. The cancellation restores the gauge Ward identities of the effective theory; it does not mean that the fermion charge traces vanish. The result is an effective gauge theory with higher-dimensional interactions, whose cutoff and high-energy completion must still be specified.

Before Higgs mixing, the two transverse vector polarizations and the tensor's scalar degree of freedom form a massive vector with three polarizations. A separate physical pseudoscalar requires additional phases. We now identify those phases in the GUT model.

\section{The effective ultralight field}
\label{sec:HOTW}
Once a symmetry-breaking curvature is specified and $H\lesssim m_\chi$, the approximately harmonic coherent field has the scale-dependent effective sound speed \cite{HOTW,Marsh}
\begin{equation}
 c_{s,\chi}^2(k,a)=\frac{k^2}{4m_\chi^2a^2+k^2},\qquad
 \lambda_{\rm dB}=\frac{2\pi}{m_\chi v}
 \simeq1.2\,\mathrm{kpc}
 \left(\frac{10^{-22}\,\mathrm{eV}}{m_\chi}\right)
 \left(\frac{100\,\mathrm{km\,s^{-1}}}{v}\right).
 \label{eq:axihiggs-effective-sound-speed}
\end{equation}
These wave effects depend on the supplied mass and coherent fraction. Expanding the periodic interaction with $U=m_\chi^2\chi^2/2+\lambda_\chi\chi^4/4!+\cdots$ gives
\begin{equation}
 m_\chi^2=\frac{C_1+4C_2}{F^2},\qquad
 \lambda_\chi=-\frac{C_1+16C_2}{F^4}.
 \label{eq:quartic}
\end{equation}
A dominant first harmonic therefore has an attractive self-interaction $\lambda_\chi=-m_\chi^2/F^2$. Both parameters enter comparisons with ultralight-field constraints.

\section{Conditional coherent abundance}
\label{sec:coherent-abundance}
The dilute particle-production channels derived in Sec.~\ref{sec:portal-vertices} do not supply the very large cold occupation required for an ultralight component. A classical displacement of the periodic field can produce the required energy density without treating the condensate as a gas of particles. Its evolution and redshift during radiation domination follow the standard misalignment calculation \cite{HOTW,Marsh}; Ref.~\cite{GatewayPortal} applies the same constant-mass solution to the hidden-$SU(2)$ example. The model-specific quantity here is the phase width $F$ derived from the hidden Higgs and St\"uckelberg system. We assume that the physical phase already exists, that the periodic potential has reached its zero-temperature form before oscillations start and that the first harmonic dominates. With this normalization, the canonical displacement is
\begin{equation}
 \chi=F\Theta.
 \label{eq:misalignment-field-definition}
\end{equation}
The homogeneous phase equation and the early-time initial conditions are
\begin{equation}
 \ddot\Theta+3H\dot\Theta+m_\chi^2\sin\Theta=0,
 \qquad \Theta(t\ll t_{\rm osc})=\Theta_i,
 \qquad \dot\Theta(t\ll t_{\rm osc})=0.
 \label{eq:misalignment-eom}
\end{equation}
For a dominant first harmonic the periodic identification and our representative interval are
\begin{equation}
 \Theta\equiv\Theta+2\pi,\qquad -\pi<\Theta\leq\pi.
 \label{eq:misalignment-angle-period}
\end{equation}
No thermal correction to the potential is introduced. This constant-mass approximation isolates the consequence of the phase normalization derived in Eq.~\eqref{eq:F}.

The onset occurs well inside radiation domination for the masses considered here. In the harmonic approximation the restoring force is linear in the angle. The expansion rate and dimensionless time variable are
\begin{equation}
 H=\frac1{2t},\qquad x=m_\chi t.
 \label{eq:misalignment-radiation-variables}
\end{equation}
The solution regular at the origin of this time variable is \cite{Marsh}
\begin{equation}
 \Theta(x)=2^{1/4}\Gamma\!\left(\frac54\right)\Theta_i
 x^{-1/4}J_{1/4}(x).
 \label{eq:RD-bessel}
\end{equation}
It approaches
\begin{equation}
 \Theta(x)\simeq C_R\Theta_i x^{-3/4}
 \cos\!\left(x-\frac{3\pi}{8}\right),
 \qquad
 C_R=\frac{2^{3/4}\Gamma(5/4)}{\sqrt\pi}\simeq0.86.
 \label{eq:RD-late}
\end{equation}
The $x^{-3/4}$ envelope is the expected $a^{-3/2}$ decrease of a coherently oscillating massive field. Averaging its kinetic and potential energies over the rapid oscillations gives
\begin{equation}
 \overline\rho_\chi=\frac{C_R^2}{2}m_\chi^2F^2\Theta_i^2
 (m_\chi t)^{-3/2}.
 \label{eq:RD-density}
\end{equation}
Evolving this density with the measured radiation density gives the useful normalization
\begin{equation}
 \Omega_\chi h^2\simeq0.066
 \left(\frac{F}{10^{17}\,\GeV}\right)^2
 \left(\frac{m_\chi}{10^{-22}\,\mathrm{eV}}\right)^{1/2}
 \Theta_i^2\,{\cal A}(\Theta_i).
 \label{eq:misalignment-abundance}
\end{equation}
The harmonic result corresponds to a unit anharmonic factor. The numerical normalization uses the standard radiation history and the cosmological parameters of Ref.~\cite{Planck2018}. The weak $m_\chi^{1/2}$ dependence follows because a heavier field begins to redshift as matter earlier.

For a cosine potential, large initial angles delay the first descent and enhance the abundance. Direct integration of Eq.~\eqref{eq:misalignment-eom} gives
\begin{equation}
 {\cal A}(1)\simeq1.1,\qquad
 {\cal A}(3.0)\simeq4.2,\qquad
 {\cal A}(3.1)\simeq6.6.
 \label{eq:anharmonic-factors}
\end{equation}
These factors multiply the explicit $\Theta_i^2$ in Eq.~\eqref{eq:misalignment-abundance}. They apply to a settled first-harmonic cosine. For the last initial angle, the distance from the hilltop is
\begin{equation}
 \Delta\Theta_i=\pi-3.1\simeq0.042.
 \label{eq:hilltop-angle-distance}
\end{equation} In this region the delayed onset is sensitive to the detailed shape of the maximum: a second harmonic comparable with the first generically changes ${\cal A}(\Theta_i)$ by order unity and can move or split the hilltop. The shape, onset and abundance must then be recomputed from the full two-harmonic potential.

The numerical possibilities can now be stated within the four-dimensional model. We define the abundance fraction by
\begin{equation}
 f_\chi=\frac{\Omega_\chi}{\Omega_{\rm DM}}.
 \label{eq:misalignment-fraction-definition}
\end{equation}
For equal vevs in the saturated regime, we use
\begin{equation}
 F\simeq\frac{v_0}{\sqrt2},\qquad
 m_{\chi,\rm ref}=3\times10^{-19}\,\mathrm{eV}.
 \label{eq:misalignment-reference-point}
\end{equation}
The chosen mass is a normalization reference at the lower edge of the dwarf-galaxy limit discussed below. At this mass, the approximate displacements giving the observed dark-matter density in the assumed homogeneous cosine evolution are
\begin{equation}
 \begin{aligned}
 \Theta_i=1:&\quad F\simeq1.7\times10^{16}\,\GeV,
              &v_0\simeq2.5\times10^{16}\,\GeV,\\
 \Theta_i=3.0:&\quad F\simeq3.0\times10^{15}\,\GeV,
              &v_0\simeq4.2\times10^{15}\,\GeV,\\
 \Theta_i=3.1:&\quad F\simeq2.3\times10^{15}\,\GeV,
              &v_0\simeq3.2\times10^{15}\,\GeV.
 \end{aligned}
 \label{eq:full-DM-benchmarks}
\end{equation}
For a first harmonic with $v_G=v_0$ and large $M$, the required explicit-breaking coefficients are respectively $a_0\simeq7.4\times10^{-89}$, $2.5\times10^{-87}$ and $4.3\times10^{-87}$. The positive dressed-square completion has $\alpha\simeq4.5\times10^{-56}\,\GeV^2$ at this mass, independently of the common vev in the saturated limit. The two larger angles delay the descent from the potential maximum. Their perturbation spectra depend on that initial displacement and must be evolved with the actual potential. These abundance equalities alone do not establish observational viability. For a ten-percent component, each quoted width and vev is multiplied by $\sqrt{0.1}$, giving equal-vev scales of approximately $7.8\times10^{15}$, $1.3\times10^{15}$ and $1.0\times10^{15}\,\GeV$. At $v_0=10^{15}\,\GeV$ and this new reference mass, the same angles give $f_\chi\simeq0.0016$, $0.056$ and $0.095$.

The dependence on the parameters can be summarized without choosing one numerical example:
\begin{equation}
 v_{0,\rm req}\propto
 f_\chi^{1/2}m_\chi^{-1/4}
 \bigl[|\Theta_i|\sqrt{{\cal A}(\Theta_i)}\bigr]^{-1}.
 \label{eq:required-vev-scaling}
\end{equation}
Changing the reference mass from $2\times10^{-20}\,\mathrm{eV}$ to $3\times10^{-19}\,\mathrm{eV}$ at fixed fraction and angle gives
\begin{equation}
 \frac{v_{0,\rm req}(3\times10^{-19}\,\mathrm{eV})}
 {v_{0,\rm req}(2\times10^{-20}\,\mathrm{eV})}
 =15^{-1/4}\simeq0.51.
 \label{eq:misalignment-vev-rescaling}
\end{equation}
The coefficient $a_0$ must be increased when the mass is raised at the smaller required vev. Equations~\eqref{eq:coefficient-for-fixed-mass} and \eqref{eq:required-vev-scaling} specify this adjustment.

For a canonical harmonic field constituting all of the dark matter, the Lyman-$\alpha$ analysis in Ref.~\cite{RogersPeiris} gives the 95\% limit
\begin{equation}
 m_\chi>2\times10^{-20}\,\mathrm{eV}.
 \label{eq:canonical-lyman-alpha-reference}
\end{equation}
The measured sizes and stellar kinematics of Segue~1 and Segue~2 give the stronger bound $m_\chi>3\times10^{-19}\,\mathrm{eV}$ at 99\% confidence in the all-dark-matter wave-heating calculation of Ref.~\cite{DalalKravtsov}. It assumes that stellar heating by soliton fluctuations and other sources is negligible. Nonlinear Schr\"odinger--Poisson simulations support the conservative character of that heating estimate \cite{MayDalalKravtsov}. The same analysis finds $m_\chi>8\times10^{-18}\,\mathrm{eV}$ at 95\% confidence if Ursa Major~III/UNIONS~1 is a dark-matter-dominated galaxy. These results constrain the all-dark-matter examples and require the mass to exceed the appropriate lower limit. A near-maximum displacement or a substantial second harmonic requires its own perturbation calculation. The bounds for a field forming all of the dark matter cannot be assigned directly to the small coherent fraction used in the optional matter benchmark. Mixed-fraction limits and the actual transfer function then enter the comparison \cite{LiuMixedFDM}.

Black-hole superradiance supplies a separate constraint on the existence of a light scalar through extraction of spin into a bound field cloud. The relevant condition is $\omega<m\Omega_H$, where $m$ is the azimuthal mode number and $\Omega_H$ the horizon angular velocity. Its rate depends on $G M_{\rm BH}m_\chi$, the measured black-hole spin and the available growth time. For the first-harmonic potential, $\lambda_\chi=-m_\chi^2/F^2$ in the convention $U\supset\lambda_\chi\chi^4/4!$. The resulting attractive self-interaction can change cloud occupations and suppress spin extraction \cite{BaryakhtarSelfSuperradiance}. A small cosmological abundance does not remove this production mechanism. The supermassive-black-hole likelihoods of Ref.~\cite{HoofSuperradiance} probe the mass range of the illustrative examples and include the dependence on the interaction scale. Applying those likelihoods with the derived $F$ and the actual periodic potential remains part of a complete observational test; the homogeneous abundance solutions above are conditional normalizations.

If the phase is established before inflation, its fluctuations add an independent constraint. For a fraction $f_\chi$, the isocurvature amplitude is approximately \cite{HTW}
\begin{equation}
 {\cal P}_S^{1/2}\simeq
 f_\chi\left|\frac{\partial\ln\Omega_\chi}
 {\partial\Theta_i}\right|\frac{H_I}{2\pi F}.
 \label{eq:isocurvature-fraction}
\end{equation}
For the measured adiabatic amplitude, the Planck isocurvature limit gives \cite{PlanckInflation}
\begin{equation}
 \beta_{\rm iso}<0.038,\qquad
 {\cal P}_S^{1/2}\lesssim9\times10^{-6}.
 \label{eq:misalignment-isocurvature-bound}
\end{equation}
For the three all-dark-matter normalization examples in Eq.~\eqref{eq:full-DM-benchmarks}, numerical differentiation of the anharmonic solution gives $|\partial\ln\Omega_\chi/\partial\Theta_i|\simeq2.2$, $3.6$ and $8.6$. The corresponding upper scales are $H_I\lesssim4.5\times10^{11}$, $4.6\times10^{10}$ and $1.5\times10^{10}\,\GeV$. The enhancement near the maximum strengthens the inflationary condition. If the phase forms after inflation, this primordial isocurvature limit is absent, but the abundance requires a spatial average over initial angles and any defects of the full vacuum manifold.

\section{The antisymmetric representation index}
\label{app:group-index}
The one-loop running calculation used $T(\mathbf{351})=75$. It can be derived from the antisymmetric tensor construction without importing a representation table. Let $R$ have dimension $d$, traceless generators $t^a$ and normalization $\tr_R(t^at^b)=T(R)\delta^{ab}$. On $R\otimes R$, the generator is $\mathbb T^a=t^a\otimes1+1\otimes t^a$. Taking the product-space trace gives $2dT(R)\delta^{ab}$ because the cross terms contain $\tr t^a=0$.

Let $P$ exchange the two factors. The projector on the antisymmetric representation is $(1-P)/2$, and the identity $\Tr[P(A\otimes B)]=\tr(AB)$ gives
\begin{equation}
 \Tr[P\mathbb T^a\mathbb T^b]=4T(R)\delta^{ab}.
\end{equation}
Each of the four terms in the product contributes $T(R)\delta^{ab}$. Denoting the index of the antisymmetric tensor representation by $T_{\rm antisym}$, we obtain
\begin{equation}
 T_{\rm antisym}=\frac12[2dT(R)-4T(R)]=(d-2)T(R).
 \label{eq:antisymmetric-index}
\end{equation}
For $d=27$ and $T(\mathbf{27})=3$, this is $25\cdot3=75$. The symmetric product has index $(d+2)T(R)=87$; subtracting the index three of its 27-dimensional summand leaves 84 for the distinct symmetric 351-dimensional representation. This checks why the prime convention matters to running.

\section{Full gauge orbit and transverse quartic restriction}
\label{app:full-orbit}
The explicit tensor action of Ref.~\cite{E6Tensors} provides a reproducible calculation beyond the selected radial and phase coordinates. Represent a fundamental vector by three matrices $L,M,N$ transforming as $(\mathbf3,\overline{\mathbf3},\mathbf1)$, $(\mathbf1,\mathbf3,\overline{\mathbf3})$ and $(\overline{\mathbf3},\mathbf1,\mathbf3)$. The invariant cubic is proportional to $\det L+\det M+\det N-\tr(LMN)$. In a canonical wedge basis define
\begin{equation}
 w_\alpha=L_{\alpha3}\wedge N_{3\alpha},\qquad
 e=\frac{w_1+w_2+w_3}{\sqrt3},\qquad
 q_a=\frac{2w_1-w_2-w_3}{\sqrt6},\qquad
 q_b=M_{11}\wedge M_{22}.
 \label{eq:explicit-full-orbit-tangents}
\end{equation}
Each wedge has unit norm. These formulas select the stated $(\mathbf1,\overline{\mathbf3},\mathbf3)$ component and fix the convention used for the calculation. A conjugate tensor convention conjugates the generators and vev together and leaves the orbit spectrum unchanged. Both $q_a,q_b$ are orthogonal to $e$ and to all real gauge tangents $iT_ae$. They are physical common-orientation directions of the two multiplets.

The 78 compact Hermitian generators were constructed from the $SU(3)^3$ action and the 54 generators connecting the three fundamental blocks. Their trace normalization and quadratic Casimir obey
\begin{equation}
 \tr_{27}(T_aT_b)=3\delta_{ab},\qquad
 \sum_{a=1}^{78}T_a^2=\frac{26}{3}\mathbf1_{27}.
 \label{eq:full-generator-checks}
\end{equation}
The numerical residuals are below $2\times10^{-15}$. The generators preserve the cubic, and 200 sampled commutators close in their span with residual norms below $10^{-15}$. Acting on $e$ gives the Gram matrix in Eq.~\eqref{eq:full-E6-orbit-spectrum}. An independent construction with entries in $\mathbb Q(i,\sqrt3)$ gives its rational blocks and their eigenvalues exactly; its trace is $50/3=C_2(\mathbf{351})$. This gives rank 63 and the full connected stabilizer in Eq.~\eqref{eq:full-E6-stabilizer}. Adding the opposite $X$ charges and the nonzero St\"uckelberg shift gives gauge rank 64. Possible global discrete quotients are not fixed by this Lie-algebra calculation.

For completeness, the scalar obstruction can be checked without a numerical search. Use antisymmetric matrix entries $A_{ij}$ with canonical norm $\lVert A\rVert^2=\sum_{i<j}|A_{ij}|^2=\tr(AA^\dagger)/2$, and choose the real symmetric tensor normalization $d_{imn}d_{jmn}=10\delta_{ij}$. Define
\begin{equation}
 \mathcal N_{\alpha\beta}(A)=d^{\mu\nu\alpha}d^{\sigma\tau\beta}
                  A_{\mu\sigma}A_{\nu\tau},\qquad
 \mathcal K_\gamma(A)=2d_{\alpha\beta\gamma}\mathcal N_{\alpha\beta}(A).
 \label{eq:full-quartic-covariants}
\end{equation}
A basis of the six balanced self-quartic contractions is \cite{E6Tensors,CFIT}
\begin{equation}
 \begin{aligned}
 J_1&=\lVert A\rVert^4,&
 J_2&=\tr[(AA^\dagger)^2],\\
 J_3&=d^{\mu\nu\lambda}d_{\xi\eta\lambda}
        A_{\mu\sigma}A_{\nu\tau}\bar A^{\xi\sigma}\bar A^{\eta\tau},&
 J_4&=\sum_{\alpha,\beta}|\mathcal N_{\alpha\beta}|^2,\\
 J_5&=d^{\mu\nu\alpha}d^{\sigma\beta\gamma}
       d_{\xi\eta\beta}d_{\lambda\alpha\gamma}
       A_{\mu\sigma}A_{\nu\tau}\bar A^{\xi\lambda}\bar A^{\eta\tau},&
 J_6&=\sum_\gamma|\mathcal K_\gamma|^2.
 \end{aligned}
 \label{eq:full-six-quartic-basis}
\end{equation}
Constant normalizations of these contractions are absorbed into their coefficients. At unit $e$ their values are $(1,2/3,2/3,4,2/3,0)$. The external charges require an ordinary quartic to contain equal numbers of each species and its conjugate: the $E_6$ centre first requires two tensors and two conjugate tensors, and $X$ neutrality then fixes the species counts. On parallel fields every mixed contraction therefore reduces to a balanced self-quartic evaluated on their common direction. All its coefficients can be included in the $c_i$ used in Sec.~\ref{sec:full-multiplet-stability}.

Let $n_j(s)=(e+s q_j)/\sqrt{1+s^2}$ for $j=a,b$. Direct exact contraction gives
\begin{equation}
 \begin{aligned}
 (1+s^2)^2\,[J_i(n_a(s))-J_i(e)]
   &=\begin{cases}
       (s^4+4\sqrt2s^3+8s^2)/3,&i=2,3,5,\\
       0,&i=1,4,6,
      \end{cases}\\
 (1+s^2)^2\,[J_i(n_b(s))-J_i(e)]
   &=\begin{cases}
       (4s^4-4s^2)/3,&i=2,3,5,\\
       0,&i=1,4,6.
      \end{cases}
 \end{aligned}
 \label{eq:exact-orientation-polynomials}
\end{equation}
The canonical coordinate at the origin is $\zeta_j=vs$, since the common-orientation kinetic term is $v^2(\partial s)^2/2+O(s^2)$. Twice the quadratic coefficients divided by $v^2$ give Eq.~\eqref{eq:full-orientation-obstruction}. A positive semidefinite physical Hessian requires $\mathcal C=0$. In that case the complete restricted potential is constant on both curves, to every order in $s$, for the stated degree-four interactions and the displayed norm/bilinear phase potential. Since $J_2$ itself changes on each curve, neither curve is a gauge orbit. This proves the failure of vacuum isolation within the specified class. It does not constrain different alignments, higher-dimensional operators or quantum corrections.

Direct differentiation of the full norm-and-alignment potential reproduces the masses in Eqs.~\eqref{eq:symmetric-joint-vacuum-masses} and \eqref{eq:full-relative-orientation-mass} and the flat directions of Sec.~\ref{sec:full-multiplet-stability}. Independent second differences verify the directional curvatures. As a check, take $\mu^2=E_0^2$, $u=0.7$, $\lambda_{12}=0.1$, $\alpha=0.02E_0^2$, $g_X=0.1$ and $M=7E_0$. Changing $\kappa$ from $0.4$ to $-0.1$ makes the relative-orientation modes unstable while preserving both the reduced potential and boundedness from below. The change in stability comes entirely from fluctuations away from parallel alignment.

\section{The neutral symmetric-Higgs benchmark}
\label{app:stable-benchmark}
The scalar completion is specified by its canonical tensor basis, potential and coefficients. Use the fundamental ordering $L_{11},\ldots,L_{33},M_{11},\ldots,M_{33},N_{11},\ldots,N_{33}$ of Appendix~\ref{app:full-orbit}, and let $u$ denote the unit vector along $M_{33}$. The symmetric invariant has $d_{imn}d_{jmn}=10\delta_{ij}$. Define the rank-ten symmetric matrix $J_{ij}=d_{ij k}u_k$ and a diagonal matrix $D$. Its entries are $+2$ on $L_{\alpha3}$ and $N_{3\alpha}$, $-3$ on $M_{11},M_{12},M_{21},M_{22}$, and zero on the other fundamental components. Then
\begin{equation}
 \begin{gathered}
 P=uu^{\rm T},\qquad T=\frac{DJ}{\sqrt{60}},\qquad
 \tr(PP^\dagger)=\tr(TT^\dagger)=1,\\
 \tr(PT^\dagger)=0,\qquad d_{ijk}P_{jk}=d_{ijk}T_{jk}=0.
 \end{gathered}
 \label{eq:benchmark-symmetric-tensors}
\end{equation}
$P$ is the $SO(10)$ singlet direction and $T$ is the Pati--Salam singlet in the $\mathbf{54}$. The former has a 45-dimensional stabilizer algebra, the latter a 21-dimensional one. Together with the two antisymmetric vevs they leave the fifteen-dimensional algebra in Eq.~\eqref{eq:full-E6-stabilizer}. These statements follow from the full 78-generator action.

The benchmark is specified by the following independent dimensionless coefficients:
\begin{equation}
\begin{aligned}
u_A&=\lambda_A=\kappa_A=1,
&u_\Sigma&=2.1,
&\beta&=0.12,\\
\eta&=-0.053,
&a_A&=8.5\times10^{-5},\\
b_\Sigma&=-0.017,
&c_\Sigma&=0.025,\\
d_\Sigma&=-0.020,
&\delta_\beta&=9.6\times10^{-4},
&\frac\tau{E_0}&=\frac{\sqrt2}{4}.
\end{aligned}
\label{eq:benchmark-independent-inputs}
\end{equation}
Here $\rho=-2(\beta-\delta_\beta)$. Every decimal in Eq.~\eqref{eq:benchmark-independent-inputs} is an exact input defining a rational number, so no additional digits are needed to reconstruct this point. The three tadpole equations at $(v,s,w)=(1,4,1)E_0$ determine $m_A^2,m_\Sigma^2$ and $\sigma$. Their two-figure approximate values are
\begin{equation}
\frac{m_A^2}{E_0^2}\simeq-0.050,
\qquad\frac{m_\Sigma^2}{E_0^2}\simeq-35,
\qquad\frac\sigma{E_0}\simeq-0.011.
\label{eq:benchmark-dependent-inputs}
\end{equation}

Differentiating the following exact restriction gives the defining stationarity equations:
\begin{equation}
 \begin{aligned}
 V_{\rm sing}(v,s,w)={}&m_A^2v^2+\frac{m_\Sigma^2}{2}(s^2+w^2)
 +\left(\frac{u_A}{2}+\frac{\lambda_A}{4}+\frac{a_A}{3}\right)v^4
 +\frac{u_\Sigma}{4}(s^2+w^2)^2
 +\frac\eta2v^2(s^2+w^2)\\
 &+b_\Sigma\left(\frac{s^4}{4}+\frac{7w^4}{240}\right)
 +c_\Sigma\left(\frac{s^2w^2}{2}+\frac{7w^4}{240}\right)
 +d_\Sigma\left(s^2w^2+\frac{w^4}{4}\right)\\
 &+\frac{2\beta+\rho}{30}v^2w^2
 -\frac{\tau}{\sqrt2}v^2s+\frac{3\sigma}{2\sqrt2}sw^2.
 \end{aligned}
 \label{eq:benchmark-singlet-potential}
\end{equation}
The canonical real-amplitude coordinates are $\sqrt2\,\delta v$, $\delta s$ and $\delta w$. The full gradients vanish because these are the only real singlets that can acquire a linear term at the exchange-symmetric, real configuration; the imaginary gradients vanish for the real coefficients and phases chosen here. Direct differentiation in all field components confirms the vanishing.

The boundedness certificate uses elementary tensor inequalities. Antisymmetric singular values occur in pairs, so
\begin{equation}
 \begin{gathered}
 0\leq\tr(R_r^2)\leq2I_r^2,\qquad
 0\leq\tr(R_\Sigma^2)\leq z^2,\qquad
 0\leq T_3\leq10z^2,\qquad 0\leq T_4\leq100z^2,\\
 \left|\tr(A^{(1)}\Sigma^\dagger A^{(2)}\Sigma^\dagger)\right|
 \leq\frac12\sum_r\tr(R_rR_\Sigma).
 \end{gathered}
 \label{eq:benchmark-quartic-bounds}
\end{equation}
The last inequality follows from Cauchy--Schwarz and the symmetry of $\Sigma$. In the benchmark, $\beta-|\rho|/2\simeq9.6\times10^{-4}>0$, so the mixed trace terms have a nonnegative sum. The other positive terms give
\begin{equation}
 V_4\geq\frac34W^2+U_\Sigma z^2+\eta Wz,\qquad
 U_\Sigma=u_\Sigma-|b_\Sigma|-100|d_\Sigma|
 \simeq0.083.
 \label{eq:benchmark-coercive-quartic}
\end{equation}
Here $I_1^2+I_2^2+I_1I_2\geq3W^2/4$ and the positive $c_\Sigma T_3$ has been retained through its lower bound. The smallest eigenvalue of the quadratic form in $(W,z)$ exceeds $0.081$, proving Eq.~\eqref{eq:stable-extension-boundedness}. The cubic terms satisfy $|V_3|\leq C(W+z)^{3/2}$ for a finite constant $C$ by the same $d$-tensor bounds. Thus $V_0$ is coercive on the entire tensor space.

For the complete mass calculation, choose the 351 independent antisymmetric coefficients of each $A$ and an orthonormal basis of the symmetric $d$-traceless tensors. The latter is obtained by projecting the 378 canonical symmetric components orthogonally to the 27 independent $d$ traces. Real and imaginary components are canonical after the usual factor $1/\sqrt2$. Diagonalizing the CP-even and CP-odd blocks gives the spectrum and positive gap reported in Sec.~\ref{sec:stable-extended-vacuum}. The numerical checks compare the analytic curvatures with finite differences, including along the lightest modes. Restoring $b$ and including the quadratic contribution of Eq.~\eqref{eq:stable-extension-phase-square} gives 2043 positive eigenvalues and 64 gauge zero eigenvalues for this vacuum.

The positive spectrum occupies an open region around the exact input point. To examine its width, we vary one independent coefficient at a time and solve the three tadpole equations again, retaining the specified vevs. Conservative slices of the connected interval with a positive physical Hessian are
\begin{equation}
\begin{aligned}
-1.6\times10^{-3}&<\delta b_\Sigma<7.6\times10^{-4},\\
-1.6\times10^{-3}&<\delta c_\Sigma<5.7\times10^{-4},\\
-5.0\times10^{-4}&<\delta d_\Sigma<5.2\times10^{-4},\\
-1.3\times10^{-5}&<\delta a_A<3.4\times10^{-5}.
\end{aligned}
\label{eq:benchmark-local-stability-slices}
\end{equation}
Each row is a separate one-parameter variation. The short exact input point lies inside a finite interval in every displayed direction; the smallest heavy masses remain sensitive to the smaller coefficient $a_A$. The boundedness certificate remains positive on these displayed slices. A sufficient simultaneous neighbourhood follows from Weyl's eigenvalue inequality. With $\mathcal H_\perp$ the Hessian on the physical heavy-scalar subspace and $\theta_i$ the twelve independent dimensionless inputs, the affine dependence after imposing the tadpoles gives
\begin{equation}
\|\delta\mathcal H_\perp\|_2
\leq\sum_i |\delta\theta_i|
\left\|\frac{\partial\mathcal H_\perp}{\partial\theta_i}\right\|_2.
\label{eq:benchmark-stability-perturbation}
\end{equation}
The sum of the derivative norms is smaller than $2.0\times10^2E_0^2$. Thus $|\delta\theta_i|<4.0\times10^{-8}$ for every independent input retains a positive heavy spectrum with a remaining gap exceeding $8.0\times10^{-6}E_0^2$. The individual intervals are much wider in several directions. These statements characterize the classical parameter neighbourhood.

The vacuum-energy comparison can be completed throughout the full subspace preserving $H$, the connected group associated with Eq.~\eqref{eq:full-E6-stabilizer}. Each antisymmetric multiplet has one complex $H$ singlet and $\Sigma$ has two. At the benchmark, $\tau>0$, $\rho<0$ and $\sigma<0$ allow the minima of all phase-dependent terms to be attained simultaneously. At fixed $x=(r_1^2+r_2^2)/2$, the remaining dependence on $p=r_1r_2$ has the form $D_p p^2+R_p p$ on $0\leq p\leq x$, with $D_p\simeq-0.25<0$ and $R_p=\rho w^2/30-\tau s/\sqrt2\leq0$. Its minimum is at $r_1=r_2$. The residual problem is Eq.~\eqref{eq:benchmark-singlet-potential}. At fixed $s$, it is strictly convex in $v^2,w^2$ on the nonnegative quadrant. Exhausting the four boundary and interior stationary branches and isolating their real roots with rational intervals gives
\begin{equation}
 \frac{V_*}{E_0^4}\simeq-1.5\times10^2,
 \qquad
 \frac{V_{\rm next}-V_*}{E_0^4}\simeq0.78,
 \label{eq:benchmark-singlet-energy-comparison}
\end{equation}
with the remaining admissible branch lying approximately $1.1E_0^4$ above $V_*$. The exact comparison establishes the lowest energy in the complete complex $H$-preserving subspace. The full Hessian separately establishes local stability against every $H$-breaking fluctuation.

The phase calculation also uses all vev components. In the coordinate order $(\eta_1,\eta_2,\eta_0,\eta_{54},b)$, the two neutral gauge directions are
\begin{equation}
 \boldsymbol k_N=\sqrt{\frac23}g_G(v,v,-2s,w,0),\qquad
 \boldsymbol k_X=(-2g_Xv,2g_Xv,0,0,M).
 \label{eq:benchmark-neutral-gauge-directions}
\end{equation}
The cubic potentials fix $\theta_1+\theta_2+\theta_0$ and $\theta_0+2\theta_{54}$, where $\theta_0=\eta_0/s$ and $\theta_{54}=\eta_{54}/w$. The direction orthogonal to both gauge vectors and both massive phase combinations gives Eq.~\eqref{eq:stable-extension-chi}.

The $E_6$ part of the gauge mass matrix decomposes into five orthogonal sectors. Exact tensor contractions give the projectors and show that the cross term between the two symmetric-Higgs expectation values vanishes in this matrix. Equation~\eqref{eq:stable-extension-vector-spectrum} gives the resulting masses and their multiplicities, including the external $X$ mode. The benchmark thresholds follow by inserting the expectation values of Eq.~\eqref{eq:stable-extension-vevs}.

\end{document}